\documentclass[structabstract]{aa}  
\usepackage{graphicx}
\usepackage{txfonts}
\usepackage{lscape}
\usepackage{longtable}

\def\HeI{He\,{\sc i}}
\def\HeII{He\,{\sc ii}}

\def\HI{H\,{\sc i}}

\def\CII{C\,{\sc ii}}
\def\CIII{C\,{\sc iii}}
\def\CIV{C\,{\sc iv}}
\def\NI{N\,{\sc i}}
\def\NII{N\,{\sc ii}}
\def\NIII{N\,{\sc iii}}

\def\OI{O\,{\sc i}}
\def\OII{O\,{\sc ii}}

\def\NaI{Na\,{\sc i}}
\def\MgII{Mg\,{\sc ii}}
\def\FeII{Fe\,{\sc ii}}
\def\FeIII{Fe\,{\sc iii}}
\def\FeIV{Fe\,{\sc iv}}

\def\NiII{Ni\,{\sc ii}}
\def\NiIII{Ni\,{\sc iii}}
\def\NiIV{Ni\,{\sc iv}}
\def\SiIV{Si\,{\sc iv}}
\def\SiIII{Si\,{\sc iii}}
\def\SiII{Si\,{\sc ii}}
\def\SII{S\,{\sc ii}}
\def\SIII{S\,{\sc iii}}
\def\AlII{Al\,{\sc ii}}
\def\AlIII{Al\,{\sc iii}}

\def\CrIII{Cr\,{\sc iii}}
\def\CrIV{Cr\,{\sc iv}}

\def\CaII{Ca\,{\sc ii}}
\def\Lstar{$L_{\ast}$}
\def\kms {km~s$^{-1}$}

\def\Teff{$T_{\rm eff}$}
\def\logg{$\log g$}

\def\Rstar{$R_{\ast}$}

\def\Rsun {$R_{\odot}$}

\def\Mstar{$M_{\ast}$}
\def\Mdot{${\dot M}$}

\def\vinf {$v_{\rm \infty}$}
\def\vpho {$v_{\rm phot}$}

\def\vsini {$v_{\rm e}\sin i$}
\def\vmac {$v_{\rm macro}$}
\def\Vt {$v_{\rm{turb}}$}

\def\kms{\mbox{\rm km$\;$s$^{-1}$}}
\def\vesini{\ensuremath{v_{\rm e}\sin{i}}}

\begin{document}
\title{On the nature of the galactic  early-B hypergiants} 

\author{J.~S.~Clark\inst{1}
\and F.~Najarro\inst{2}
\and I.~Negueruela\inst{3}
\and B.~W.~Ritchie\inst{1}
\and M.~A.~Urbaneja\inst{4}
\and I.~D.~Howarth\inst{5}}
\institute{
$^1$Department of Physics and Astronomy, The Open 
University, Walton Hall, Milton Keynes, MK7 6AA, UK\\
$^2$Departamento de Astrof\'{\i}sica, Centro de Astrobiolog\'{\i}a, 
(CSIC-INTA), Ctra. Torrej\'on a Ajalvir, km 4,  28850 Torrej\'on de Ardoz, 
Madrid, Spain\\
$^3$Departamento. de F\'{i}sica, Ingenier\'{i}a de Sistemas y
  Teor\'{i}a de la Se\~{n}al, Universidad de Alicante, Apdo. 99, E03080
  Alicante, Spain\\
$^4$Institute for Astronomy, University of Hawaii, 2680 Woodlawn Drive, Honolulu, Hawaii 96822, USA \\
$^5$Department of Physics \& Astronomy, University College London, Gower Street, London, WC1E 6BT, UK\\
}

\abstract{}{Despite their importance to a number of astrophysical
fields, the lifecycles of very massive stars are 
still poorly defined. In order to address this shortcoming, we present
a detailed quantitative study of the physical properties of four early-B
hypergiants (BHGs) of spectral type B1-4 Ia$^+$; Cyg~OB2~\#12, $\zeta^1$ Sco, HD~190603 and BP Cru.
 These are combined with an analysis of their long-term  spectroscopic and photometric behaviour  in order to 
determine their evolutionary status.}
{Quantitative analysis of UV--radio photometric and spectroscopic
datasets was undertaken with a non-LTE model atmosphere code in
order to derive physical parameters for comparison with  apparently
closely related objects,  such as B supergiants (BSGs) and luminous
blue variables (LBVs),  and theoretical evolutionary predictions.}
{The long-term photospheric and spectroscopic datasets compiled for
the early-B HGs revealed that they are remarkably stable over long periods ($\geq$40~yrs), with the possible exception of 
$\zeta^1$~Sco prior to the 20$^{th}$ century; in contrast to the typical excursions that
characterise LBVs. Quantitative analysis of $\zeta^1$~Sco, HD~190603 and BP~Cru  yielded
physical properties intermediate between BSGs and LBVs; 
we therefore suggest that BHGs are the immediate descendants and
progenitors (respectively) of such stars, for
initial masses in the range $\sim30-60M_{\odot}$. Comparison of the
properties of $\zeta^1$~Sco with the stellar population of its host
cluster/association NGC~6231/Sco~OB1 provides further support for such
an evolutionary scenario. 
In contrast, while the wind  properties of  Cyg~OB2~\#12 are consistent with this  hypothesis, the combination of extreme
luminosity and spectroscopic mass ($\sim 110M_{\odot}$) and comparatively low temperature means it   cannot be accommodated
 in such a scheme. Likewise, despite its co-location with
several LBVs above the Humphreys-Davidson (HD) limit, the  lack of long
term variability and  its unevolved chemistry apparently excludes such an identification. 
Since such massive stars are not expected to evolve to such cool temperatures,
instead traversing an O4-6Ia$\rightarrow$O4-6Ia$^+\rightarrow$WN7-9ha
pathway, the properties of Cyg~OB2~\#12 are therefore difficult to understand under current evolutionary paradigms.
Finally, we note that as with AG Car in its cool phase, despite exceeding the HD limit, the properties of 
Cyg OB2 \#12 imply that it lies below the Eddington limit - thus we conclude that the HD limit 
does not define a region of the HR diagram inherently inimical to the presence of massive stars.

}{}
\keywords{stars:evolution - stars:early type - stars:supergiant}

\maketitle

\section{Introduction}

\begin{figure}
\label{fig-halphas}
\includegraphics[angle=0,width=7cm]{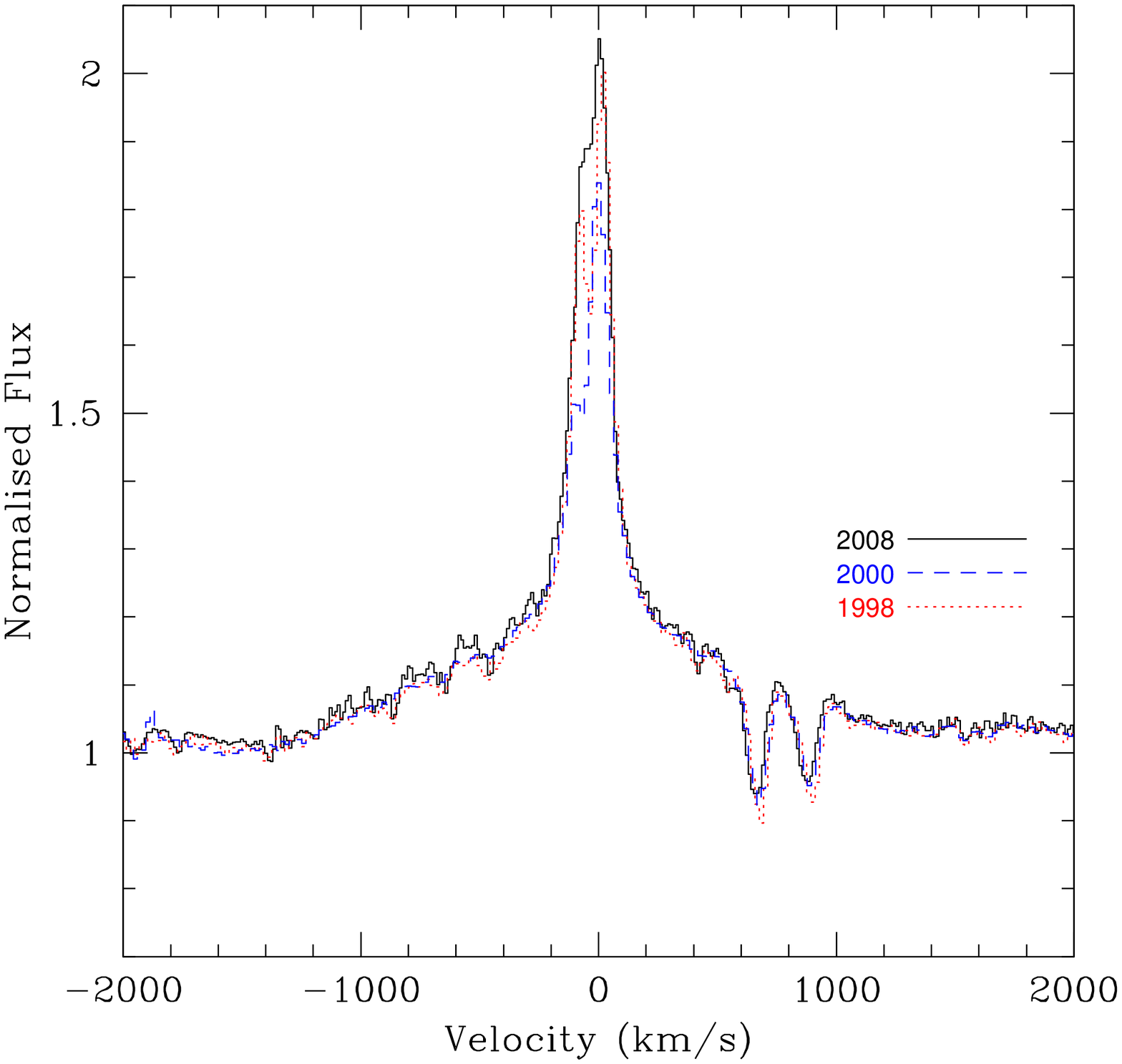}\\
\includegraphics[angle=0,width=7cm]{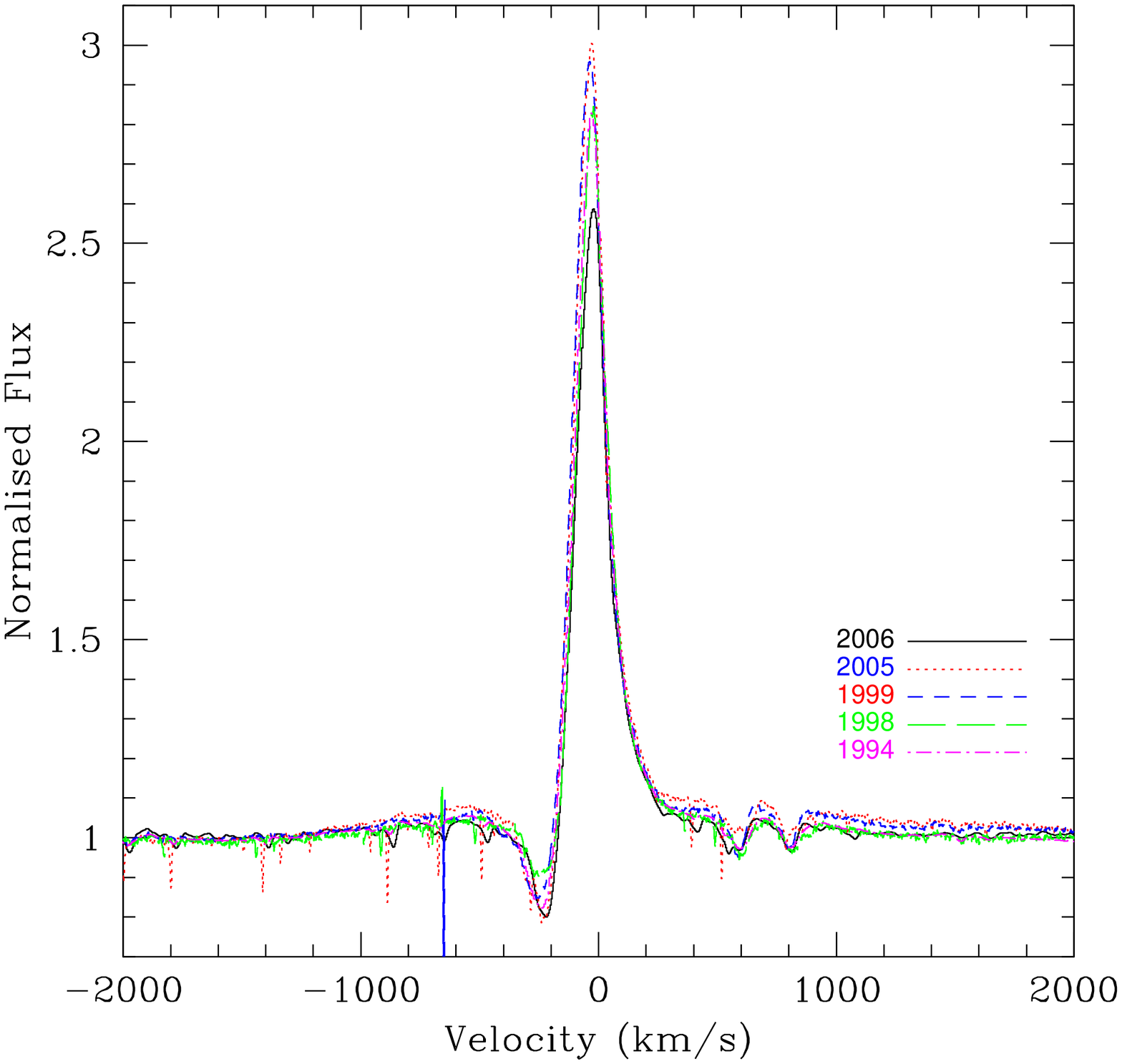}\\
\caption{Comparison of three (1999-2008) and five (1994-2009) epochs
  of H$\alpha$ observations for Cyg OB2 \#12 (upper panel) and
  $\zeta^1$ Sco (lower panel) demonstrating line profile variability at low projected 
velocities. Note the corresponding lack of variability in the
  line wings and the C\,{\sc ii}$\lambda{\lambda}$ 6578, 6582 doublet
  in the red wing.  Due to their lower resolution, spectra
  of Cyg OB2 \#12 from 1992+5 are not presented, but are
  consistent to these spectra when all are convolved to a
  matching resolution. }
\end{figure}

Very massive stars are of considerable astrophysical interest given
their role in driving galactic evolution via the copious production of
ionising radiation and the deposition of chemically processed material
and mechanical energy into the interstellar medium. They are thought
to be the progenitors of Type Ibc and  II supernovae (SNe) 
and, in low metallicity environments, gamma ray bursts (GRBs). 
The diverse nature of their core collapse SNe is also mirrored in the 
nature  of their relativistic remnants which, for a given progenitor
mass, may be either a neutron star or black hole depending on their
pre-SNe (binary mediated?) mass loss history (e.g. Ritchie et
al. \cite{magnetar}).  Moreover, with $M_V \sim -11$ during particular
phases of their post-main sequence  evolution, they have the potential
to probe the distance to - and chemical composition and star formation
history of -  external galaxies out to distances of $\sim10$ to 30~Mpc
with the current and next generation of ground based telescopes
(Kudritzki \cite{kd}).

Unfortunately, despite considerable theoretical efforts (e.g. 
Meynet \& Maeder \cite{meynet}, Heger et al. \cite{heger})
 we currently lack a comprehensive theoretical framework
to fully understand and exploit all aspects of 
these stars; a problem compounded by
comparatively weak observational constraints on a number of key
physical processes that drive massive stellar evolution, such as the
influence of (metallicity dependent) mass loss, the role of rotational
mixing and the effect of binary interactions. In this regard, a
critical test of evolutionary theory is the accurate reproduction of
the properties of stars within the short lived
transitional phase between the main sequence (MS) and H-depleted
Wolf-Rayets (WRs), 
which is populated by a diverse `zoo' of disparate
objects such as blue and yellow hypergiants (B/YHGs)
 and luminous blue variables (LBVs)/P-Cygni supergiants.  For
example, a crucial observational finding that must be replicated is
the apparent dearth of cool evolved stars at high luminosities (the
empirical Humphreys-Davidson (HD) limit; Humphreys \& Davidson \cite{hdlimit});
in this regard the YHG/RSG populations of M31 (Drout et
al. \cite{drout}) and the Galactic cluster Westerlund 1 (Clark et
al. \cite{clark10}) already pose problems for current theoretical
models.

Another manifestation of this uncertainty is an inability to
incorporate the various members of the post-MS `zoo' into a coherent
evolutionary scheme as a function of initial stellar mass. Various
authors (e.g. Langer et al. \cite{langer}, Crowther et
al. \cite{pacevol}, Martins et al. \cite{martins07}, \cite{martins08})
have proposed scenarios which have increasingly been informed by
quantitative non-LTE model atmosphere analysis of different stellar
populations, latterly located within coeval 
young massive Galactic clusters. In order to build on this approach we
present an analysis of the properties of the 
BHGs of early B spectral type, which hitherto have escaped systematic study.

The Ia$^+$ luminosity class was first applied to a group of four
highly luminous ($M_V <-8$) B `super-supergiants' within the LMC by
Keenan (\cite{keenanBHG}), which were later described as `hypergiants'
by van Genderen et al. (\cite{vGBHG}). These are further distinguished
from normal BSGs by the presence of (P Cygni) emission in the Balmer
series (cf. HD~190603; Lennon et al. \cite{lennon}). Currently, to the best of our
knowledge only 16 (candidate) BHGs have been identified within the
Galaxy, of which eight are of early ($\leq$B4) and eight of late
($\geq$B5) spectral type; these are summarised in Table 1.  Of the  early BHGs, Kaper et
al. (\cite{kaper06}) have already presented the results of a
quantitative analysis of BP Cru (= Wray 977; the mass donor in the High Mass X-ray
Binary GX301-2); in this paper we present the result of equivalent
analyses of Cyg~OB2~\#12, $\zeta^1$~Sco(=HD~152236)
and HD~190603. Foreshadowing Sect. 4, while the latter two objects
have previously been subject to similar studies (e.g. Crowther et
al. \cite{pacBSG}, Searle et al. \cite{searle}), we employ more
extensive multiwavelength datasets that enable significantly more
accurate determinations of physical parameters; for example the lack
of wind contamination in the higher Balmer and Paschen lines sampled
here permitting a more robust determination of the surface gravity.
Insufficient data exist to undertake comparable modeling of the remaining early B hypergiants, although 
we are currently in the process of obtaining the requisite observations.

Of these stars, Cyg~OB2~\#12 
is of particular interest as it has long been recognised as one of the
intrinsically brightest, and potentially most luminous stars in the
Galaxy (=Schulte 12 ; Sharpless \cite{sharpless}, Schulte \cite{schulte}), lying well above
 the HD limit and so should provide a stringent test of current
evolutionary theory.  Moreover, BHGs have previously been associated
with the LBV phenomenon by various authors (e.g. Clark et
al. \cite{clark05}) and LBVs in turn have been implicated as both
critical to the formation of WRs and as the immediate precursors of
type II SNe (Smith \& Conti \cite{sc08}).
Finally, Cyg~OB2~\#12 and $\zeta^1$~Sco are generally thought to be located 
within the Cyg OB2 and Sco OB1 associations respectively, which have
both benefited from multiple (recent) studies\footnote{e.g. Cyg OB2:
  Massey \& Thompson (\cite{massey}), Kn\"{o}dlseder
  (\cite{knodlseder}), Comer\'{o}n et al.  (\cite{comeron}), Hanson
  (\cite{hanson}) \& Negueruela et al. (\cite{iggy}) and Sco OB1:
  Reipurth (\cite{reipurth}), Sana et al.
  (\cite{sana06},\cite{sana07}, \cite{sana08}) and Raboud et
  al. (\cite{raboud}).} and hence in principal should help in the
assessment of both the nature of their progenitors as well as their
placement in a post-MS evolutionary sequence.

The paper is structured as follows. In Sect. 2 we briefly
detail the new observations of the 3 BHGs analysed  in this work (Cyg~OB2~\#12, $\zeta^1$~Sco
and  HD~190603; we utilise
 the dataset and analysis of Kaper et al. (\cite{kaper06}) for BP~Cru),
while summarising
the consolidated observational datasets in Sect. 3. This enables us
to assess the degree of short and long-term variability demonstrated
by the program stars as well as providing the spectra and
optical--radio spectral energy distributions (SEDs) for the non-LTE model atmosphere quantitative
analysis. The results of this are presented in Sect. 4, discussed in
Sect. 5 and conclusions presented in Sect. 6.
 Finally, an extensive summary of the (historical) observational properties of Galactic 
BHGs is documented in Appendix A, the spectropolarimety of  Cyg OB2 \#12 and $\zeta^1$ Sco 
discussed in Sect. 4 is presented in Appendix B and Appendix C contains an analysis of the properties
of the host cluster (NGC~6231) and association (Sco OB1) of  $\zeta^1$ Sco.

\begin{table*}
\begin{center}
\caption[]{Summary of Galactic BHG properties.}
\begin{tabular}{lccccc}
\hline
\hline
Star & Spectral & \multicolumn{2}{c}{Co-ordinates} & V band & Cluster/   \\
     & Type     &     R.A. & Decl. & (mean)                 & Association \\
\hline
BP Cru       & B1 Ia$^+$   & 12 26 37.56 & $-$62 46 13.2 & 10.68 & - \\
HD 169454    & B1 Ia$^+$   & 18 25 15.19 & $-$13 58 42.3 & 6.65 & - \\
$\zeta^1$ Sco& B1.5 Ia$^+$ & 16 53 49.73 & $-$42 21 43.3 & 4.78 & Sco OB1 \\
HD 190603    & B1.5 Ia$^+$ & 20 04 36.17 & +32 13 07.0 & 5.66 & - \\
HD 80077     & B2.5 Ia$^+$ & 09 15 54.79 & $-$49 58 24.6 & 7.57 & Pismis 11 \\
Cyg OB2 \#12 & B3-4 Ia$^+$ & 20 32 40.96 & +41 14 29.3 & 11.47 & Cyg OB2 \\
Wd1-5        & WNL/B Ia$^+$& 16 46 02.97 & $-$45 50 19.5 & 17.49 & Wd1 \\
Wd1-13       & WNL/B Ia$^+$& 16 47 06.45 & $-$45 50 26.0 & 17.19 & Wd1 \\
Wd1-7        & B5 Ia$^+$   & 16 46 03.62 & $-$45 50 14.2 & 15.57 & Wd1 \\
Wd1-33       & B5 Ia$^+$   & 16 47 04.12 & $-$45 50 48.3 & 15.61 & Wd1 \\
HD 183143    & B7 Iae      & 19 27 26.56 & +18 17 45.2 & 6.92  & - \\
HD 199478    & B9 Iae      & 20 55 49.80 & +47 25 03.6 & 5.73 & - \\
HD 168625    & B8 Ia$^+$   & 18 21 19.55 & $-$16 22 26.1 & 8.44 & - \\
Wd1-42a      & B9 Ia$^+$   & 16 47 03.25 & $-$45 50 52.1 &  - &  Wd1 \\
HD 168607    & B9 Ia$^+$   & 18 21 14.89 & $-$16 22 31.8 & - & -\\
HD 160529    & B8-A9 Ia$^+$& 17 41 59.03 & $-$33 30 13.7 & - & -\\     
\hline
\end{tabular}
\end{center}
{Note that no $V$-band magnitude is 
 available for Wd1-42a, while both HD 168607 and 160529 are known large 
amplitude photometric variables (Sect. A.6).}
\end{table*}

\section{Observations and data reduction}

In order to accomplish the goals of the paper we have compiled extensive 
datasets for Cyg~OB2~\#12, $\zeta^1$~Sco and HD~190603, utilising both new 
and 
published observations. A full presentation of the data for these (and other) 
BHGs may be  found in Appendix A, while we briefly describe the new 
observations undertaken and data reduction employed below. In appendix B we
also present the results of analysis of previously unpublished 
spectropolarimetric observations of Cyg~OB2~\#12 and  $\zeta^1$~Sco; while 
these are not directly employed in the quantitative modeling of these stars they serve as a useful check on the geometry of the cirumstellar environment of 
these stars.

\subsection{Cyg~OB2~\#12}

New high resolution and
S/N spectra of Cyg~OB2~\#12 were obtained through the William Heschel
Telescope (WHT) service programme on 2008 July 28, employing the ISIS
double-beam spectrograph equipped with the R1200B and R1200R gratings;
a summary of the resultant wavelength ranges is given in Table
2. These observations were supplemented with a further five epochs of
previous unpublished archival observations dating from 1992-2007;
these too are summarised in Table 2. All spectra were reduced in a
consistent manner using the Starlink packages Figaro and Kappa, with
selected regions presented in Figs. 1-2 and A.1. Continuum normalisation was 
accomplished via spline fitting and division, yielding uncertainties of  a few 
percent at most. Likewise, the RMS on the fits to the arc spectra are of the 
order of $\sim$0.03 pixels, resulting in a negligible error for wavelength 
callibration.

Unfortunately the distance and reddening to Cyg~OB2~\#12 preclude UV
observations, so the 4000--9000\AA\ 
spectrum obtained in 2008 formed the primary dataset employed in this
study. This was supplemented with the 1$\mu$m spectrum of Conti \&
Howarth (\cite{conti}), which encompasses
the important He\,{\sc i} 1.083$\mu$m transition, as well as the flux
callibrated ISO-SWS spectrum presented by
Whittet et al. (\cite{whittet}) and an archival Spitzer Space
Telescope IRS spectrum (Houck et al. \cite{houck}). An optical --
radio spectral energy distribution (SED) was constructed from
continuum flux measurements from the literature (Tables
A.1--A.3). While these observations were not contemporaneous, as
demonstrated in Sect. A.1 there is no evidence for substantial
variability/evolution over the timeframe spanned by these
observations. Hence we are confident that such an approach is well
justified and this is supported {\em a posteriori} by the excellent fits to the
combined dataset.

\subsection{$\zeta^1$~Sco}

A total of 18 new spectra have been utilised for the analysis of the
long-term behaviour of $\zeta^1$~Sco. Unpublished ESO 2.2m/FEROS
spectra from 1998-9 (Program ID 063.H-0080; PI:Kaufer) and 2005
(Program ID 075.D-0103(A); PI:Dufton) were kindly provided by Otmar
Stahl (2010; priv. comm.) and Phillip Dufton (2010; priv. comm.)
respectively. Data reduction was accomplished via the custom pipeline
described in Stahl et al. (\cite{feros}). Continuum normalisation was
accomplished by first dividing the spectra by the instrumental
response 
curve, with a subsequent division by a spline function; as before, the errors
in this procedure are of the order of a few percent. The spectrum
from 2006 February 17 was obtained by us with the ESO NTT/EMMI;
standard reduction procedures were employed and are discussed in
Negueruela et al. (in prep.). Finally new blue and red end spectra
from 2009 March 06 were obtained from the ESO archive (Program ID
082.C-0566(A); PI:Beletsky). These were obtained with the VLT/UVES
with cross disperser gratings CD1, 2 and 3 and were reduced using the
standard ESO pipeline.  A full summary of these new observations is
provided in Table 2 and selected spectra are presented in Figs. 1-2
and A.1.  The full dataset used for modeling also includes archival
IUE and ISO spectra, as well as published photometry summarised in
Tables A1-3.

\subsection{HD~190603}

Finally, we make  use of solely archival data for the quantitative
analysis of HD190603. As with the other objects, the sources of data
used in constructing the SED are given in Tables A1-3. Spectroscopic
data were taken from Lennon et al. (3950-5000{\AA}+ H$\alpha$;
\cite{lennon}), Rivinius et al. (4050-6800{\AA}; \cite{rivinius}),
Andrillat et al. (8390-8770{\AA}) as well as one from the IACOB
database (3710-6800{\AA}; Sergio Sim\'on-D\'{\i}az, priv. comm. 2011).

\begin{figure}

\vspace{-.3cm}

\includegraphics[angle=90,width=9.3cm]{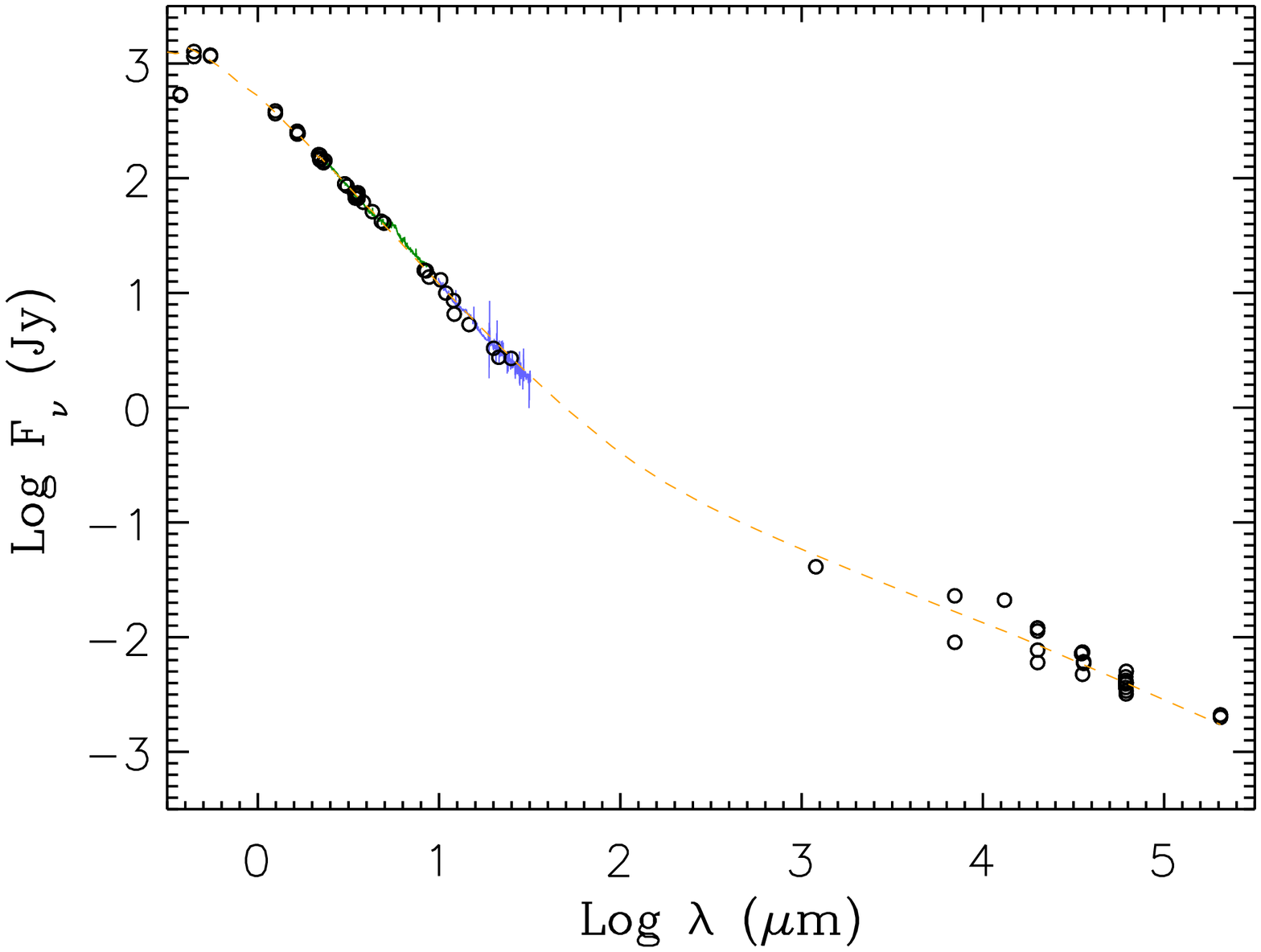}

\vspace{-.7cm}

\includegraphics[angle=90,width=9.3cm]{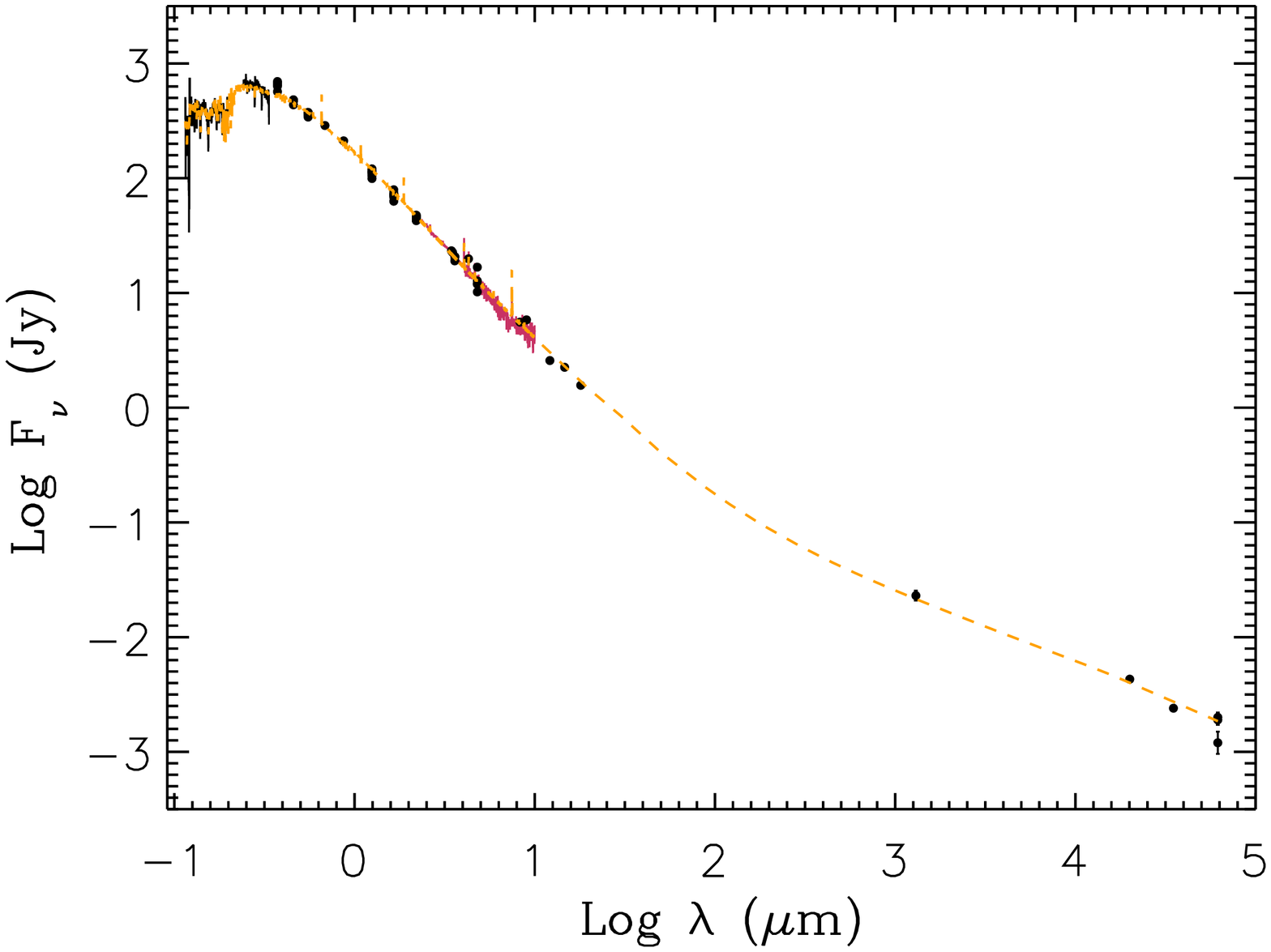}

\vspace{-.7cm}

\includegraphics[angle=90,width=9.3cm]{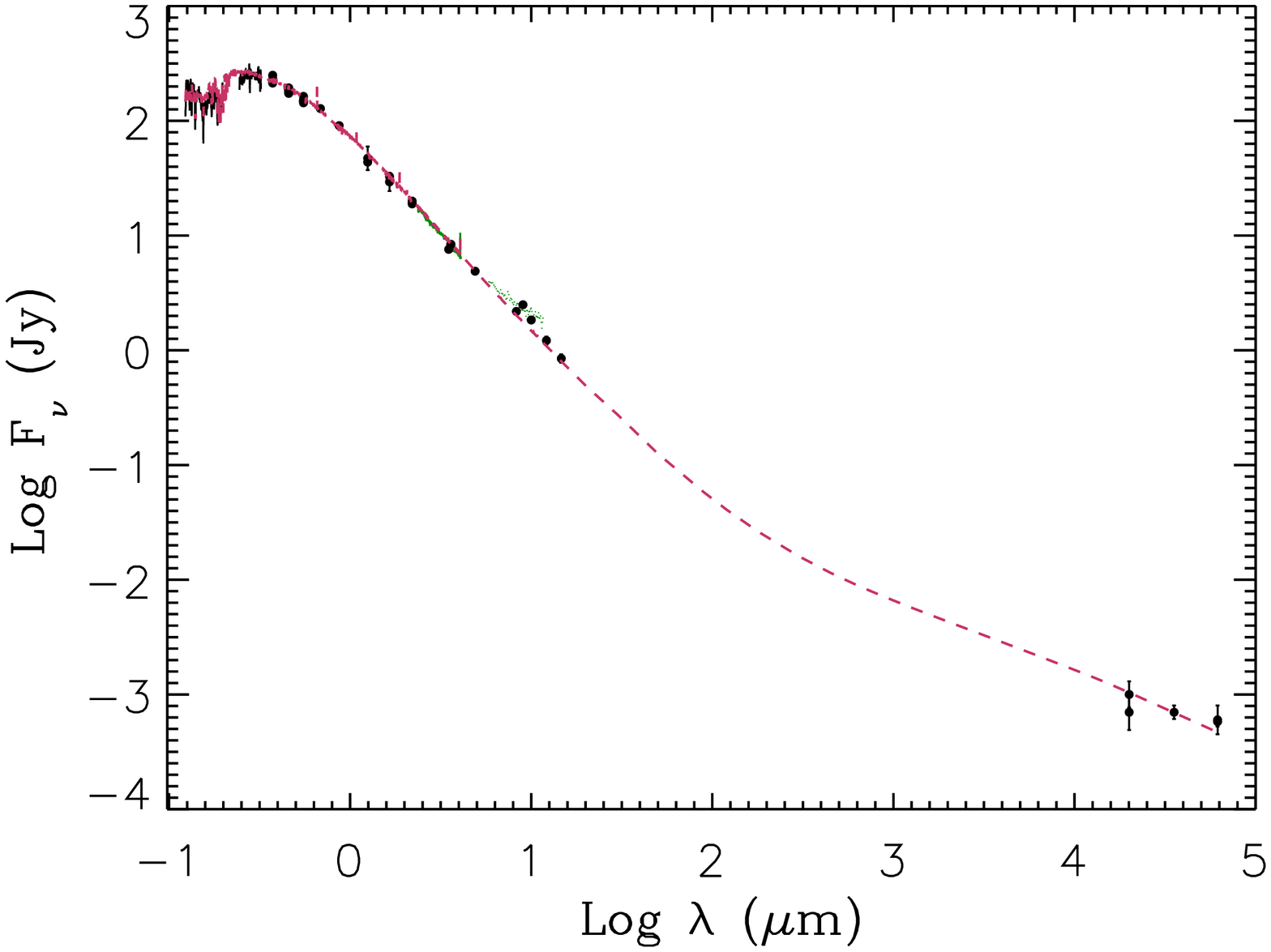}

\vspace{-.2cm}

\caption{Comparison of synthetic spectra (yellow dashed line) with the  optical to radio SEDs of CygOB2 \#12 (top panel),
$\zeta^1$~Sco (middle panel) and HD~190603 (bottom panel). Note the lack of a near-mid IR excess signalling the absence of significant amounts
 (warm) circumstellar dust. Physical parameters employed in the modeling -  such as $L_{\rm bol}$ and  $E(B-V)$ - 
may be found in Table 4 and Sects. 4.1 and  4.2.}
\label{fig-cont-fits}
\end{figure}

\begin{figure}
\resizebox{\hsize}{!}{\includegraphics[angle=270]{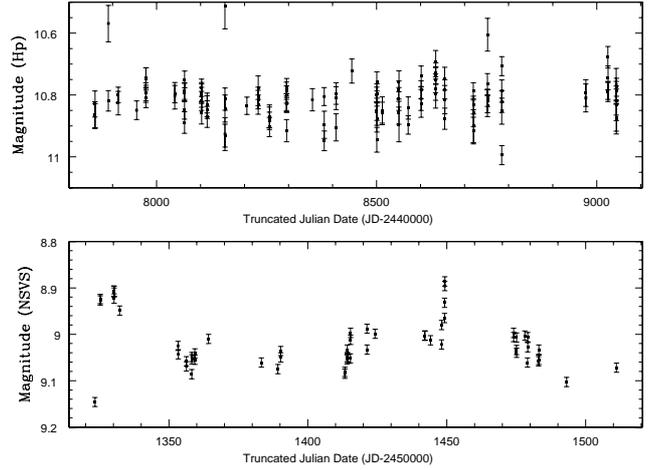}}
\caption{Hipparcos (Top panel:1989-1993) and NSVS (Bottom
  panel:1999-2000) lightcurves of Cyg~OB2~\#12.
Both panels plot the (non standard) 
instrumental magnitudes (references to these are provided in the text) and we caution that the scale of both axes differ between the 
two panels.}
\end{figure}

\begin{table}
\begin{center}
\caption[]{Summary of unpublished spectral observations.}
\begin{tabular}{lccc}
\hline
\hline
Date & Telescope & Wavelength & Resln.\\
     &            & Range({\AA})   &      \\  
\hline
\multicolumn{4}{c}{\bf Cyg\#12} \\
1992 July 17     & INT        & 4036-4836 &  4000  \\
                 &            & 5765-9586$^a$ &  4000\\
1995 July 17     & WHT        & 6172-6980  & 4000 \\
1998 Aug. 09     & WHT        & 3940-5737 &  8000 \\
                 &            & 6366-6772 &  8000\\
2000 July 17     & INT        & 3940-5737 & 8000 \\
                 &            & 6327-6727 &  8000\\
2007 August 22   &  WHT       & 7600-8900 &  10000\\
2008 July 22     &  WHT       & 4000-4700 &  4500\\
                 &            & 4500-5270 &  4500\\
                 &            & 6450-7150 &  7000\\
                 &            & 8350-8900 &  10000\\
& & & \\
\multicolumn{4}{c}{\bf $\zeta^1$~Sco}  \\
1998 October 7   & ESO 2.2-m          & 4000-8950$^b$ &48000 \\
1999 July--Aug.     & ESO 2.2-m$^c$    &  4000-8950$^b$ &  48000\\   
2005 April 24    & ESO 2.2m          & 3800-6800 & 48000\\
2006 February 17     & NTT        & 3933-7985 & 9840 \\
2009 March 06    & VLT        & 3060-5600   &   \\
\hline
\end{tabular}
\end{center}
{$^a$Note this wavelength coverage was obtained via 5 overlapping observations.
$^b$Approximate usable wavelengths. $^c$14 spectra obtained between July 17 to August 4.} 
\end{table}

\section{The observational properties of BHGs}

BHGs form a spectroscopically homogeneous class of objects that are
distinguished from BSGs by the presence of Balmer emission lines. However, they 
have been implicated in the LBV phenomenon (Sect. 1) and so it
is of interest to compare their  0.4-4.1$\mu$m spectra with
those of {\em bona fide} LBVs in both hot and cool phases. Suitable
comparison spectra across this wavelength range are provided by,
amongst others, Stahl et al. (\cite{stahl}), Hillier et al.
(\cite{316}), Groh et al. (\cite{groh}), Clark et al. (\cite{clark03},
\cite{clark09}, \cite{liege}) and Lenorzer et al.
(\cite{lenorzer}). Significant differences are evident;
specifically the weakness of the H\,{\sc i} and He\,{\sc i} lines in the BHG spectra, as well
as the  absence of  emission in  low excitation metallic
lines such as Fe\,{\sc ii} and {\sc iii} that characterise the spectra
of LBVs.

A second observational characteristic of {\em some} LBVs is the
presence of circumstellar dust -- indeed the presence of a detached circumstellar nebula has been
advanced to support the LBV nature of the late-B hypergiant HD~168625. 
Of the early BHGs in question, only BP~Cru is known to show 
evidence for warm dust (e.g. Fig. 2; Moon et al. \cite{moon}), although
with $M_{\rm dust}\sim5.2{\times}10^{-8}M_{\odot}$, considerably
less than is associated with either LBVs or supergiant B[e] stars
(Egan et al. \cite{egan2}, Clark et al. \cite{clark03}, Kastner et al. \cite{kastner}).

However, the defining characteristic of LBVs is their dramatic spectroscopic and photometric variability over timescales
of months--years. 
Motivated by reports of spectral variability amongst both early- and
late-B HGs we undertook an exhaustive 
literature search for these objects, which we present in Appendix A,
and briefly discuss here. As highlighted by previous authors, there is
strong observational evidence for a physical association between 
 LBVs and low-luminosity, late-B HGs
such as HD~160529 and HD~168607; however, the link is far
from proven for earlier, more luminous BHGs.  Owing to their
luminosities extensive spectroscopic and photometric datasets exist
for such stars, with the majority extending back to the mid-20th
century, for Cyg~OB2~\#12 and HD~190603 to the turn of last century and
(sparse) photometric data for $\zeta^1$~Sco potentially many centuries
before that (Appendix A)\footnote{Unfortunately, in many cases early
observations are presented in the literature without precise
observational dates. Nevertheless, the similarity of these data to
more recent observations supports the lack of long term  variability in
such cases.}

Cyg~OB2~\#12 has long been suspected of being spectroscopically
variable, being variously classified as B3-8 Ia$^+$ in the
literature (e.g. Table A.4 and refs. therein). 
However, upon close examination of the spectra and classification
criteria employed, we conclude that long-term  evolution of the 
stellar {\em temperature}  between 1954-2008 is probably 
absent (Appendix A.1.2) and that the reports of such behaviour result from
comparison of low S/N and resolution spectra and, crucially, the
sensitivity of the commonly employed He\,{\sc i} 4471/Mg\,{\sc ii}
4481 criterion to the properties of 
 the photosphere/wind transition zone {\em as well as 
photospheric temperature} (Sect. 4.1). This conclusion is bolstered by the lack
of secular photospheric variability dating back to the 1890s, although
low level, apparently aperiodic variability does appear present
(e.g. Gottlieb \& Liller \cite{gottlieb}).

Based on the datasets discussed in Appendix A, similar conclusions may be drawn 
for the remaining early- (BP~Cru, HD~80077, HD~190603, HD~168454, Wd1-5 
and 13) 
and mid-B HGs (Wd1-7 and 33), 
with the potential exception of $\zeta^1$~Sco. Spectroscopically, it is
the best sampled of all the BHGs and shows no evidence for evolution
between $\sim$1891-2009 (Table A.5).  Likewise, it appears to have
remained photometrically stable since at least 1949 (Table A.1), 
although isolated historical photometric observations dating back many
centuries are available in the literature, from which Sterken et
al. (\cite{sterken}) infer possible LBV like variability in the
18-19th century.

However, while long-term variability appears rare or absent amongst
the early-B HGs, rapid ($\sim$day to day) line profile 
variability (LPV) in both wind dominated emission and photospheric
absorption lines appears ubiquitous, persistent and well documented
(e.g.  Fig. 1 and Appendix A and references therein). These behaviours
are thought to arise from time variable wind structure and
photospheric pulsations respectively, 
with the former also potentially giving rise to variable radio continuum emission
(e.g. Gonz\'{a}lez \& Cant\'{o} \cite{gonzalez}), although changes in the  
the degree of ionization of the wind are another possible cause of this 
phenomenon.

Likewise, rapid, low amplitude photometric variability is present
amongst all the BHGs (e.g. Fig. 3 and Appendix A). Such behaviour has
long been recognised as being characteristic of luminous stars of all
spectral types (the `$\alpha$ Cygni' variables;
e.g. Burki \cite{burki78}, van Leeuwen et al. \cite{vanL98}, Clark et al. \cite{clark10}) and again
is typically attributed to photospheric pulsations.

We conclude that the temporal behaviour of the early-B HGs 
appears entirely typical of the wider population of luminous,
early-spectral-type 
non-LBV stars, with the potential exception of $\zeta^1$~Sco in the
18-19th centuries and a possible sudden $\Delta$m$_{\rm B}\sim$0.4~mag
`glitch' in the lightcurve of Cyg~OB2~\#12 in the mid-1940s (Gottlieb
\& Liller \cite{gottlieb}).

\section{Quantitative modeling}

\begin{table}
\begin{center}
\caption[]{Model Atoms  included in our calculations.}
\label{tab-atoms}
\begin{tabular}{llllll}
\hline
\hline
 Species &
\# full levels &
\# super levels & \# of transitions \\ 
\hline
\HI        & 30   & 30  &  435  \\
\HeI       & 59   & 59  & 592   \\
\HeII      & 5    & 5   & 10    \\
\CII       & 80   & 45  & 648   \\
\CIII      & 38   & 22  & 149   \\
\CIV$^{*}$ &  8   & 5   & 22   \\
\NI        & 63   & 25  & 344  \\
\NII       & 179  & 86  & 1936  \\
\NIII      & 20   & 11  & 61    \\
\OI        & 75   & 23  & 450   \\
\OII       & 102  & 46  & 992   \\
\NaI       & 44   & 18  & 345   \\
\MgII      & 50   & 37  & 597   \\
\AlII      & 20   & 12  & 41    \\
\AlIII     & 45   & 17  & 362   \\
\SiII      & 72   & 35  & 491   \\
\SiIII     & 83   & 47  & 532   \\
\SiIV      & 28   & 17  & 129   \\
\SII       & 85   & 26  & 745   \\
\SIII      & 63   & 31  & 331   \\
\CaII      & 32   & 25  & 157   \\
\CrIII     & 209  & 36  & 2444  \\
\CrIV      & 242  & 30  & 4732  \\
\FeII      & 709  & 233 & 18011 \\
\FeIII     & 812  & 87  & 21383 \\
\FeIV      & 417  & 62  & 8353  \\
\NiII      & 254  & 33  & 3354  \\
\NiIII     & 336  & 34  & 5181  \\
\NiIV      & 294  & 56  & 4174  \\
\hline
Total & 4160 & 1137  & 72827\\
\hline
\end{tabular}
\end{center}
{\small{$^{*}$ Not used in Cyg~OB2~\#12.}}
\end{table}

\begin{table*}
\begin{center}
\caption{Model parameters for the BHGs Cyg~OB2~\#12, $\zeta^1$~Sco, HD190603 \& BP Cru (upper panel) and 
Galactic B {\em supergiant} 
comparison (lower panel; Searle et al. \cite{searle})} 
\label{tab:model}
\begin{tabular}{lllcccccccccc}
\hline\hline
Star    &Spec.       & DM    & $M_V$    & log(\Lstar) & \Rstar     & \Teff  & \Mdot              & \vinf  & $\beta$ &  {\em f}  & \logg   & \Mstar \\
        & Type       &       &        & $L_{\odot}$       &  \Rsun     &  kK    &  $10^{-6} M_{\odot}$yr$^{-1}$ & \kms   &         &           &  &  $M_{\odot}$    \\
\hline

Cyg OB \#12 &B3-4     & 11.21 & $-$9.85  &  6.28       &  246.0     &  13.7  &  3.0               &  400   &  3.0    &  0.04 &  1.70  & 110  \\

$\zeta^1$~Sco & B1.5 & 11.07 & $-$8.50  &  5.93       & 103.0      &  17.2  &  1.55              &  390   &  2.25   &  0.06 &  1.97  & 36  \\
              &      & 11.50 & $-$8.93  &  6.10       & 125.5      &  17.2  &  2.07              &  390   &         &       &     & 53     \\

HD~190603      &B1.5  & 10.98 & $-$7.53  &  5.58       &  63.0      &  18.0  &  1.09              &  485   &  2.25   &  0.25 &  2.10    &  18    \\
              &      & 12.38 & $-$8.93  &  6.14       &  120.0     &  18.0  &  2.87              &  485   &         &       &          &  65   \\

BP Cru      &B1    & 12.41 & $-$7.47  &  5.67       &  70.0      &  18.1  &  10.0              &  305   &  2.25   &   1.0 &  2.38    &  43   \\

\hline 
 & B1  &  & &5.44 & 36.5& 22.0& & &  & &2.41   &  12\\ 
 & B1.5 &  & &5.44 & 44.5& 19.9& & &  & & 2.41&    18 \\
 & B3   &  & &5.37 & 60.4& 16.4& & &  & & 2.16&    19\\ 
 & B4   &  & &5.34 & 63.5& 15.8& & &  & & 2.06&     16 \\
\hline
\end{tabular}
\end{center}
\end{table*}

\begin{table*}
\begin{center}
\begin{tabular}{llcccccccc}
\hline
\hline
Star & Spec & \Vt   & \vsini & \vmac & $v_{\rm phot}$ & H/He & N/N$_{\odot}$ & C/C$_{\odot}$ & O/O$_{\odot}$ \\
     & Type & \kms  &  \kms  & \kms   & $v_{\rm sound}$    & &               &               &               \\
\hline
Cyg OB \#12 & B3-4 &  12.5 &  38    &  50 &0.02   &  10   &  2.7         &  0.21         & 0.55                \\
$\zeta^1$ Sco & B1.5 &  12.5 &  45   &  50  & 0.5  &  5    &  5.5         &  0.33         & 0.32            \\
HD~190603     &B1.5  &    15   &  49    &  55    &  0.5  &5  &  5.5         &  0.33         & 0.40         \\
BP Cru & B1          &   10   &  55    &  --    & 0.5 &  3.5  &  4.8         &   2.0         & 0.40          \\
\hline
\end{tabular}
\end{center}
{
Note that \Rstar\ corresponds to $R(\tau_{\rm Ross}=2/3)$, the H/He ratio is given by number and $v_{\rm phot}$ in terms of $v_{\rm sound}$. Abundances are relative to solar values 
from Anders \&\ Grevesse (\cite{anders}) and have uncertainties of, typically,
$\sim$0.2dex; 
if we use the values from Asplund et al. (\cite{asplund}) as a reference the derived ratios need to be
scaled by 1.49, 1.86 and 1.86 for C, N and O respectively. The results presented here assume a 
distance of 1.75~kpc for Cyg~OB2~\#12 ($DM=11.21$). 
Apart from the assumed distance of 1.64~kpc for $\zeta^1$~Sco
($DM=11.07$, Sana et al 2006) we also provide 
corresponding stellar properties for the other distance estimate (1.99~kpc, $DM=11.50$) assumed in previous spectroscopic studies.
For HD190603 we display the stellar properties assuming a  distance of 1.57~kpc ($DM=10.98$) as well as those which would result if the object
had the same $M_V$ as  $\zeta^1$~Sco. Finally a distance of 3.04kpc (Kaper et al 2006) is assumed for BP~Cru}
\end{table*}

\begin{figure*}
\resizebox{\hsize}{!}{\includegraphics[angle=0]{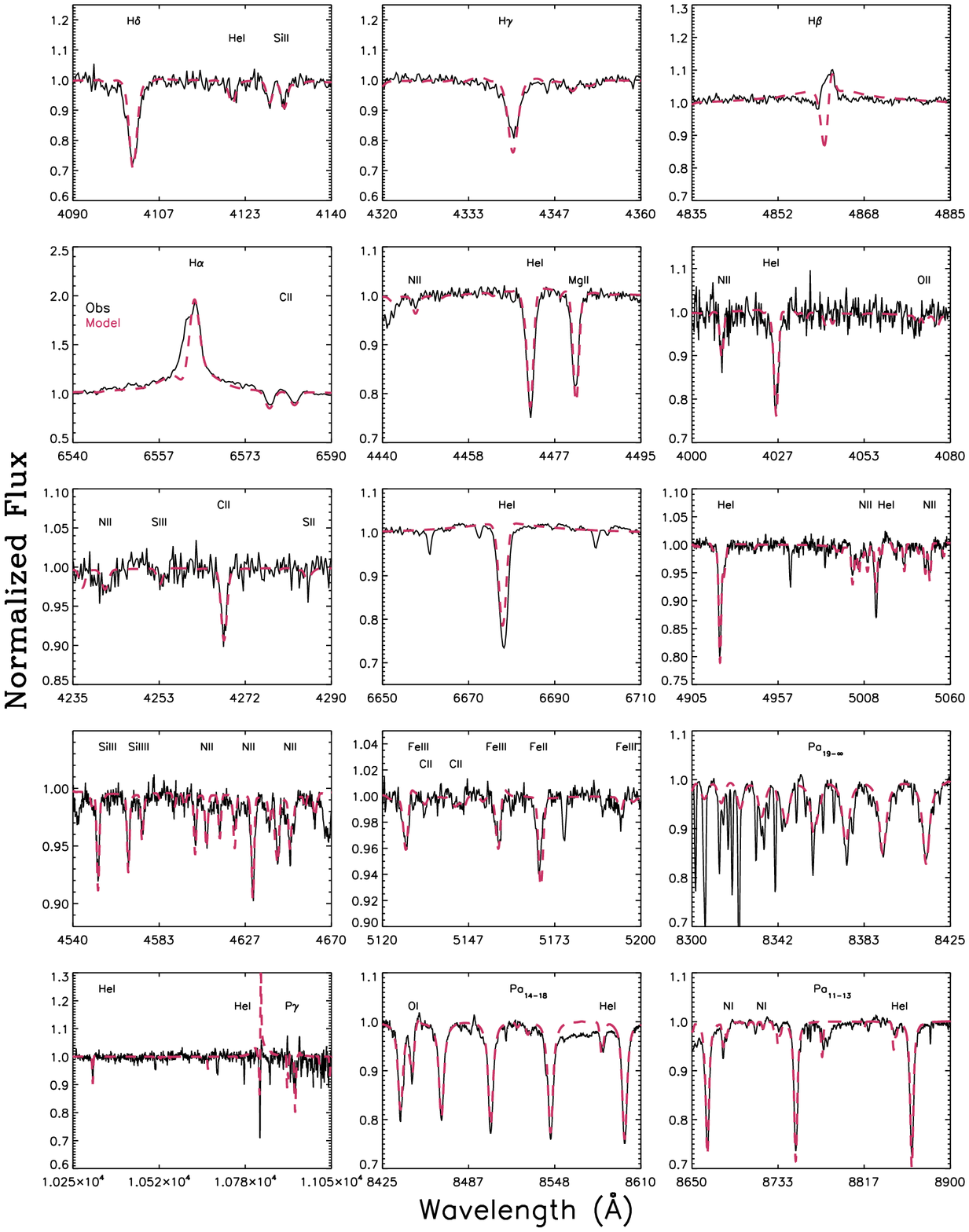}}
\caption{Comparison of the synthetic spectrum of Cyg~OB2~\#12  (red
  dashed line) with observational data for various 
transitions between 4090-11050{\AA}. Note that the narrow absorption lines which have not 
been labeled are telluric in origin.}
\label{fit-cygspec}
\end{figure*}

\begin{figure}
\includegraphics[angle=90,width=8.9cm]{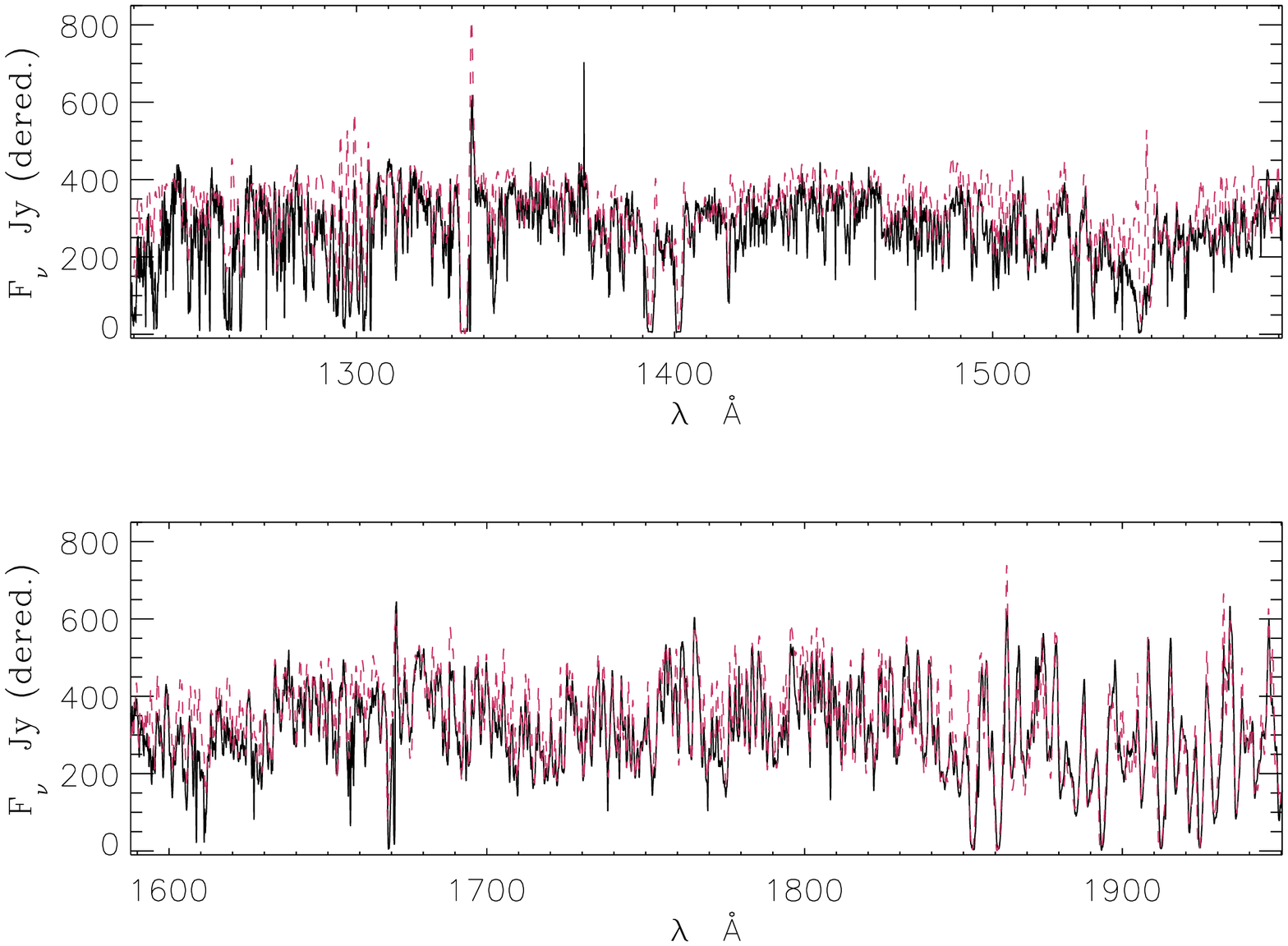}

\vspace{-.2cm}

\includegraphics[angle=90,width=8.9cm]{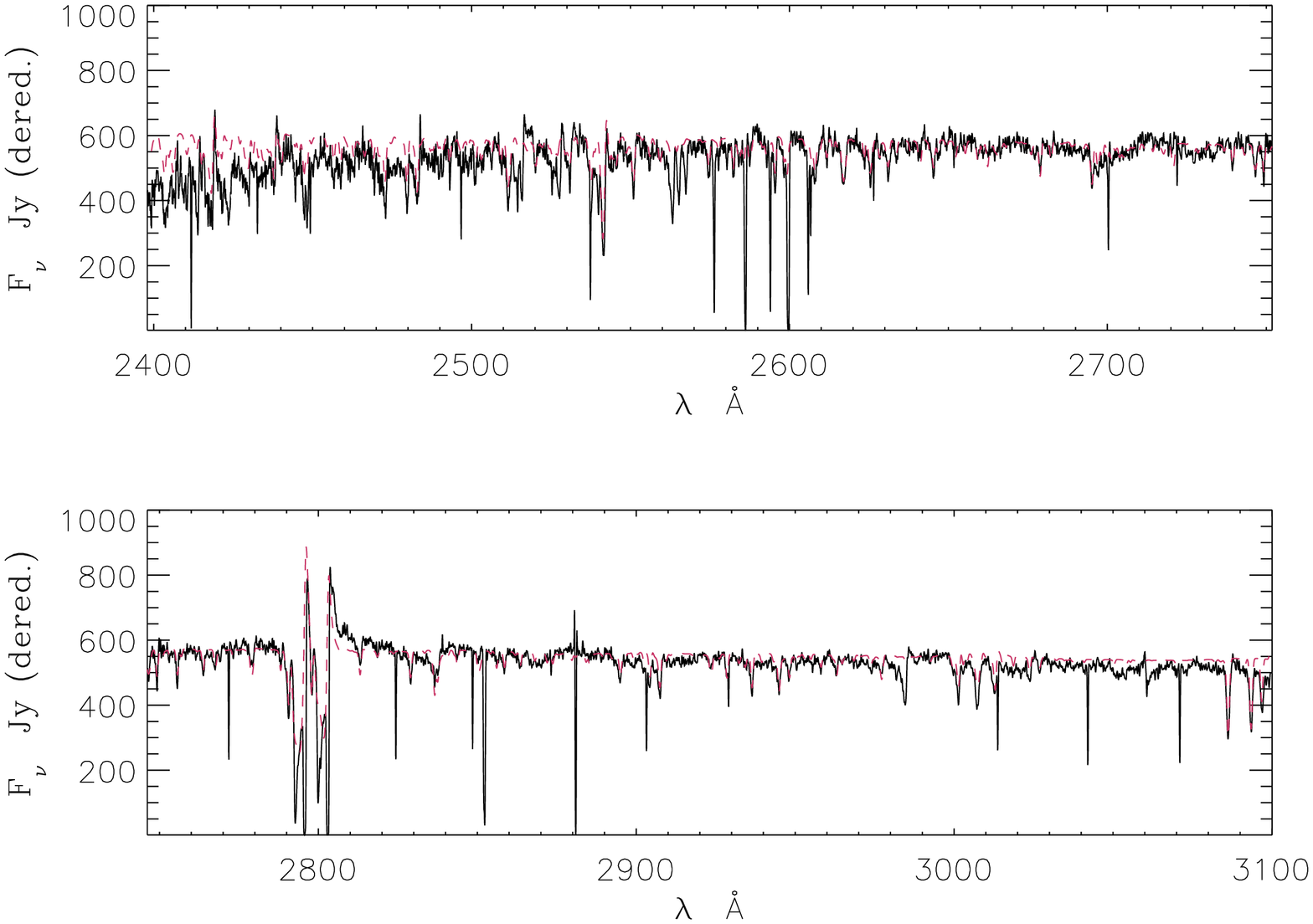}\\
\caption{Comparison of the synthetic  (red dashed line) with the observational (black line) UV spectrum
 of $\zeta^1$ 
Sco. Note that the narrow lines that are not replicated in the model are interstellar in origin.}
\label{fit-zetuvspec}
\end{figure}

\begin{figure*}
\resizebox{\hsize}{!}{\includegraphics[angle=0]{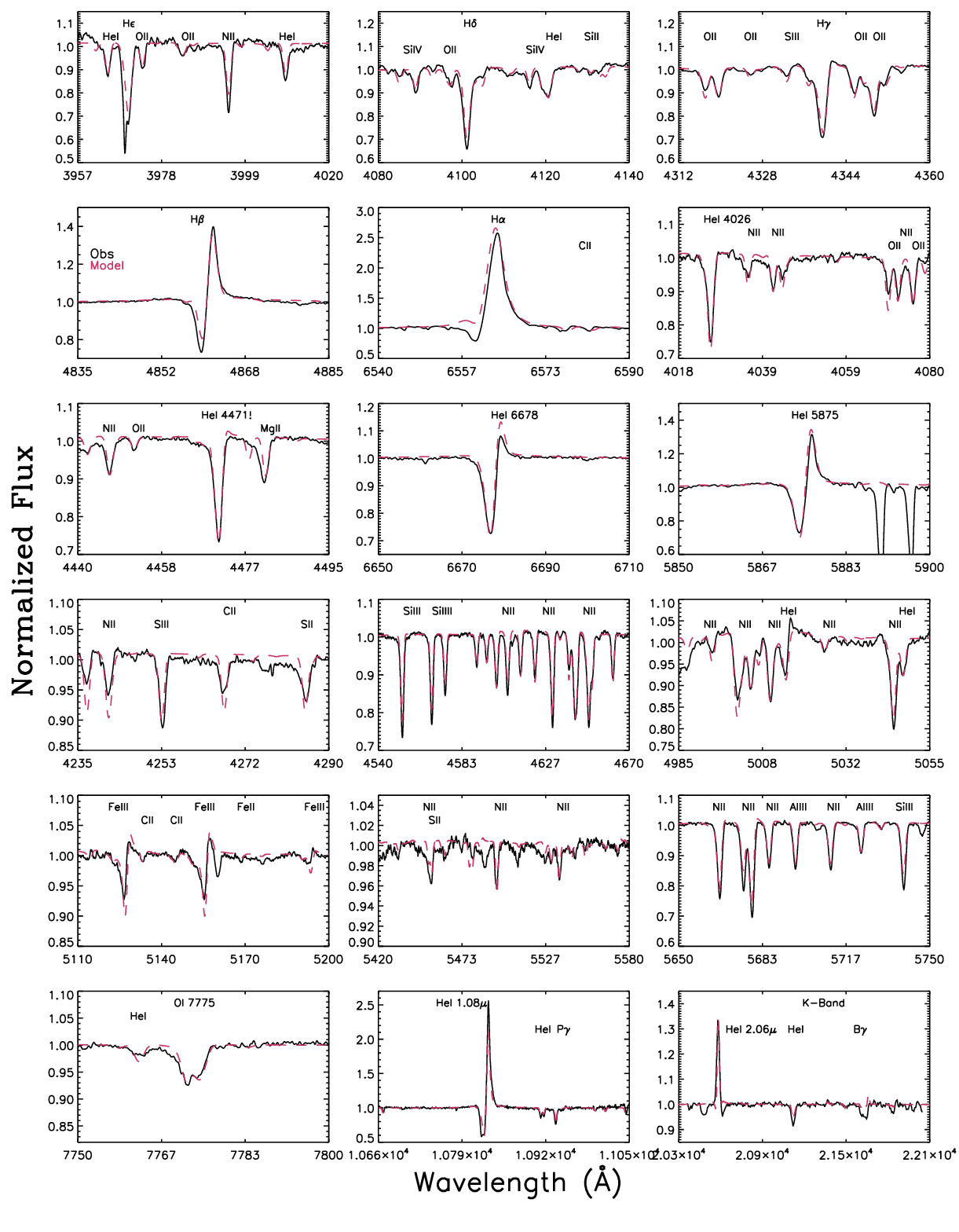}}
\caption{Comparison of the synthetic spectrum of $\zeta^1$~Sco  (red dashed line) with observational data for various transitions between 4090-22100{\AA}. 
Optical data correspond to the 2006 ESO NTT/EMMI run. Note that narrow unlabeled absoption lines are 
telluric in origin.}
\label{fit-zetopspec}
\end{figure*}

\begin{figure*}
\resizebox{\hsize}{!}{\includegraphics[angle=0]{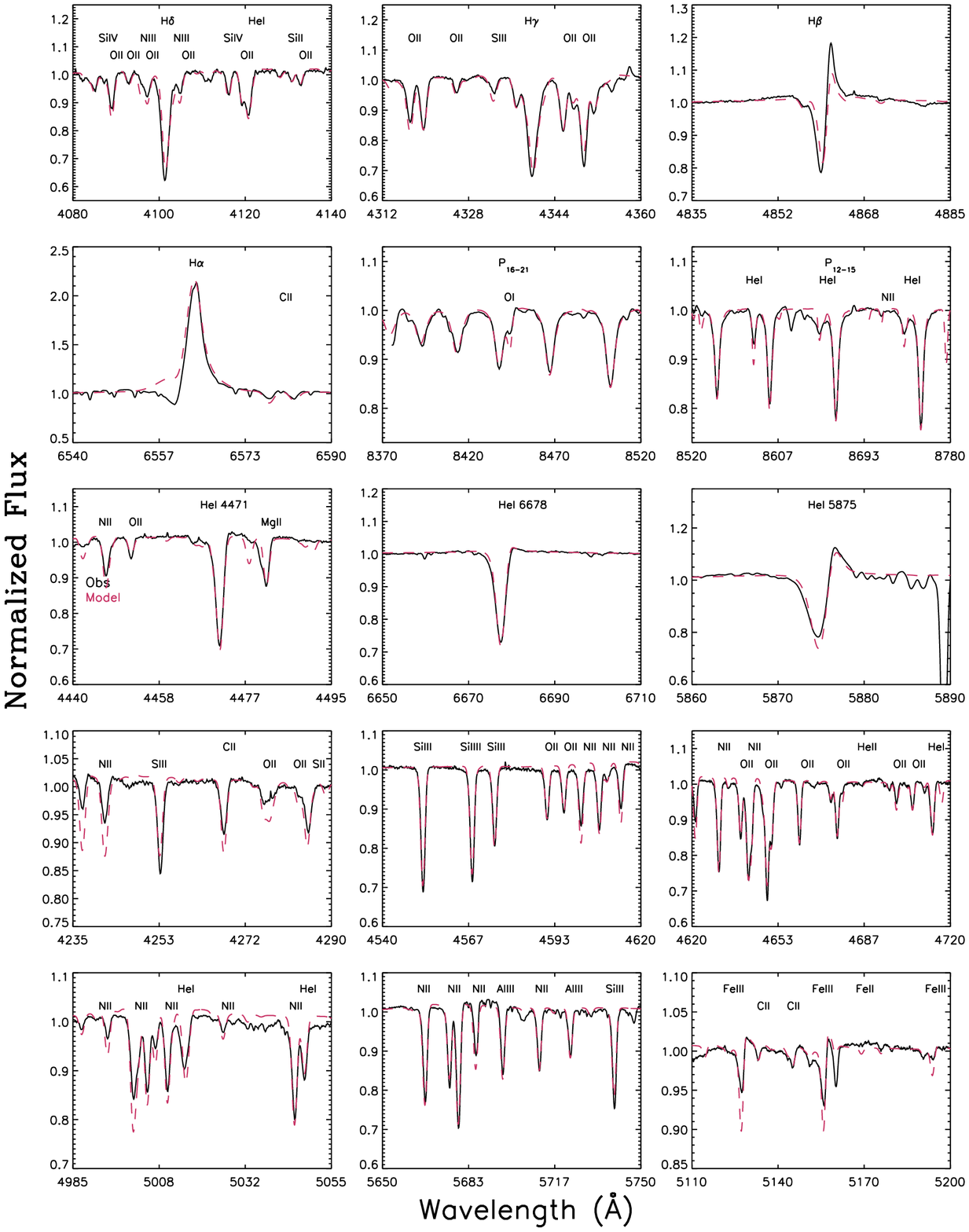}}
\caption{Comparison of the synthetic spectrum of HD190603 (red dashed
  line) to observational data.  Optical data correspond to the FEROS
  data kindly provided by Otmar Stahl. As with $\zeta^1$~Sco (Figs. 7
  and C.3), no long-term variability is present between different
  epochs (not shown). Note that narrow unlabeled absorption lines are telluric in origin.}
\label{fit-hd190603-stahl}
\end{figure*}

In order to  more fully address 
the physical properties of Cyg~OB2~\#12, $\zeta^1$~Sco and HD~190603
we have employed the non-LTE model atmosphere code CMFGEN (Hillier \&
Miller \cite{hil98}, \cite{hil99}). This solves the radiative-transfer
equation for a spherically symmetric wind in the co-moving frame under the
constraints of radiative and statistical equilibrium. 
Since the code does not solve for the wind structure from first
(physical) principles, a velocity structure must be chosen; we adopted
a standard $\beta$-type law\footnote{The $\beta$-type law is given by 
$v(r)=v_{\infty}\times(1-b/r)^{\beta}$, where 
$b=r_{\rm phot}\times(1-(v_{\rm phot}/v_{\infty})^{(1/{\beta})})$ and $r_{\rm phot}$ and
$v_{\rm phot}$ are the radius and velocity at the connection point between the $\beta$-law and the hydrostatic structure.}.
  Given the relatively high wind densities
of BHGs, the role of the hydrostatic structure and the transition
region between photosphere and wind becomes crucial to interpret the
spectra (e.g. Sect. 4.1 and Fig. 8).  Because of this, we used a hydrostatic density structure at depth
 (the `pseudo-photosphere') matched to a $\beta$-law wind, requiring
 the density to be continuous at the transition.  The location of the
 transition is fixed by the adopted base velocity of the wind,
 \vpho\ (a free parameter in the fitting), which specifies the
 density through the equation of mass continuity.  This approach is
 similar to that of, e.g., Santolaya-Rey et al. (\cite{sph97}),
 except that in our implementation the flux-weighted mean opacity is
 used in preference to the Rosseland mean opacity
 Several comparisons using `exact' photospheric
structures from TLUSTY (Hubeny \& Lanz \cite{hub95}) showed excellent
agreement with our method.\footnote{The latest version of CMFGEN,
  which allows the user to 
compute the exact hydrostatic structure, showed full consistency
with our method. At this stage, we prefer
our approach as it allows for more flexibility to study the crucial
transition region.}.

The main atomic processes and data sets are discussed in Hillier \&
Miller (\cite{hil98}, see Dessart \& Hillier \cite{dess10} for an
updated description of the atomic data).  The list of model atoms
utilized in our calculations is provided 
in Table~\ref{tab-atoms}, including for each ion the number of super
and full levels and number of bound-bound transitions.  The CMFGEN model 
is then prescribed by the stellar radius, \Rstar, the stellar
luminosity, \Lstar, the mass-loss rate, \Mdot, the velocity field,
$v(r)$ (defined by $\beta$, \vpho  and the terminal velocity, \vinf), the volume filling
factor, {\em f},  characterizing the clumping of the stellar wind, and elemental
abundances. The interstellar reddening parameters $E(B-V)$
and $R_V$ are also obtained by fitting the model SEDs to the available
photometric data. Finally, the equatorial rotation
 velocity, \vsini, and the atmospheric macroturbulent velocity,
 \vmac, were estimated in Fourier space (Sim{\'o}n-D{\'{\i}}az \&
 Herrero \cite{simondiaz}) from a selected sample of photospheric
 lines.

Given the number of free parameters required to specify a fit it is
not possible to survey the full parameter space systematically
 in order to establish robust error estimates;
hence errors quoted in this paper represent the range of values for
which an acceptable fit to the data may be obtained. Nevertheless, we
discuss error estimates for key parameters for each of the stars
analysed below.  The validity of this technique has been demonstrated
by calibration to stars of similar temperature and luminosity for
which UV, optical and near-IR data were available (Najarro et
al. \cite{paco99}, Najarro \cite{paco01}).

A comparison of the predicted SEDs to our data for the three stars is
presented in Fig.~\ref{fig-cont-fits}, while selected regions of the
synthetic spectra are overplotted on the observed 4090-11050{\AA}
spectrum of CygOB2 \#12 in Fig.~\ref{fit-cygspec}, the  UV--near IR
($\sim$1200--22100{\AA}) spectrum of $\zeta^1$~Sco in
Figs.~\ref{fit-zetuvspec} \& \ref{fit-zetopspec} and the optical 
($\sim$4000--7000{\AA}) spectrum of HD~190603 in Fig. 7.
 The optical data
shown in Fig.~\ref{fit-zetopspec} correspond to the 2006 ESO EMMI/NTT
run. Fits to the earlier FEROS spectra supplied by Otmar Stahl - which
provide a more extended wavelength coverage - are shown in the
Appendix (electronic version only) in Figs.\ref{fit-zetopspec-stahl1}
and \ref{fit-zetopspec-stahl2}.  The latter were used to obtain the
final stellar properties as they encompass the high Balmer and Paschen
lines which constrain the surface gravity.  A
summary of the results of the analyses for the 3 stars is presented in
Table~\ref{tab:model}, along with the parameters of the B1 Ia$^+$ star
BP~Cru, obtained via an identical methodology by Kaper et
al. (\cite{kaper06}).

\begin{figure*}
\includegraphics[angle=0,width=9.2cm]{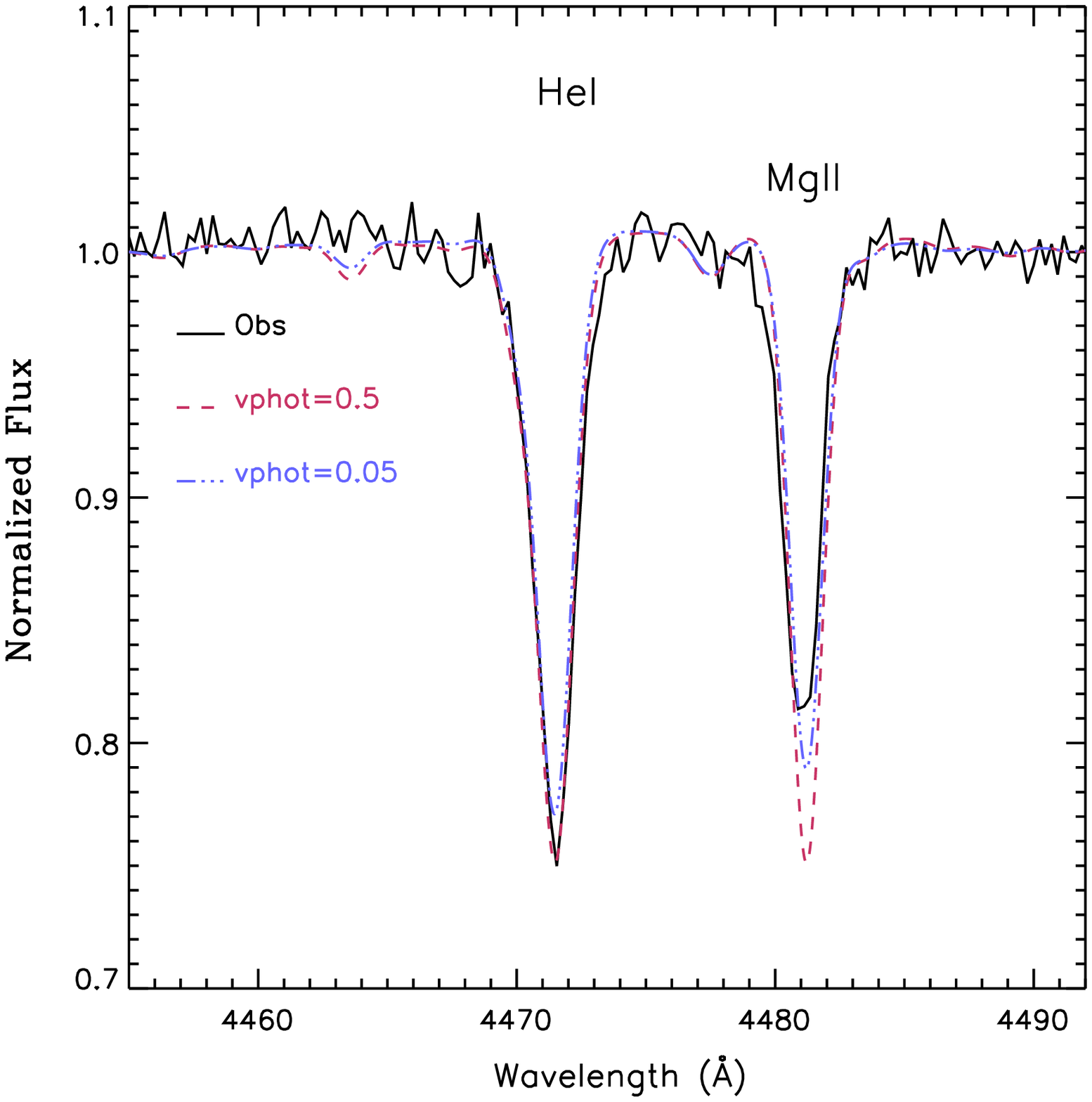}
\includegraphics[angle=0,width=9.2cm]{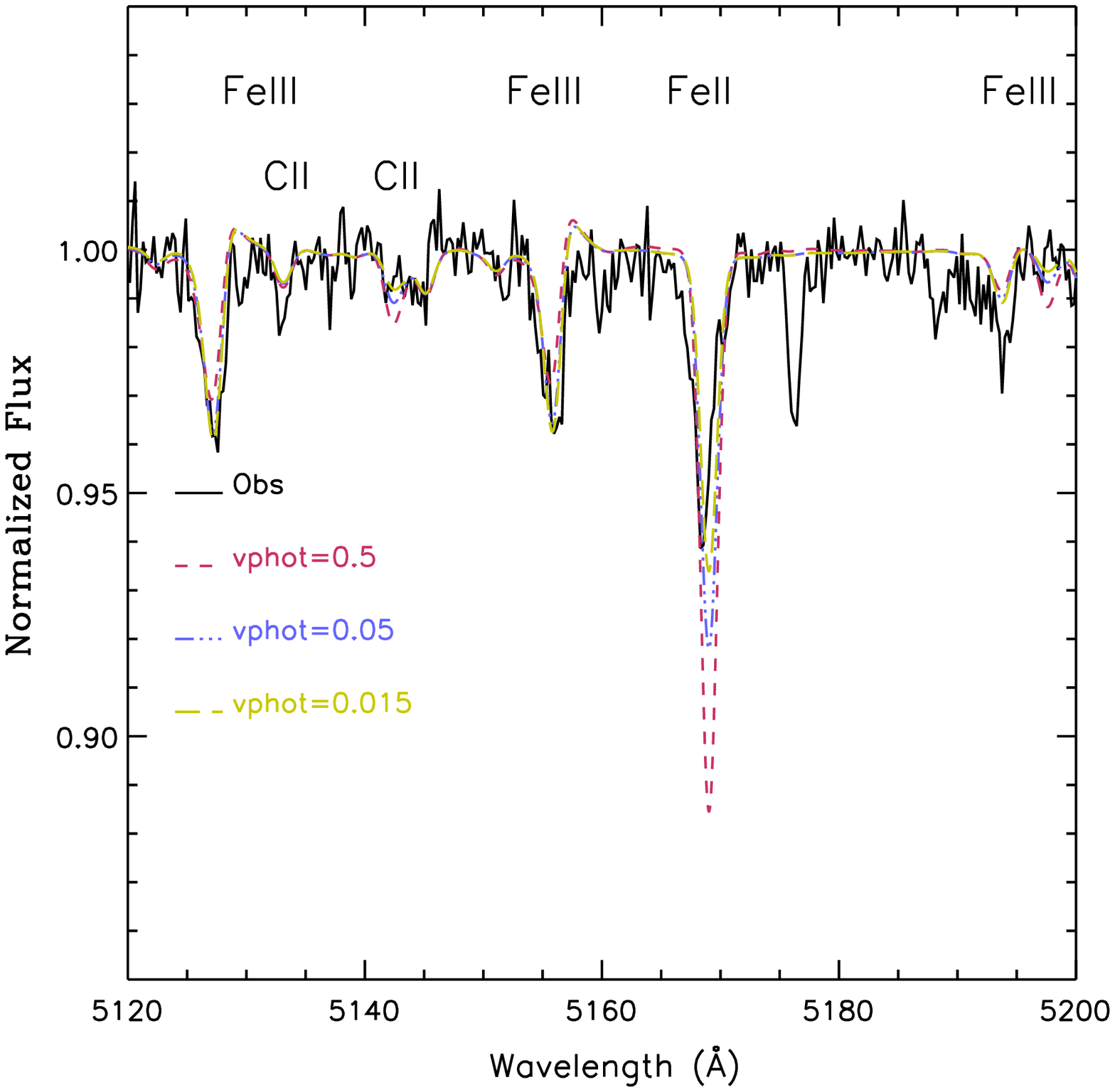}
\caption{ Importance of the wind/photosphere transition region on  the emergent spectrum of Cyg~OB2~\#12. 
{\bf{Left:}} Influence of the transition velocity (in units of the sound speed, $v_{\rm sound}$ ) 
on the spectral typing diagnostic lines \HeI~4471 and \MgII~4481{\AA}. {\bf{Right:}} Influence
on the \FeIII/\FeII\ ionization equilibrium diagnostic lines. The \FeII~5170~{\AA} line is severely affected by the transition 
velocity while the \FeIII\ lines remain roughly unaltered.}
\label{fig-vphot}
\end{figure*}

\subsection{Cyg~OB2~\#12}
\label{sub-cyg12}

For the purposes of modeling Cyg~OB2~\#12 we employed the dataset
described in Sect. 3 and Appendix A. We have also assumed membership
of CygOB2 and hence a distance, d$\sim$1.75~kpc (e.g. Negueruela et al. 
\cite{iggy} and refs. therein); we return to this
issue below.

\paragraph{Temperature,  gravity and Luminosity}
The available optical and NIR spectra allowed us to make use of
several ionization equilibria to estimate the effective temperature, 
and thus constrain the ionization structure of Cyg~OB2~\#12. 
Hence, we were able to utilize simultaneously \SiIII/\SiII, \OII/\OI,
\NII/\NI\ and \FeIII/\FeII ~line ratios. The \SIII/\SII\ equilibrium
was only used for a consistency check, as the weak diagnostic
\SIII~4253{\AA} line suffers from a poorer S/N ratio in the blue spectral region
 due to reddening. We refer the reader to Sec~\ref{app-sub-spec-class} for a detailed
description of the available diagnostic lines.

When analyzing the ionization equilibria, we found that while lines
belonging to higher ionization stages react sensitively to effective
temperature and to lesser extent to gravity, those corresponding to
the lower ionization stage display a high sensitivity not only to
temperature and gravity, but also the location of the transition region between
photosphere and wind, characterised by \vpho\ (especially the \SiII, \FeII\ and \OI\
lines).  This sensitivity is clearly illustrated
 in Fig.~\ref{fig-vphot} (right panel), where the choice of
\vpho\, significantly affects the resulting strength of the
\FeII~5170{\AA} line, leaving the \FeIII\ lines basically
unaffected. Thus, if a lower transition velocity is chosen, a lower
effective temperature is required to match the \FeIII/\FeII~ratio.  As
we will show later, this effect is responsible for our somewhat lower
\Teff\ values when compared to other studies (see
Sect.\ref{sub-sco190}).  Interestingly, Fig.~\ref{fig-vphot} (left
panel) also shows how the ratio of the diagnostic lines \HeI~4471 and
\MgII~4481{\AA}, which are used for spectral typing, also shows a
moderate dependence on the transition region. Thus, a model with
$v_{\rm phot}=0.5v_{\rm sound}$ will yield equally strong \HeI~4471 and
\MgII~4481{\AA}, while if the transition takes place at
$v_{\rm phot}=0.05v_{\rm sound}$, \HeI~4471{\AA} clearly becomes stronger
than \MgII~4481{\AA}.

We note that this moderate dependence on the transition region becomes
important in BHGs, due to the presence of a strong, dense
wind and should be negligible in B supergiants, where the absorption lines will
form in deeper photospheric layers that are largely unaffected by
the photosphere-wind transition region.  Therefore, despite the large
number of diagnostics, our final estimate of the 
effective temperature, ($T_{\rm eff}=13.7$kK, see Table~\ref{tab:model}) is
subject to a moderate uncertainty. We find +800K and -500K as upper
and lower error bounds.

Once the effective temperature was obtained, and assuming a distance
of d$\sim$1.75~kpc, 
we proceeded to fit the observed SED of Cyg OB2
\#12 from the optical through radio (see Fig.\ref{fig-cont-fits}) and
hence derived the reddening, stellar radius and, therefore, the
stellar luminosity. We found $E(B-V)=3.84$ and a reddening parameter
$R_V=2.65$, corresponding to $A_V=10.18$.  This value shows excellent
agreement with the $A_V=10.20$ found by Torres et al. \cite{torres},
although they obtained a higher total to selective extinction
parameter ($R_V=3.04$)
 
From these values we obtained a stellar radius of 246\Rsun ~and a
final luminosity of $1.9\times10^6L_{\odot}$.  The temperature and
resultant luminosity are broadly comparable to previous qualitative 
studies (e.g. Massey \& Thompson \cite{massey}, Hanson \cite{hanson}),
confirming that it is an extraordinarily luminous, but comparatively
cool BHG.

The Paschen lines in the I Band provide the best constraints for the
surface gravity, especially the run of the line overlap among the
higher members (Fig. \ref{fit-cygspec}).  Compared to the classical
optical diagnostic lines - H$\gamma$ and H$\delta$ - the higher
Paschen series lines are significantly less affected by the stellar
wind (and consequently clumping) and the wind/photosphere transition
region.  Nevertheless, the influence of the latter once again
translates into a larger uncertainty in the (lower) error. Thus, we
find log$g=1.70^{+0.08}_{-0.15}$, corresponding to a 
spectroscopic mass of $M=110^{+23}_{-31}M_{\odot}$.

\begin{figure*}
\includegraphics[angle=0,width=6.15cm]{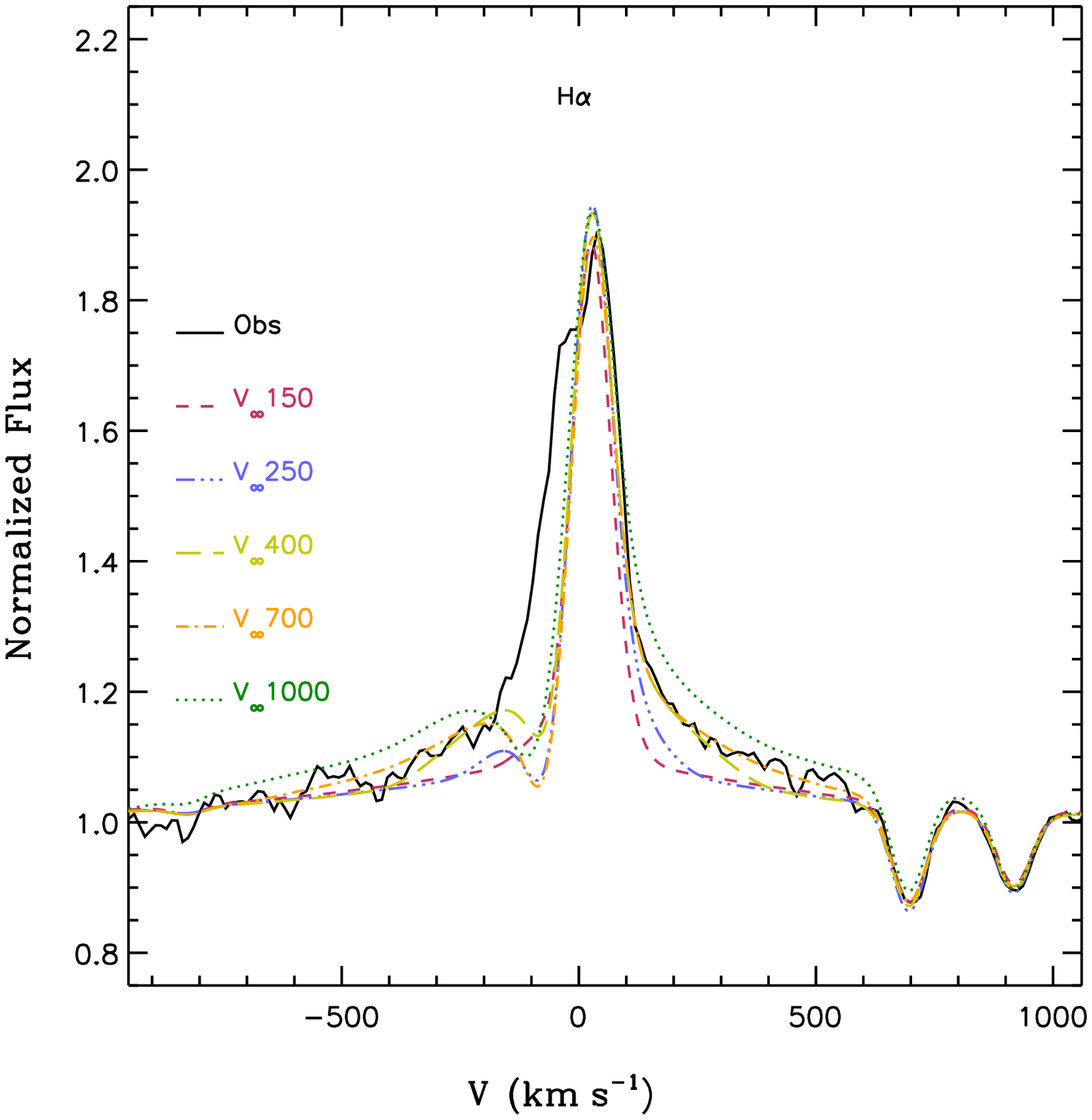}
\includegraphics[angle=0,width=6.15cm]{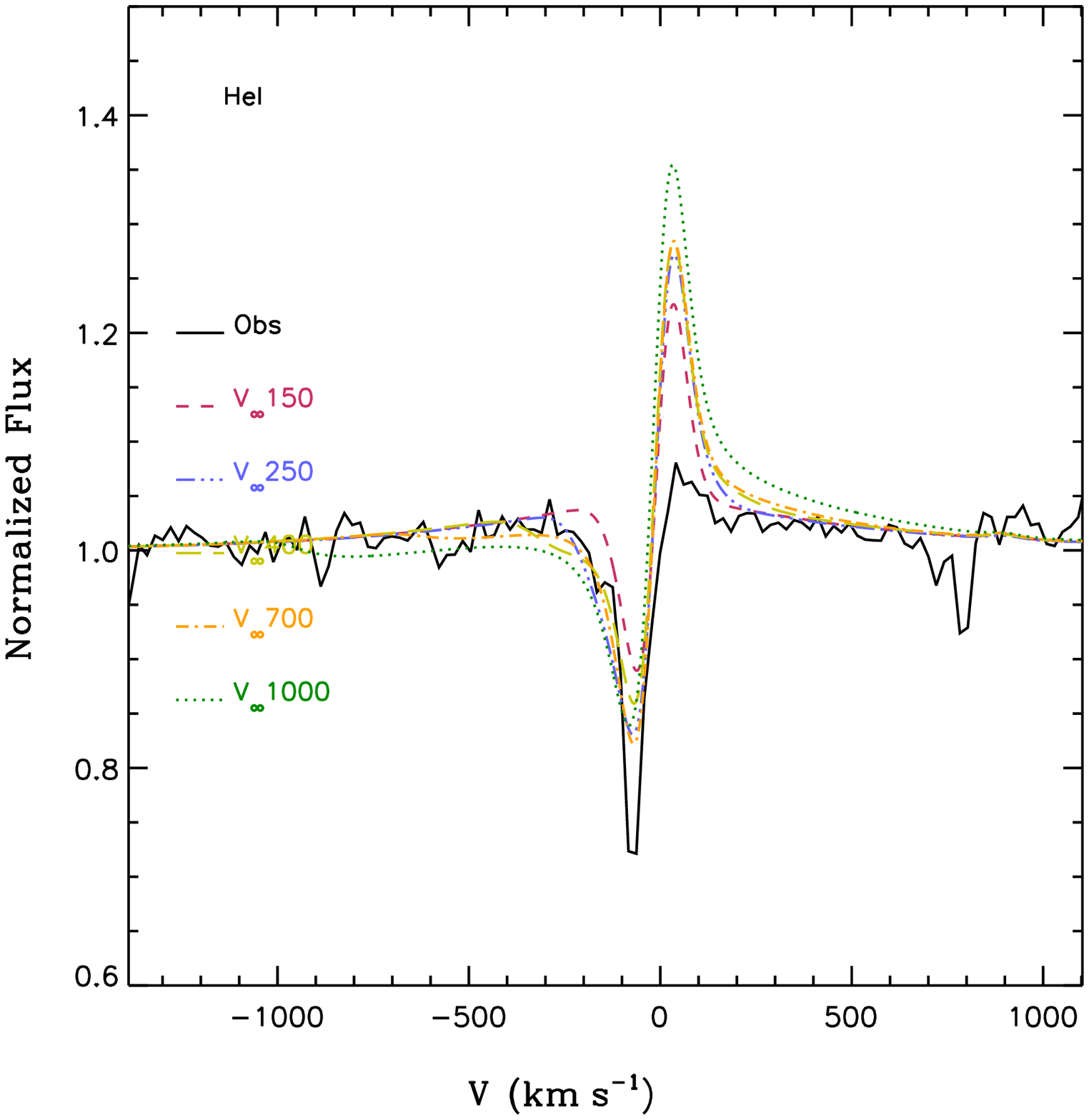}
\includegraphics[angle=0,width=6.15cm]{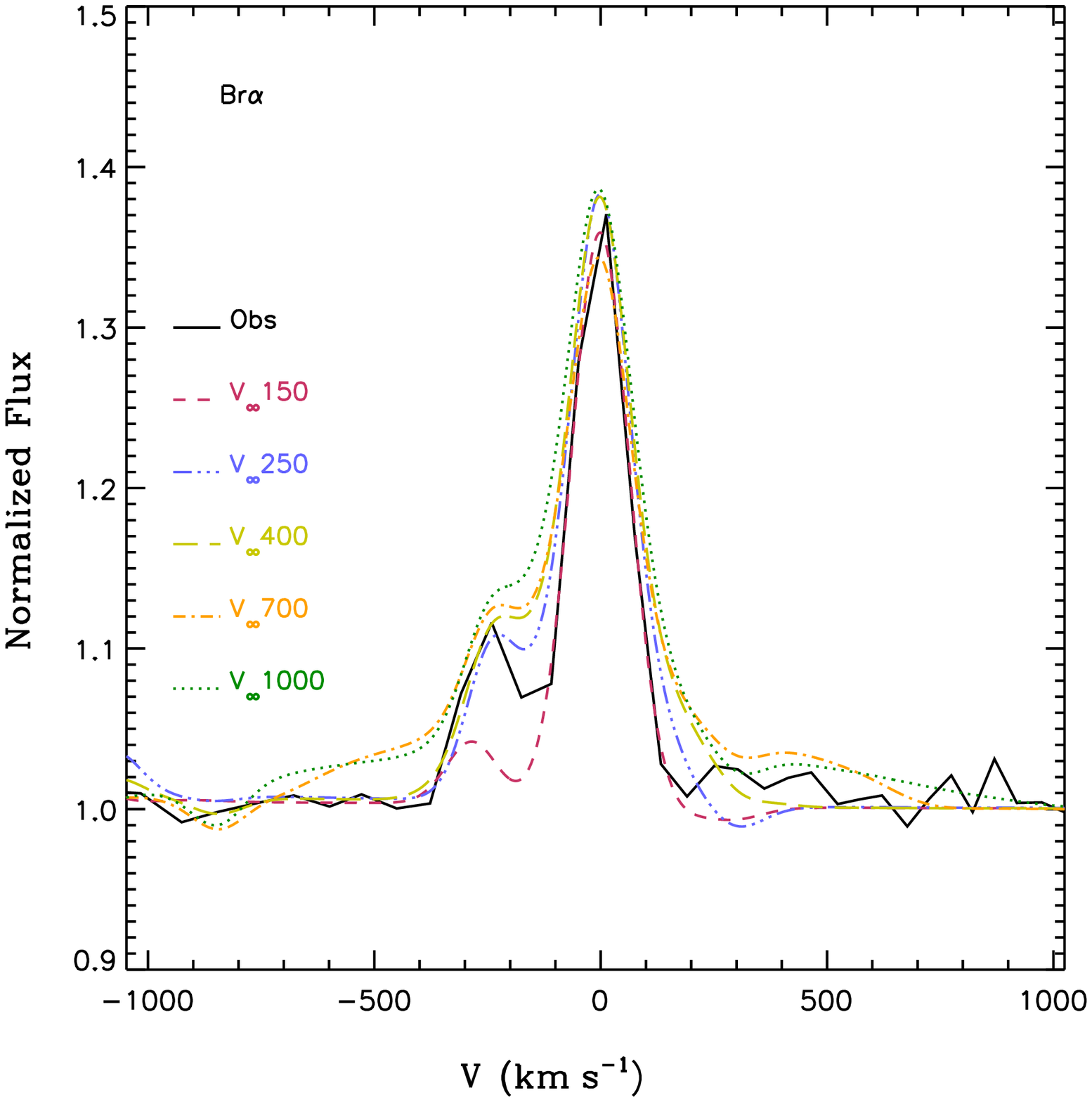}
\caption{ Estimate of \vinf\ for Cyg~OB2~\#12 
from the H$\alpha$ (left panel) and He\,{\sc i} 10830{\AA} lines (middle panel) and
B$\alpha$/\HeI~4.048$\mu$m complex (right panel).}
\label{fig-velo}
\end{figure*}

\paragraph{Wind properties and clumping}
Unfortunately, the moderate reddening affecting Cyg~OB2~\#12 prevents
us from securing UV observations from which one might derive firm
\vinf\ estimates. Furthermore, as with $\zeta^1$~Sco 
and HD~190603, Cyg~OB2~\#12's stellar wind is not sufficiently dense
that alternative \vinf\ diagnostic lines such as H$\alpha$ or
\HeI~10830{\AA} can reach their full potentiallity and unambiguously
yield the wind terminal velocity.

Highlighting this difficulty, Souza \& Lutz (\cite{souza}) suggested a
terminal velocity of 1400\kms, based on the presence of a blueshifted
absorption feature in the H$\alpha$ profile; a value which was
subsequently used to obtain the mass-loss rate from radio (Abbott et
al. \cite{abbott}) and IR measurements (Leitherer et
al. \cite{leitherer}). Subsequently, Klochkova \&\ Chentsov
(\cite{klochkova}) revised the estimate of the wind's terminal
velocity significantly downwards.  Based on higher resolution
H$\alpha$ observations they identified strong electron scattering
wings extending up to 1000\kms, as well as blueshifted absorption up
to $\sim 160$\kms\ which they attributed to the wind's true \vinf.
However our models indicate that this blueshifted feature {\em does
  not directly reflect} \vinf\ but rather results from  the run of
the density and ionization structures within the wind which shapes
H$\alpha$.

This is shown in Fig.~9, where models with \vinf\ ranging from 400 to
1000\kms yield an absorption feature around 100-200\kms\ to the blue
of H$\alpha$ in the resultant synthetic spectra. Moreover,
values of \vinf\ significantly below 400\kms\ appear too low to
reproduce the high velocity line emission while, depending on the
normalization errors of the H$\alpha$ profile, values between 400 and
1000\kms\ are potentially consistent. We also found that  the blue absorption component of
the \HeI~10830{\AA} line is likewise not able to distinguish between
$v_{\infty}=150$km$s^{-1}$ and $v_{\infty}=1000$km$s^{-1}$. We note that if the wind density of
Cyg~OB2~\#12 was a factor of two higher, the blue absorption component
would develop up to the corresponding value of \vinf.

Nevertheless, an upper limit on \vinf\ may be obtained by means of the
Br$\alpha$/\HeI~4.048$\mu$m complex. 
This is apparent in Fig.~\ref{fig-velo}, where we see that ISO
observations of this feature clearly resolve both components. Our
models indicate that if \vinf\ is above $\sim$500\kms, both components
are blended, while strong emission bluewards of 400\kms is present,
which is not seen  in the data.  On their own, these observations
lack sufficient S/N and spectral resolution to accurately constrain
\vinf, but in conjunction with H$\alpha$ they jointly provide stronger
constraints.  Following this approach, we adopt $v_{\infty}=400$kms$^{-1}$ for the
remainder of the paper, while recognising that values between 300 and
1000\kms\ cannot {\em formally} be discarded at present.  Regarding
this, we suggest that ground based high resolution spectra of
Br$\alpha$ and with sufficient S/N to trace the line wings may
constitute the best observational constraint available to estimate
\vinf.
 
Likewise, we make use of the shape of H$\alpha$ and the
Br$\alpha$/\HeI~4.048$\mu$m complex to estimate $\beta$, 
the parameter characterizing the velocity law. We obtain $\beta=3$
indicating a relatively flat velocity field. Values of $\beta$ below 2
or above 4.5 can be ruled out  from the line fits.

We are, however,  unable to fully reproduce the blue shoulder
 of the H$\alpha$ line, 
noting that  this is the case irrespective of the terminal wind
velocity adopted in the modeling. In this regard we  highlight comparable   discrepancies 
between the synthetic and observed H$\alpha$ profiles
 for $\zeta^1$ Sco and HD~190603 (cf. an 
inability to reproduce the P Cygni absorption features in these stars; Figs.
 6 and 7) while in many cases the model fits to BSGs presented by
 both Crowther et al. (\cite{pacBSG}) and Searle et al. (\cite{searle}) 
also suffer similarly. 

Regarding Cyg OB2 \#12, the lack of changes in  polarisation through the line (Appendix B) strongly argues 
against any large  scale wind asymmetries  that might have been supposed 
to explain this feature. Hence we feel confident in the application of 
CMFGEN - which adopts spherical symmetry -  to this and other stars in this study,
while  noting that  this discrepancy might indicate  a shortfall in the physics employed.
Subject to this - and the preceding - caveats, we emphasise that the
presence of this disagreement between model and observations does not affect our
ability to determine the terminal velocity of the wind following the
methodology described above\footnote{For completeness we also explored the 
possibility that the discrepancy between observations and model could be due to the presence
of (spatially unresolved) blueshifted  nebular emission. Fitting an additional  simple gaussian profile 
to the blue  emission `shoulder'  in the  H$\alpha$ and $\beta$  lines, 
scaled 
assuming typical nebular line intensity ratios (H$\alpha$/$\beta \sim$2.9 
and H$\beta$/$\gamma \sim$2.2),
 resulted in 
significantly improved line fits. However, we caution that the adoption of such a solution appears 
premature at present as no other expected nebular emission lines such as  [N\,{\sc ii}],
 [S\,{\sc  ii}] or  [O\,{\sc i}] are present in the spectrum of Cyg OB2 \#12; 
further adaptive optics or coronographic observations to search for compact nebular emission 
would be of value to determine if such an approach is physically well motivated.}.

Finally, we were able to derive a mass-loss rate of \Mdot$=3.0 \times
10^{-6}M_{\odot}$yr$^{-1}$~(Table 4). Of course this value is bound to the
adopted \vinf\ and clumping law. Thus, if \vinf\ values of 300, 700
and 1000\kms were adopted, fits of similar quality would be obtained
for \Mdot $\sim2.5$, 4.0 and $6.5\times 10^{-6}M_{\odot}$yr$^{-1}$ respectively.
 
The main observational constraints which set the run of the clumping
law are the H$\alpha-\beta$ and Br$\alpha$ emission components and the
IR and submillimeter + radio continuum. An onset of the clumping at
relatively high velocity ( $\sim 200$\kms) is required to avoid strong
emission in the above lines and too great an excess in the IR
continuum. On the other hand, we find that a final clumping value of
0.04 is required to reproduce the submillimeter and radio continuum
(see Fig.\ref{fig-cont-fits}).
 
Our derived mass loss rate presents one of the major differences with
respect to previous works.  The significantly lower {\em clumped} mass
loss rate of \Mdot$\sim3.0\times10^{-6}M_{\odot}$yr$^{-1}$
corresponds to an unclumped (\Mdot/$f^{0.5}$) rate of
$\sim1.5\times10^{-5}M_{\odot}$yr$^{-1}$; a factor of $\sim$3 lower
than that reported by Leitherer et al.  (\cite{leitherer}), due to the
considerably smaller wind terminal velocity ($v_{\infty}$) adopted
(400 versus 1400kms$^{-1}$). The effects of the high degree of wind
clumping found for Cyg~OB2~\#12 mirror recent findings for other
massive (evolved) stars (e.g. Najarro et al. \cite{paco09}, Groh et
al. \cite{agcar}, \cite{hrcar}).

Nevertheless, such a {\em clumping corrected} value is in excess of that found for the less luminous and massive 
B supergiants  studied by Crowther et al. (\cite{pacBSG}) and Searle et al. (\cite{searle}). For wind velocities 
$v_{\infty} \leq400$~kms$^{-1}$ the clumping corrected mass loss rate approaches those 
of known LBVs (Fig. 11; Sect. 5), although the terminal velocity is comparable to BSGs of similar temperature.
Adopting  $v_{\infty}\sim1000$~kms$^{-1}$ leads to a mass loss rate comparable to those of the  LBVs, but such a wind
velocity is significantly greater than those determined for  both LBVs and BSGs of equivalent spectral type. 
 With respect to this, the lack of the low excitation metallic emission lines that characterise 
cool phase LBV spectra (Sect. 3) is a result  of the extremely large  radius found for  Cyg~OB2~\#12.

\begin{figure}
\begin{center}
\resizebox{\hsize}{!}{\includegraphics[angle=0]{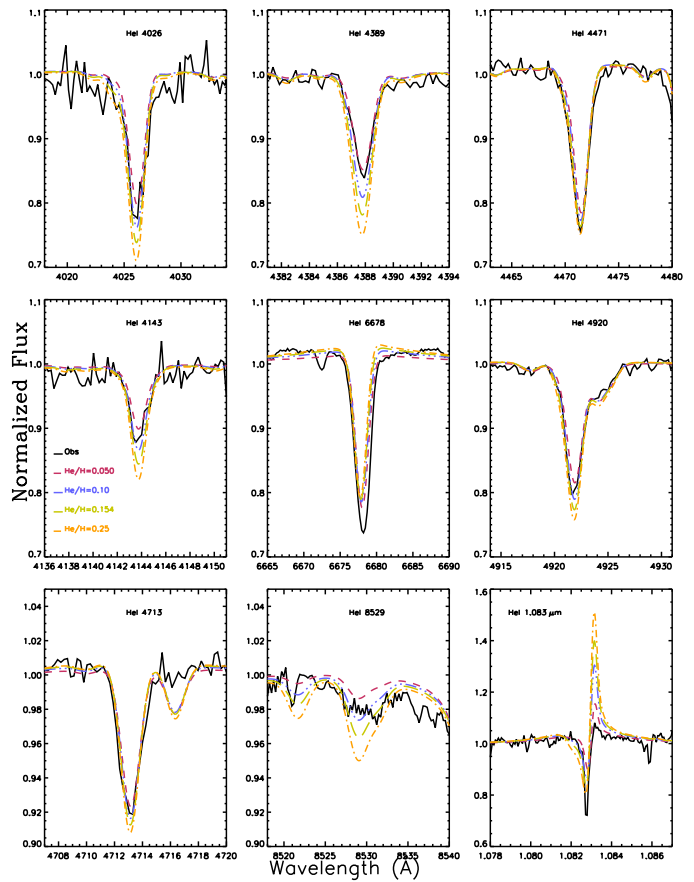}}
\caption{Montage of selected He\,{\sc i} lines in the spectrum of Cyg OB2 
\#12 with synthetic spectra computed utilising different He/H ratios 
overplotted.}
\end{center}
\end{figure}

\begin{figure}
\begin{center}
\resizebox{\hsize}{!}{\includegraphics[angle=0]{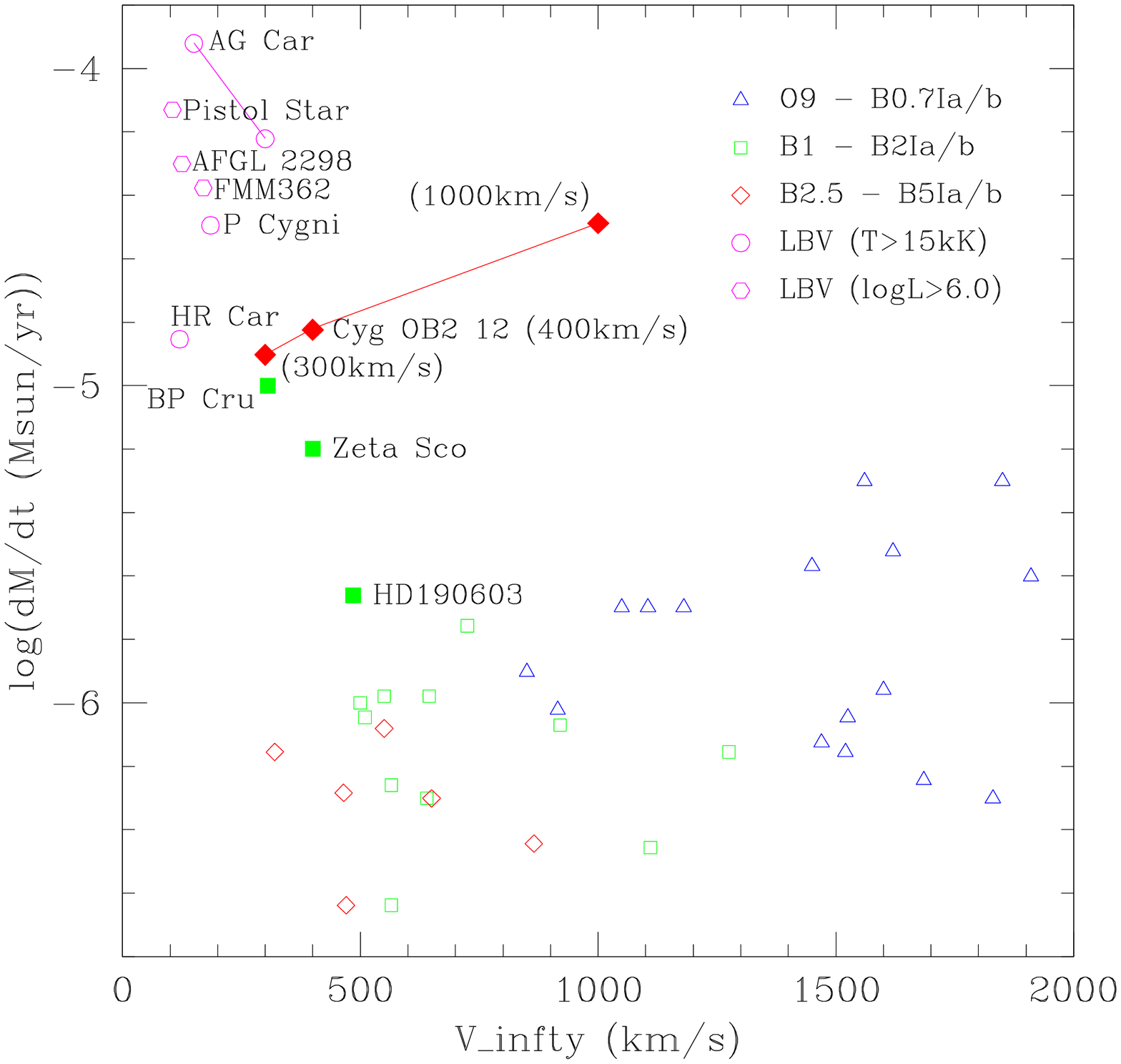}}
\caption{Comparison of the wind properties of the BHGs to field BSGs (plotted according to spectral type), LBVs 
of temperature selected to match the BSGs ($\geq$15kK) and those of comparable luminosity to Cyg~OB2~\#12 
(log$(L/L_{\odot}) >6.0$). Data from Crowther et al. (\cite{pacBSG}), Searle et al. (\cite{searle}), 
Najarro et al. (\cite{paco99},\cite{paco09}), Clark et al. (\cite{clark09}) and Groh et al. (\cite{agcar},\cite{hrcar}). Inclusion of later BSGs continues the trend to lower mass loss rates and
 wind velocities (e.g. Markova et al. \cite{markova}). Due to the uncertainty in the wind velocity for
Cyg~OB2~\#12,  we present 3 pairs of possible values for $v_{\infty} =300, 400$ and 
1000~kms$^{-1}$ (Sect. 4.1).} 
\end{center}
\label{fig-mdot-bsg}
\end{figure}

\begin{figure}
\begin{center}
\resizebox{\hsize}{!}{\includegraphics[angle=0]{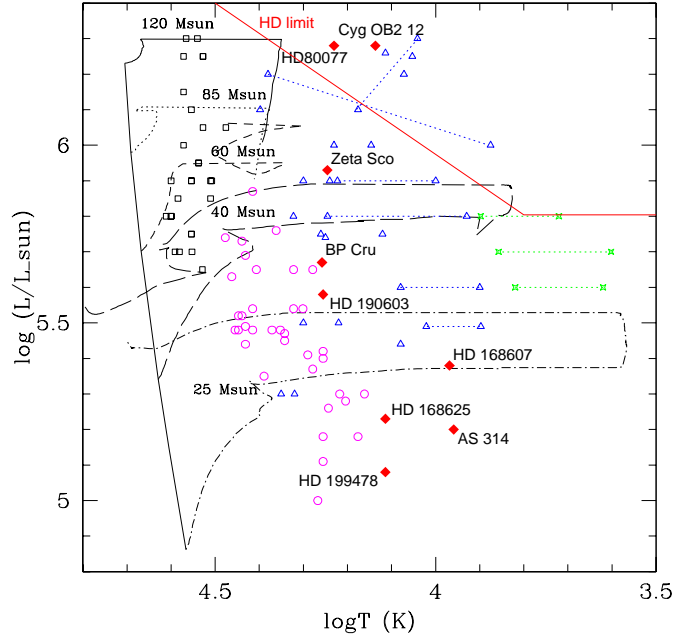}}
\caption{ HR diagram with the positions of the BHGs (red diamonds) plotted with respect to the evolutionary tracks of 
Meynet \& Maeder (\cite{meynet}) and the empirical HD limit. The position of HD~80077 {\em assumes} membership of 
Pismis 11 and hence a distance of 3.5~kpc (Marco \& Negueruela \cite{marco}).
 The  positions of other evolved stellar populations are also indicated: 
black squares -  WN7-9h, O4-6 Ia and  O4-6 If$^+$ stars within the Arches; 
purple circles - galactic 
field BSGs; blue  triangles - (candidate)LBVs and WNL stars; green stars - 
YHGs. Data from Martins et al. 
(\cite{martins07}, \cite{martins08}) Crowther et al. (\cite{pacBSG}), Searle et al. (\cite{searle}) and Clark et al. 
(in prep.).}
\end{center}
\end{figure}

\begin{figure}
\begin{center}
\resizebox{\hsize}{!}{\includegraphics[angle=0]{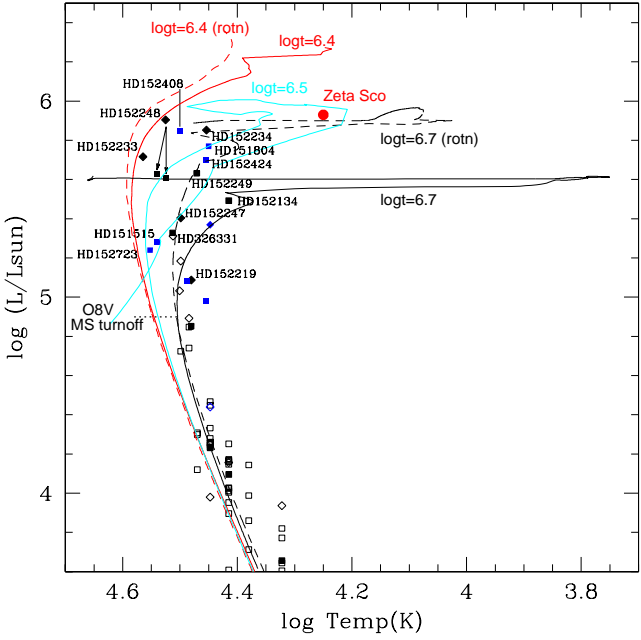}}
\caption{HR diagram for NGC6231 (black symbols) and the wider Sco OB1
  association (blue symbols) with the non rotating isochrones of
  Schaller et al. (\cite{schaller}; solid lines) and the rotating
  isochrones of (Meynet \& Maeder \cite{meynet}; dashed line)
  overplotted. Open and filled symbols represent main sequence and
  evolved stars, while diamonds indicate binarity.  For HD152248 we
  plot the position of both the individual components and also that of
  the integrated system. See Appendix B. for details of contruction.}
\end{center}
\end{figure}

\begin{figure}
\begin{center}
\resizebox{\hsize}{!}{\includegraphics[angle=0]{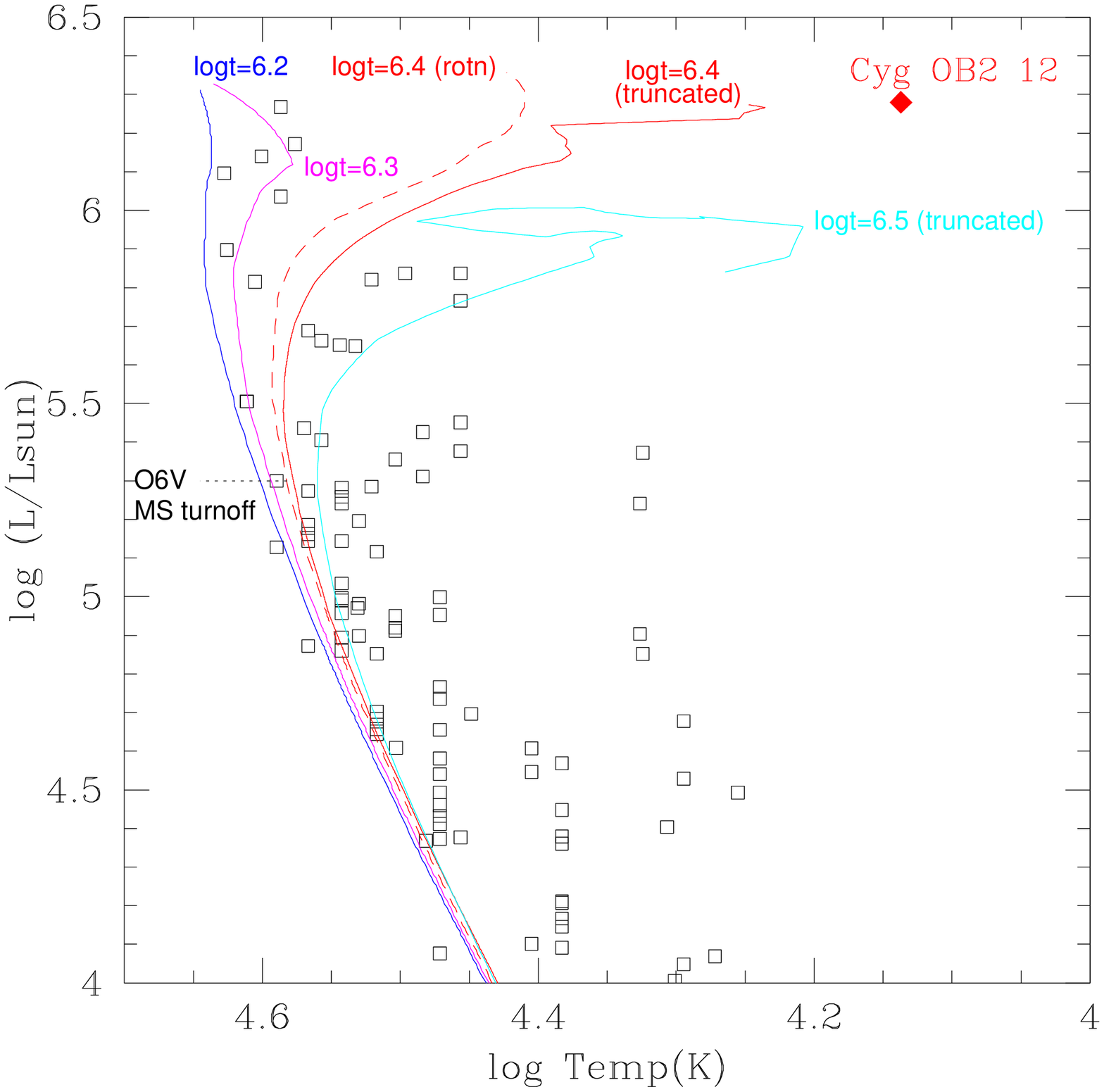}}
\caption{HR diagram for Cyg OB2 with the non rotating isochrones of
  Schaller et al. (\cite{schaller}; solid lines) and the rotating
  isochrone of (Meynet \& Maeder \cite{meynet}; dashed line)
  overplotted. Data from Negueruela et al. (\cite{iggy}) and Hanson et
  al. (\cite{hanson}); the reader is refered to these studies for the
  positions of individual stars, which have been omitted here for
  reasons of clarity.}
\end{center}
\end{figure}

\paragraph{Abundances}

Fortunately, unlike stars such as HDE~316285 (Hillier et al. \cite{316}), Cyg OB2 \#12 does not suffer from
a  $T_{\rm eff}$/He-abundance degeneracy. Since, as previously  described, 
$T_{\rm eff}$ can be accurately determined from e.g. the 
\SiIII/\SiII ~or \OII/\OI ~equilibria, we may 
therefore reasonably constrain  the He/H ratio. In Fig. 10 we overplot 
synthetic spectra constructed with a wide range of He/H abundances on selected 
He\,{\sc i} transitions; note that as expected the best constraints are provided by the weaker, non saturated, transtions. This reveals one 
 of the fundamental results of our  study - that  Cyg OB2
\#12 demonstrates a  solar H/He=10 ratio (by number).
Unlike the rest of BHGs analysed in this paper (see Table~\ref{tab:model}), 
which clearly demonstrate surface helium
enrichment, the \HeI\ lines of Cyg~OB2~\#12 are best reproduced
assuming no helium enrichment at all 
(noting that previous studies of  BSGs 
{\em adopted} H/He=5 for all stars considered; Crowther et al. \cite{pacBSG}, Searle et
al. \cite{searle}). Furthermore, even better fits to some of the strong
\HeI\ lines in the $R$- (\HeI~6678{\AA}), $I$- (\HeI~8581 and 8845{\AA})
and $Z$-bands (\HeI~10830{\AA}) are obtained if we assume He to be
underabundant. Moreover, we find that H/He$\sim$8.0 appears to provide a robust lower limit to the  abundance
ratio, with lower values resulting in unacceptably poor fits to the data 
(Fig. 10).

We note that given the current stellar temperature and
luminosity, the lack of He enrichment at the stellar surfaces is at
odds with the predictions from evolutionary models and challenges
present theory of stellar evolution of very massive stars.  We discuss
this somewhat unexpected finding, and the implications for   the evolutionary
state of Cyg~OB2~\#12 in Sect. 5.  From Table~\ref{tab:model} we see
indications of CNO processing (N enhancement and CO depletion) in Cyg
OB2 \#12, though to a lesser extent than for $\zeta^1$~Sco and
HD190603.  This result is consistent with the derived H/He ratios.

\subsubsection{The unexpected luminosity  of  Cyg~OB2~\#12}
A critical finding of this analysis is the rather extreme luminosity
of Cyg~OB2~\#12 (e.g. Fig. 12 and Sect. 5); are we overestimating this?
An obvious explanation is that a chance alignment with its `host'
association leads us to overestimate its distance.  If this were the
case, the large degree of reddening demonstrated by Cyg~OB2~\#12 would
either be due to a (narrow) sightline of anomalously high interstellar
extinction towards {\em it and only it} - but which nonetheless
yielded Diffuse Interstellar Bands (DIBs) comparable to those found
for other {\em bona fide} association members (e.g. Hanson
\cite{hanson}) - or would be circumstellar in origin (e.g. Massey \&
Thompson \cite{massey}), thus permitting a more normal run of
interstellar reddening for a lower distance.  However, our model fits
to the expanded SED show no evidence for an IR excess at wavelengths
shorter than 30$\mu$m; suggesting that any circumstellar component
would have to be significantly cooler than observed around known LBVs,
where pronouned emission from warm dust is detected at $\leq$25$\mu$m
(e.g.  Egan et al. \cite{egan2}, Clark et al. \cite{clark03}).

A second alternative is that it is a multiple system. Given the recent
suggestion of a high binary fraction within CygOB2 (Kiminki et
al. \cite{kiminki2}) and the hard, luminous X-ray emission from Cyg
OB2 \#12 (Albecete Columbo et al. \cite{albacete}, Rauw et
al. \cite{rauw}) this would appear to be physically well motivated.
To reduce the luminosity of Cyg~OB2~\#12 such that it was marginally
consistent with current theoretical predictions (Sect. 5) would
require a companion(s) of equal luminosity.  However clear
spectroscopic signatures of binarity in terms of (i) radial velocity
shifts, (ii) the presence of double lines and/or (iii) the dilution of
spectral features appear absent\footnote{In particular we note that no
  (anti-phased) radial velocity shifts indicative of binarity are
  present in the additional emission components of the H$\alpha$,
  $\beta$ and $\gamma$ lines.}.  Moreover, the excellent model fit to the 
observed spectral energy distribution also excludes a companion of comparable luminosity
but very different temperature.

 While not fatal to a binary scenario, these observational constraints
would require any putative binary companion of {\em comparable}
luminosity to have an identical spectral type - to avoid its spectral
signature being visible - and for the binary to either be very wide or
seen face on - to avoid RV shifts in the spectrum. However, the
temperature of the hard X-ray component ($kT=1.8$keV) implies, via the
strong shock condition, that the companion would have to have a wind
velocity of $\sim$1300kms$^{-1}$ - inconsistent with the emission
arising in the wind collision zone produced by such a `twin' and
necessitating the presence of a third, unseen star to yield the high
energy emission.  While we cannot exclude such a `finely tuned'
hierarchical system, we find no compelling evidence for one either.

Lastly, one might suppose that Cyg~OB2~\#12 were currently undergoing
a long-term`LBV eruption' resulting in an increased
luminosity. However, while the physical properties of such events are
ill constrained we find the current physical properties to be distinct
from those of quiescent LBVs (e.g. Sect. 5), while mass loss rates of
$>>10^{-5}M_{\odot}$yr$^{-1}$ are typically inferred for such
stars in outburst (Clark et al. \cite{clark09}).  Therefore, while we
may not formally exclude any of these possibilities, we find no
compelling observational evidence to support them either and hence
consider it likely that the primary BHG within Cyg~OB2~\#12 is indeed
unexpectedly luminous in comparison to theoretical predictions.
Moreover, even if its luminosity were reduced via one of these
scenarios such that it was consistent with current evolutionary
tracks, we note that the revised mass loss rate (\Mdot $\propto
d^{1.5}$) and (unchanged) chemistry would still remain discrepant with
respect to field BSGs and BHGs.

\subsection{$\zeta^1$~Sco and HD~190603}
\label{sub-sco190}
For the analysis of $\zeta^1$~Sco we adopted a distance
$d\sim1.64$~kpc, based upon the revised estimate for NGC6231/Sco OB1
(Appendix C); for convenience we also show results for the slightly
larger distance of $d\sim1.9$~kpc adopted in previous works to enable
a direct comparison(Crowther et al. \cite{pacBSG}). Unfortunately, HD~190603 is not associated with a
host cluster making the distance significantly more uncertain. For
consistency to prior studies we adopt an identical distance
($d\sim1.57$~kpc; Crowther et al. \cite{pacBSG}), while also presenting results under the assumption
that it has an identical intrinsic visual magnitude to $\zeta^1$~Sco.

As in the case of Cyg~OB2~\#12, several ionization equilibria could be
used to determine the stellar temperature. Interestingly we could use
three ionization stages of silicon (\SiIV/\SiIII/\SiII) for both
objects. In the case of HD190603, our hottest object, the \FeII\ and
\NI\ lines are not detected so we used them in our models as
indicators of a lower limit for \Teff. On the other hand,
\HeII~4686{\AA} was weakly detected on HD190603 allowing to use as
well the \HeII/\HeI\ ionization criterium. 

Since UV spectra are available for both objects, the terminal velocity, \vinf, could be
accurately determined for both objects (e.g. Fig. 5). Given that we are fitting non-simultaneous observations from the 
UV to the IR we might expect to find some transitions in this wavelength range which are not well fitted by our synthetic spectra. This 
appears to be the case for C\,{\sc iv} 1548-1551{\AA}, which is significantly underestimated by our models. Moreover these, as well as the Si\, {\sc iv} 1394-1403{\AA}
UV lines, are very sensitive to X-rays in this parameter domain; thus they should be regarded with caution.
 
The main stellar properties of both objects are presented in
Table \ref{tab:model}. Uncertainties in stellar parameters are similar
to those derived for Cyg~OB2~\#12, except for the effective
temperature, which is slightly better constrained $\Delta {T_{\rm 
eff}}={\pm}
500$K and the terminal velocity with $\Delta{v_{\infty}}={\pm} 50$kms$^{-1}$. Errors on the H/He ratio are of the order of $\sim$20\%. Interstellar reddening was also determined;  for $\zeta^1$~Sco
we found $E(B-V)=0.66$ and   $R_V=3.3$ and for HD~190603 $E(B-V)=0.70$ and  
 $R_V=3.1$.

We found both stars to be less extreme than Cyg~OB2~\#12 in terms of
luminosity, although $\zeta^1$~Sco is still significantly in excess of
the range of luminosities spanned by Galactic O9-B5 Ia/b supergiants
(Searle et al. \cite{searle}, Crowther et al. \cite{pacBSG}), even
with the downwards revision of the distance. HD~190603 lies in the
upper reaches of this distribution, but we caution that this could be
subject to revision given the uncertainties in its distance estimate. As expected
from their earlier spectral types, 
they are hotter than Cyg~OB2~\#12 , although the temperatures we find  
are slightly lower ($<$1kK) than previous estimates for these objects.
These lower temperature estimates are due in part to the effect of the
wind transition region (see Sect.~\ref{sub-cyg12}) and to the use of
all available ionization equilibria criteria.  Following from these
results, spectroscopic mass estimates are significantly lower than
that of Cyg~OB2~\#12 for both stars.

Terminal wind velocities are similar to Cyg~OB2~\#12, but lie at the
lower range of values found for BSGs of comparable spectral type; in
contrast the {\em clumping corrected} mass loss rate for HD~190603
lies at the upper range of those found for B1-2 Ia/b supergiants while
that of $\zeta^1$~Sco is considerably in excess of this range
(Fig. 11). These compare favourably with previous estimates for these
stars (Searle et al. \cite{searle}, Crowther et al. \cite{pacBSG},
Markova \& Puls \cite{markova}); although as with Cyg~OB2~\#12, the
high degree of clumping required to fit the spectra provide a
significant downwards revision of {\em absolute} mass loss rates
between this and prior studies.  Values of log$g$ are also lower
than previous estimates of both stars, due to the greater accuracy
afforded us by the inclusion of the higher Balmer and Paschen lines in
the spectroscopic datasets.

Finally, in contrast to Cyg~OB2~\#12, but in line with estimates for
field Galactic BSGs, we find a moderate H-depletion
(or He-enrichment; H/He$\sim$5  by number) for both stars as well as 
evidence for stronger CNO processing (higher N).

\section{Discussion}

While the galactic BHG population spans over an order of magnitude in luminosity (Fig. 12), when 
combined with previous modeling results for BP~Cru (Table 4; Kaper et
al. \cite{kaper06}), our analyses show the early B1-4 Ia$^+$
hypergiants to be, as expected, rather luminous evolved stars, with
 HD~80077 and CygOB2 \#12
appearing to lie above the empirical HD limit (Fig. 12; Marco \& Negueruela \cite{marco}).
In comparison to Galactic BSGs of the same spectral subtypes 
they are overluminous and support significantly higher mass loss
rates, with wind terminal velocities at the lower end of the range
populated by the BSGs.

\subsection{The evolutionary status of BHGs}

Subject to these observational constraints how may {\em single}
(early) BHGs be understood in evolutionary terms?  Several authors
have presented evolutionary sequences for massive ($>30M_{\odot}$)
stars, including Langer et al.  (\cite{langer}), Crowther et
al. (\cite{pacevol}), Meynet \& Maeder (\cite{meynet})  and Martins et al. (\cite{martins07},
\cite{martins08}). After the initial O-type main-sequence
 phase, the recent {\em empirical} evolutionary scheme of the latter authors may be summarized as:\newline

$\bullet \sim60-120M_{\odot}$: O Ia $\rightarrow$ OIaf$^+$ $\rightarrow$ WNL + abs $\rightarrow$ WN7
\newline

$\bullet \sim30-60M_{\odot}$: O/B Ia $\rightarrow$ WN9-11h $\rightleftharpoons$ LBV $\rightarrow$ WN8 $\rightarrow$ WN/C
\newline

While not explicit in the above scheme, the properties of LBVs such as AG Car (Groh et al. \cite{groh11})  and 
the Pistol star (Najarro et al. \cite{paco09}) strongly imply that even very massive $M_{\rm ini}>60M_{\odot}$
stars may experience an LBV phase. Moreover, it is likely that a further
subdivision will occur for $M_{\rm ini}<40M_{\odot}$ to accommodate 
the high luminosity (log$(L/L_{\odot}) \sim 5.8$) cool hypergiants
present in clusters such as Westerlund 1 (Wd~1; Clark et
al. \cite{me}):\newline

$\bullet \sim30-40M_{\odot}$: O $\rightarrow$ OB Ia $\rightarrow$ cool Ia$^+$ $\rightarrow$ WN 
$\rightarrow$ WC\newline

 \noindent
noting that the evolution of spectral classification through the red
loop and the final state reached prior to SN is particularly
uncertain.

Given that the physical properties of the BHGs appear more extreme
than the field BSGs and that they appear co-located with the LBVs on
the HR diagram (Fig. 12) a close physical association is suggested and
we propose that the early BHGs - {\em with the notable exception of
  CygOB2 \#12} - are the immediate evolutionary progenitors of LBVs
for stars of $\sim 40-60M_{\odot}$. Thus, they would form lower mass
analogues of the O4-6 Iaf$^+$ stars in the Arches, which Najarro et al.  (\cite{paco04}) and
Martins et al. (\cite{martins08}) demonstrate are intermediate between the O4-6
Ia and WN7-9ha stars both in terms of the degree of (increasing)
chemical evolution and also wind properties, with advancing
evolutionary state leading to a reduction in terminal wind velocities
coupled with an increase in mass loss rate.

The BHGs appear to  follow an analogous evolutionary template, with wind
properties approaching those of the LBVs (Fig. 11).  Moreover, the
BHGs also appear less chemically evolved than (candidate)
LBVs\footnote{HD~316285 (H/He$\sim$1.5), AG Car (H/He$\sim$2.3), W243
(H/He$\sim$5), P Cygni (H/He$\sim$3.3), FMM362 (H/He$\sim$2.8) and
the Pistol Star (H/He$\sim$1.5; Hillier et al. \cite{316}, Groh et
al. \cite{agcar}, Ritchie et al. \cite{ritchie}, Najarro et
al. \cite{paco01}, \cite{paco09}, respectively).}, again suggesting
that they occupy an earlier evolutionary phase. Such an hypothesis
appears bolstered by the apparent lack of long-term(secular)
variability demonstrated by the early-B HGs
(Sect. 3), although we are aware that {\em bona fide} LBVs such as P
Cygni may also encounter long periods of quiescence.

We emphasise that this evolutionary scheme is appropriate for single
stars, but it also appears possible that early-B HGs 
may form via binary interaction and, through such a channel,
potentially from lower mass stars. Wellstein \& Langer
(\cite{wellstein}) present the results of simulations that show the
properties of the known binary BP~Cru are consistent with quasi-conservative mass
transfer in a $26M_{\odot}+25M_{\odot}$ progenitor 
system. Likewise, the BHG/WNVL transitional object Wd1-13 is the
primary of a $23.2M_{\odot}+35.4M_{\odot}$ 9.27~day period binary
(Ritchie et al.  \cite{magnetar}). 
This configuration is too compact for the primary to have passed
through a BSG phase, and hence presumably formed via an episode of
binary mediated mass loss (Sect. A.5), with a likely progenitor mass
of $\sim35M_{\odot}$. In this respect we highlight that Wd1-13
appears spectroscopically distinct from the BHGs considered here,
showing pronounced (variable) emission lines in the Paschen series
(Fig. A.2 and Ritchie et al. \cite{magnetar}).

Finally, is it possible to place the later (lower luminosity) BHGs
into comparable sequences? The consistency of the population of B1-3
Ia and cooler B5-9 Ia$^+$ stars within Wd~1
with theoretical evolutionary isochrones (Negueruela et
al. \cite{iggy10}) suggests an extension
of the paradigm that BHGs are the more physically extreme direct
descendants of hotter O/BSGs to lower progenitor masses
($\sim35M_{\odot}$). However, the placement of these BHGs in a pre-
or post-RSG/LBV phase is currently uncertain in the absence of
tailored abundance analyses - ESO/VLT observations currently underway
will allow this to be directly addressed. Finally, following previous
suggestions in the literature (Sect. A.5) - and again in the absence
of quantitative analyses - it appears likely that lower luminosity
(log$(L/L_{\odot})<5.5$) BHGs are post-RSG objects possibly
encountering an LBV phase.

\subsection{Constraints from the host stellar aggregate}

Do other observations support such an evolutionary scheme? Following
the above discussion, the stellar populations of both the Arches and
Wd~1 are clearly consistent. However we would also predict the presence of
early BHGs within the Quintuplet cluster, which is likely to be
intermediate in age between these clusters (Figer et
al. \cite{figer}). While neither these authors nor Liermann et
al. (\cite{liermann}) report the presence of BHGs, we note that
e.g. LHO96, 100 and 146 - which are classified as O6-8 If by Liermann
et al. - have K band spectra identical to that of $\zeta^1$~Sco
(Fig. 6). Simply applying the bolometric correction derived from our
analysis of $\zeta^1$~Sco yields an estimated
log$(L/L_{\odot})\sim6.0-6.3$ for these stars; directly comparable
to the Pistol Star and FMM362 (Najarro et al. \cite{paco09}) and hence
consistent with a pre-LBV BHG identification.  Further high S/N and
resolution observations of these stars  would permit
quantitative testing of this assertion by virtue of an accurate
determination of temperature, luminosity and wind properties.

Encouragingly, stars with a similar K band spectral morphology are
also present in the Galactic Centre cluster, where they too are found
to be of higher temperature and possess faster, lower mass loss rate
winds than the more evolved WNL - and candidate LBV - stars present
(Najarro et al. \cite{paco97}, Martins et al. \cite{martins07}).

Importantly, are the properties of $\zeta^1$~Sco and Cyg~OB2~\#12 and
their host stellar aggregates consistent with this evolutionary
hypothesis? Unfortunately, observations of Cyg OB2 and NGC6231/Sco OB1
raise the possibility of the stars in either  association being non-coeval
(Negueruela et al. \cite{iggy}; Appendix C). Nevertheless, the
physical properties of $\zeta^1$~Sco appear consistent with the
population of (initially) rapidly rotating, 5~Myr-old stars present
within NGC6231 (Fig. 13 and Appendix B; noting that the low {\vesini}
of $\zeta^1$~Sco may be understood as a combination of spindown and/or
an unfavourable inclination) {\em or} a somewhat younger
($\sim3.2-4$~Myr) population of less rapidly rotating objects that
{\em may} also be present (Appendix B). This would imply a progenitor
mass of $\sim40-60M_{\odot}$ for $\zeta^1$~Sco, depending on age
and initial rotation, broadly consistent with the current
spectroscopic mass of $\sim36M_{\odot}$.

Finally, we address Cyg~OB2~\#12. As highlighted previously, this star
appears difficult to accomodate in the current evolutionary scheme due
to the combination of high luminosity and low temperature displacing
it from any theoretical isochrones applicable to Cyg OB2 (Fig. 14;
Negueruela et al. \cite{iggy}). While its co-location in the HR
diagram with {\em bona-fide} LBVs such as the Pistol Star, FMM 362 and
AFGL 2298 (Fig. 12) suggests a similar nature, the long-term stability
combined with its relatively unevolved chemsitry - and critically
solar H/He ratio - appear to exclude such an identification. We are
thus currently unable to place Cyg~OB2~\#12 into a coherent
evolutionary scheme, noting that with a {\em current} spectroscopic
mass of $\sim 110M_{\odot}$ - significantly in excess of both the
Pistol Star and FMM 362 (Najarro et al. \cite{paco09}) - one might
expect it to evolve through a much hotter O4-6I/Iaf$^+$/WNLha
evolutionary sequence. 

In this respect, the results of a quantative
analysis of HD~80077 would be of particular interest, given its
apparent similarly extreme luminosity (Marco \& Negueruela
\cite{marco}). Such an analysis has been performed for the M33
 B1 Ia$^+$ star [HS80] 110A by Urbaneja et al. (\cite{urbaneja}) which yielded stellar
 parameters that, like Cyg OB2 \#12,  both place it above the HD-limit and appear to distinguish 
it from known LBVs\footnote{log$(L/L_{\odot})\sim6.5$, $T_{\rm eff} \sim 21$kK, $v_{\infty} \sim750$kms$^{-1}$ and 
{\Mdot}$=1.06{\times}10^{-5}M_{\odot}$yr$^{-1}$.}. This finding appears to  suggest that there might indeed be an hitherto unidentified (but rarely 
traversed?) evolutionary pathway that leads to the production of very luminous but comparately cool BHGs at very high stellar masses.

\section{Conclusions}

In order to determine the physical properties and hence evolutionary
state of Galactic early-B hypergiants, 
we have undertaken new quantitative analyses of CygOB2 \#12, $\zeta^1$
Sco and HD~190603 and employ previous model results for BP~Cru.  
Synthetic spectra and SEDs were calculated and compared to
comprehensive UV/optical--radio spectroscopic and photometric datasets
compiled from the literature and supplemented by new and previously
unpublished observations. Building on this effort, we also constructed
exhaustive spectroscopic and photometric histories for all Galactic
BHGs in order to test the assertion that they are physical
identifiable with LBVs.

Turning first to stellar variability and both the early- and late-B HGs 
demonstrated the rapid ($\sim$days)
spectroscopic and photometric variability that
is symptomatic of the aspherical wind substructure and photospheric
pulsations that characterises luminous early stars. While low-luminosity, late-B HGs
have long be known to undergo LBV excursions (e.g. HD~160529) we found
no evidence for such secular variability amongst the early, high
luminosity BHGs, with the possible exception of $\zeta^1$~Sco  in the sparse photometric data
obtained prior to the 20$^{th}$ Century. In this regard it is interesting 
 that both Cyg~OB2~\#12  and HD~80077 appear to lie above the empirical HD limit.

Model results reveal that the early BHGs have physical properties -
luminosity, wind terminal velocity, mass loss rate and chemical
abundances - intermediate between Galactic BSGs and LBVs of comparable
temperature, suggesting they provide a link between both evolutionary
phases for stars of inital mass between $\sim40-60M_{\odot}$. In
this respect, they would play a similar role to the more massive and
luminous O4-6Iaf$^+$ objects for stars in the $\sim60-120M_{\odot}$
range (Najarro et al. \cite{paco04}, Martins et al. \cite{martins08}). The simultaneous presence of
populations of early B1-3 Ia supergiants and mid-late B5-9 Ia$^+$ BHGs
within Wd~1 suggest that BHGs are potentially the immediate
descendents of stars with initial masses as low as
$30-35M_{\odot}$. However, the presence of early BHGs within this
cluster indicates that a parallel channel may also lead to this phase
via close binary evolution.

Nevertheless, Wd~1 and the Quintuplet provide direct observational tests of the
evolutionary sequence proposed; in the former cluster the BHGs should be more
chemically evolved than the BSG population but less than the RSGs that
are present, while in the latter cluster one would expect to find a population
of early BHGs; observations aimed at verifying both hypotheses will be
undertaken later this year.

The apparent physical association of $\zeta^1$~Sco with NGC~6231/Sco~OB1 also allows us to test this  
scenario. Encouragingly, we find that the physical properties derived from our analysis are consistent with the observed 
stellar population of this region and in turn with theoretical isochrones for stars of between 3.2-5~Myr age - further 
discrimination is difficult in the absence of an absolute rather than projected rotational velocity. 

Conversely, the 
combination of  extreme luminosity and cool temperature of CygOB2 \#12  is inconsistent with theoretical isochrones,
 {\em assuming membership of Cyg OB2}. Given its co-location with a handful of {\em bona fide} LBVs above the HD limit 
it might be supposed to be a similar object, but a combination of (i) a lack of the characteristic LBV variability, (ii) extreme {\em current} 
stellar mass and (iii) lack of chemical evolution differentiate it from these. We are therefore currently unable to place
this star in a consistent evolutionary scheme, noting  that Negueruela et al. (\cite{iggy}) also  find that a number of 
the hottest and most luminous stars within  CygOB2 are difficult to accommodate in a formation scenario for the 
association where the most recent epoch of star formation occurred $\sim$2.5~Myr ago.

Nevertheless, we note that the stability of CygOB2 \#12 for the last century 
would appear to indicate that the   HD limit does  not appear to define a region of the HR diagram 
utterly inimical to the presence of  massive stars, but a combination of (pulsational?) instabilities
 and extreme  mass loss rates presumably prevent stars residing  above it for large fractions of their lifetime. 
Indeed, from Fig. 12 we may see that in addition to highly luminous BHGs such as Cyg~OB2~\#12 the {\em empirical} HD limit appears to 
be delineated by LBVs and other closely related (variable) stars  such as the cool hypergiants and the WNVLs, further suggesting that it 
should not be regarded as a firm barrier to stellar stability, nor the location at which  the onset of instabilities occurs. In this regard the 
B1 Ia$^+$ star [HS80] 110A and the cool F-hypergiant B324 (both located in M33) are of considerable interest; like Cyg OB2\#12, the 
combination of  temperatures and 
luminosities they demonstrate are not replicated by current evolutionary tracks and place both stars  well above the HD limit (Urbaneja et al. 
\cite{urbaneja}, Clark et al. \cite{clark12}). Quantitative modeling of the Galactic BHG HD~80077 would therefore be of considerable 
interest to see 
whether this too violated the HD limit.

Moreover, despite residing above the HD limit,  we find an Eddington 
parameter, $\Gamma_{\rm Edd}$, of only  $\sim0.41$  (and  
$\Gamma_{\rm Total}\sim0.56$  at $R(\tau_{\rm Ross}=2/3)$ for  Cyg OB2 \#12; 
below the Eddington limit.  
Unfortunately, in the absence of an inclination we may not determine how 
close Cyg OB2 \#12 is to the rotationally modified Eddington limit, 
although we note that the Eddington parameter for it is  lower 
than Groh et al. (\cite{groh11}) found for AG Car at any epoch. Given 
that AG Car is also a known rapid rotator but has not been observed at,  
or to have exceeded, the rotationally modified Eddington limit we suspect 
the same to be true for Cyg OB2 \#12. From this we might therefore 
conclude that the  HD-limit is not the direct result of a star (b)reaching 
the rotationally  modified Eddington limit. Indeed,  Groh et 
al. (\cite{groh11}) show that AG Car most closely approaches the rotationally modified Eddington
 limit during its  hot, compact phases when, in contrast to its cool extended state, it lies 
{\em beneath} the HD limit (Fig. 12).

\begin{acknowledgements}

JSC acknowledges support from an RCUK fellowship.
This research is partially supportedby the Spanish Ministerio de
Ciencia e Innovaci\'on (MICINN) under
grants AYA2008-06166-C03-02/03,  AYA2010-21697-C05-01/05  and CSD2006-70.
MAU acknowledges support by NSF under grant AST-10088798
We also thank John Hillier for providing the CMFGEN code and Dan Kiminki,
Otmar Stahl, Philip Dufton and Sergio Sim\'on-D\'{\i}az for providing 
electronic versions of published and unpublished stellar spectra.

\end{acknowledgements}

{}

\appendix

\section{Summary of historical data for Galactic BHGs}

In order to understand the role of the  BHG phase in the lifecycles 
of massive stars it is invaluable to contrast the long term
(in)stability of such stars to other, potentially related evolutionary
phases such as the LBVs. In order to accomplish this we undertook a
literature review of all known Galactic examples. This has 
allowed the construction of the comprehensive spectroscopic and
photometric datasets used in the quantitative modeling descibed in
Sect. 4.  To the best of our knowledge there are 16 known or candidate BHGs
within the Galaxy, of which 8 have early (B1-4) spectral types.
In constructing this list we have explicitly included 3 high
luminosity stars of late-B
spectral type which, although classified as luminosity class Iae, have
prominent Balmer emission lines that appear characteristic of {\em
  bona fide} BHGs.

\subsection{The early-B HGs: Cyg~OB2~\#12}

\subsubsection{Photometric variability}

Gottlieb \& Liller (\cite{gottlieb}) provide an extensive optical photometric history for Cyg OB2  \#12, comprising, where available,
 annual mean $B$ band magnitudes  from $\sim$1890-1980. They report low level annual variability
 ($m_{B}\sim14.5-15.0$mag.) albeit with no apparent secular evolution.
Despite being somewhat limited, the compilation  of multi-epoch  
$UBVRIJHK$ photometry presented in Table A.1 is consistent with little
photometric activity being present\footnote{We note that in common with
  early observations of all the BHGs considered, interpretation is
  complicated by the lack of a reported date for some observations - a
  problem also present for the spectroscopic observations.}.

 While  systematic  photometric 
monitoring of Cyg~OB2~\#12 ceased after 1980 - fortuitously at a point when 
spectroscopic observations increased in frequency (Sect. A.1.2) - two further long-term photometric datasets are available. 
The first, acquired via the  Hipparcos mission (ESA \cite{ESA}, Perryman et al. \cite{perry}), provides data from 1989 November to 
1993 March, with the passband, $H_p$,  spanning 3400-9000{\AA} (van Leeuwen et al. \cite{vanL97}). Subsequently, the 
Northern Sky Variability Survey (NSVS; Wo\'{z}niak et al. \cite{wozniak}) yielded a further year long dataset (1999 
April - March 2000); this  operated without filters, resulting in a  wide optical band with a response defined by 
that of the CCD,  extending from 450 to 1000nm. 

These data are summarised in Fig. 3 
and although the non-standard passbands do not allow a direct comparison to previous 
data they do constrain variability over the periods in question. As with the results of Gottlieb \& Liller 
(\cite{gottlieb}), these datasets indicate that  low level ($\sim$0.3mag) non secular and aperiodic variability is
 present, which occurs  over rather short ($\sim$days) timescales. 

While only single epoch mid-IR observations are available (Table A.2), multiple epochs of  radio continuum observations are 
available in the literature, and are summarised in Table A.3. While the spectral indices derived from these data are
consistent with thermal emission from a partially optically thick stellar wind, these fluxes  demonstrate significant 
variability. As with the short period photometric variability, a possible  interpretation of this behaviour is provided in Sect. 3.

\subsubsection{Spectral classification and variability}

\label{app-sub-spec-class}

A summary of the multiwavelength spectroscopy obtained for Cyg OB2
\#12 over the past half century is presented in Table A.4. The
spectral classifications given are those reported by the works in
question; reclassification is provided only for those occasions where
the source papers assumed an earlier result (this being the case for
the IR studies, where Cyg~OB2~\#12 was typically included as a
spectral standard).  Regarding reclassification, a luminosity class of
Ia$^+$ is implictly assumed throughout the following
discussion. \newline

\paragraph{\bf 4-6000{\AA}:} Souza \& Lutz (\cite{souza}), Massey \&
Thompson (\cite{massey}), Klochkova \& Chentsov, (\cite{klochkova})
and Kiminiki et al. (\cite{kiminki}) all provide a detailed discussion
of the classification of Cyg~OB2~\#12 
from 4000-5000\AA\ spectra at high resolution and S/N,
using the He\,{\sc i} 4471{\AA}:Mg\,{\sc ii}
4481{\AA} line ratio as the primary spectral-classification criterion.
Following this methodology, spectral types in the range B3-B8 have
been inferred by these authors, with variability in the line ratio
(e.g. Kiminiki et al. \cite{kiminki}) taken as evidence for changes in
the {\em spectral type} of Cyg~OB2~\#12. 

However, we caution that extreme care must be taken in inferring changes 
in the {\em physical properties} of the star from this diagnostic.
Firstly, the Mg\,{\sc ii} 4481{\AA} transition shows a  dependence on stellar luminosity
and so callibrations of spectral type versus the  He\,{\sc i} 4471{\AA}:Mg\,{\sc ii} 4481{\AA}
line ratio derived from lower luminosity stars may not {\em a priori} be directly 
applied to  Cyg~OB2~\#12. More critically and  as demonstrated in Sect. 4.1,
this ratio is highly sensitive to the properties of the
photosphere/wind transitional zone. Consequently, changes in
the structure of the inner wind - which we may infer from line profile
variability in H$\alpha$ (see below) - at {\em constant stellar temperature} will 
cause this line ratio to vary, and in turn lead to the erroneous conclusion that  
because the spectral type has varied then the  {\em temperature has also changed}.
\newline

Given the quality of their spectra - and in conjunction with data
presented in
Walborn \& Fitzpatrick (\cite{walborn90}) - Kiminki et
al. (\cite{kiminki}) also employed the weak absorption feature at
$\sim$4542{\AA} as a classification criterion, which they attributed
to a blend of He\,{\sc ii} and Fe\,{\sc ii}.  While an absorption
feature at $\sim$4542{\AA} is observed in B1.5-4 supergiants, it is
unlikely to be due to He\,{\sc ii} (seen only for B0 and earlier
supergiants) or Fe\,{\sc ii} (B5 and later). Given these
uncertainties, we instead choose to employ the Si\,{\sc iii}
4552{\AA}/Si\,{\sc ii} 4128{\AA} line ratio as a primary temperature
diagnostic. Thus, upon consideration of the above, we suggest a
spectral classification of B5 for the 1992 July 22 spectrum and B4 for
the 1998 August, 2000 September and 2008 July spectra (Fig. A.1),
noting that the marginally later classification in 1992 is likely to
be a result of the lower S/N and spectral resolution of these data. We
thus find no evidence for changes in the spectral type of Cyg~OB2~\#12
in these data. \newline

\paragraph{\bf 6-8000{\AA}:} The spectral region surrounding H$\alpha$
is largely devoid of classification criteria (e.g. Negueruela et
al. \cite{iggy10}).  However (i) the strong central H$\alpha$ emission
peak superimposed on broad emission wings, (ii) lack of He\,{\sc ii}
6527, 6683{\AA} absorption, (iii) absence of S\,{\sc iv} 6668,
6701{\AA} emission and (iv) the presence of strong C\,{\sc ii} 6578,
6582{\AA} photospheric lines are uniformly present in spectra from
1992 onwards\footnote{The low resolution and S/N of previously
  published spectra are insufficient to comment on the presence, or
  otherwise, of photospheric features.} and are all consistent with a
highly luminous B supergiant (Fig. 1). As found by Klochkova \&
Chentsov (\cite{klochkova}), the central peak of the H$\alpha$ profile
appears to be variable at low projected velocities
($\leq140$kms$^{-1}$). \newline

\paragraph{\bf 8-9000{\AA}:} 
Despite spanning a period of over 16
years, the 3 I band spectra are essentially identical (a
representative spectrum from 2008 is presented in Fig. A.2). To classify
Cyg~OB2~\#12 from these data we employed the scheme described in Negueruela
et al. (\cite{iggy10}), supplemented by spectra of other early-mid
BSGs and BHGs of known spectral type. While the BHG spectra  show a
similar morphological progression to those of normal BSGs (Negueruela
et al. \cite{iggy10}, Fig A.2),  there is no indication of emission in
the Paschen series (noting that the lower Balmer transitions are  seen in emission). 
While such a finding precludes the separation of
BHGs from BSGs in the spectroscopic  window in which GAIA - the Global Astrometric 
Interferometer for Astrophysics - will operate, 
the lack of wind contamination aids in the quantitative determination
of the underlying stellar parameters (Sect. 4).

Regarding the classification of Cyg OB2 \#12, the presence of N\,{\sc i}
absorption lines indicates a spectral type of B3 or later, with their
comparative weakness favouring B3.  This finding is consistent with
both the strength of the \OI~8446{\AA} line, which clearly lies
between B2.5Ia and B4Ia (see Fig. A.2), and the temperature sensitive
Pa15/Pa11 and Pa16/Pa11 line ratios.  However, the differences between
B3 and B4 Ia stars are small - hence we suspect that the slight
discrepancies between classifications based on the $\sim$4-5000{\AA}
and $\sim$8-9000{\AA} windows evident for data obtained in 1992 and
2008 are unlikely to be real. Likewise, distinguishing between B5 and
B8 supergiants on the basis of I band spectra alone is particularly
difficult (Negueruela et al. \cite{iggy10}); hence we caution against
using the results of Sharpless (\cite{sharpless}) and Zappala
(\cite{zappala}) as evidence for spectroscopic variability prior to
1970. Echoing Sect. 4, changes in the physical properties of the
photosphere/wind transitional zone have negligible impact on the I
band spectral morphology - hence explaining potential discrepancies
between the spectral type of Cyg~OB2~\#12 determined from the I band
and the He\,{\sc i} 4471{\AA}:Mg\,{\sc ii} 4481{\AA}
diagnostic. \newline

\paragraph{\bf 0.98-1.10$\mu$m:} Two epochs of observations from 1990
and 1992 are presented by Conti \& Howarth (\cite{conti}), who report
no variability between the spectra. Unfortunately a paucity of
spectral features and little calibration data complicate
classification from these data, although the presence of weak He\,{\sc
  i} 1.031$\mu$m suggests a classification of B8 Ia$^+$ or earlier and
the absence of He\,{\sc ii} 1.0124$\mu$m provides an upper limit of B3
Ia$^+$. Such a result is consistent with that derived from our spectra
obtained on 1992 July 22. \newline

\paragraph{\bf 1.66-1.72$\mu$m:} A single epoch of H band spectroscopy
was presented by Hanson et al. (\cite{hansonH}), with Br11 and
He\,{\sc i} 1.700$\mu$m lines in absorption and of $\sim$equal
strength and He\,{\sc ii} 1.693$\mu$m absent. The lack of He\,{\sc ii}
1.693$\mu$m is characteristic of a B spectral type, with
EW(Br11)$\geq$EW(He\,{\sc i}) occuring for B1.5 Ia$^+$ and later (at
the resln. of the spectrum in question) and only becoming $>>$1 at B8
Ia$^+$, hence providing upper and lower limits to the spectral type
(Hanson et al. \cite{hansonH}, \cite{hansonh}).\newline

\paragraph{\bf 2.0-2.4$\mu$m:} Hanson et al. (\cite{hansonK}) present a
K band spectrum from 1994, in which Br$\gamma$ and He\,{\sc i}
2.112$\mu$m are seen in absorption. For a star of the expected
luminosity of Cyg~OB2~\#12 and with the low S/N and resolution of the
data in question, only broad constraints are possible, with the lack
of He\,{\sc ii} 2.189$\mu$m and the presence of He\,{\sc i}
2.112$\mu$m implying a classification between B0-9 Ia.\newline

\paragraph{\bf 2.35-4.10$\mu$m:} Finally, Lenorzer et
al. (\cite{lenorzer}) and Whittet et al. (\cite{whittet}) present
2.35-4.1$\mu$m spectra, which we may supplement with an additional
spectrum from 1996 April 4. All four display emission in Br$\alpha$,
Pf$\alpha$ and $\beta$, with the higher transitions seen in
absoprtion. No variability is apparent between any of these spectra;
hence we do not reproduce them here. Unfortunately, as with the
$\sim$1-2$\mu$m window, there is a dearth of classification features
in this region and we may only infer a spectral type between B3-9
Ia$^+$ from the presence and strength of the Brackett and Pfund series
and the lack of He\,{\sc i} photospheric lines. \newline

To summarise - we find that the available data provides no substantive
evidence for the long-term evolution of the spectral type of Cyg OB2
\#12 over the past half century, although this conclusion is
necessarily weaker prior to the 1990s, given the sparser dataset
available. Nevertheless, this finding is entirely consistent with the
long-term lightcurve. A classification of B3-4 Ia$^+$ appears
appropiate thoughout this period, although short term variability is
present in both wind (e.g.  H$\alpha$...$\delta$) and photospheric
lines (e.g. Kiminki et al. \cite{kiminki}).

Both phenomena are common in other luminous B super-/hypergiants
(e.g. Clark et al. \cite{clark10}) where they are assumed to be due
to, respectively, time-dependent wind structure and photospheric
pulsations resulting in changes in photospheric temperature.
Regarding the latter, changes in the He\,{\sc i}
4471{\AA}:Mg\,{\sc ii} 4481{\AA} flux ratio have led to the conclusion that
the spectral type is variable (B3-8 Ia$^+$), from which  changes in the 
photospheric temperature (16-12kK) of Cyg~OB2~\#12 have been {\em inferred}; 
such a range is fully consistent with models of the pulsating B
supergiant HD64760 (Kaufer et al. \cite{kaufer}) and the BHGs Wd1-7 and
42 (Clark et al. \cite{clark10}).  However we caution that the
detailed non-LTE model atmosphere analysis reveals that variation in
the physical structure of the photosphere/wind transitional zone can
also lead to this behaviour; hence concluding that the photospheric
temperature has been variable based on the current spectroscopic
dataset is clearly premature.

\subsection{The early-B HGs: $\zeta^1$~Sco}

\subsubsection{Photometric variability}

Sterken (\cite{sterken77}) and Sterken et al. (\cite{sterken}) have
previously undertaken long-term(differential) {\em uvby} Stromgren
and {\em V BLUW} photometric campaigns between 1973-4 and 1982-95
respectively. They report low level ($\Delta{m}\sim0.01$mag)
quasi-periodic variability over a wide range of timescales
($\sim16.5-\sim2000$~days), with additional stochastic variability
($\Delta{m}\sim0.05$mag) superimposed.  (Aperiodic) low level
($\Delta{m}=0.08$mag) photometric variations were also present in the
Hipparcos dataset between 1989-1993 (Lef\`{e}vre et al. \cite{lef})
and between 1979-80 (Burki et al. \cite{burki}).

In addition to these data Sterken et al. (\cite{sterken}) also present
historical observations from the 10$^{th}$-19$^{th}$ Century, to which
we may add a number
of more recent data from the 20$^{th}$ Century (Table A.1). 
Sterken et al. (\cite{sterken}) suggest that
$\zeta^1$~Sco is a long-term variable, reporting $m_V \sim 2.8$ in the
middle of the 18$^{th}$ Century, m$_V \sim$4.3 and 5.4 in 1875 and 1878
and $m_V \sim4.5-4.8$ between 1890-1900.  The latter values are broadly
consistent with the photometric data presented in Table A.1, which
span the 20$^{th}$ Century and provide no evidence for the long term
secular variability that characterises e.g. the LBV phase. Moreover,
these data also demonstrate that the near-IR continuum is similarly
stable, albeit over a shorter, $\sim30$~yr baseline.  We therefore
conclude that the sole evidence for significant ($\Delta{m}_V >0.3$mag)
variability is provided by historical visual estimates dating from
before $\sim$1890, for which we are unfortunately unable to quantify
the observational uncertainty.

Finally, we turn to the radio data, for which Bieging et
al. (\cite{bieging}) report possible variability. This mirrors the
findings for Cyg~OB2~\#12 and we provisionally associate this
behaviour with the same physical cause (Sect. 3).

\subsubsection{Spectroscopic classification and variability}

Given the brightness
of $\zeta^1$~Sco, spectroscopic observations dating from the turn of
the 19$^{th}$ Century are reported in the literature, although a lack
of accurate dates for the earlier observations complicates their
interpretation. Nevertheless, a summary of these observations are
presented in Table A.5; we note that the lower reddening to $\zeta^1$
Sco facilitates a more homgeneous dataset of blue end spectroscopy in
comparison to Cyg~OB2~\#12. Fortuitously, the period between 1900-1950
is well sampled and hence complements 
the sparse photometric dataset in this period.

This compilation reveals an absence of long-termvariability or
secular evolution of spectral type over a $\sim$110~yr interval. The
description of the main features of the optical spectra over this
whole period are remarkably consistent; the spectra presented in
Fig. A.1 for the period 1994-2009 demonstrating this stability.
Nevertheless, numerous authors report line profile variability in the
wind dominated P Cygni profile of H$\alpha$ and $\beta$, with the
higher H\,{\sc i} transitions repeatedly varying between emission and
absorption over at least a $\sim$50~yr interval (cf. Table A.5 and
refs. therein). Rivinius et al. (\cite{rivinius}) report the rapid
variability of both photospheric {\em and} wind lines between 1992-5;
we consider it likely that the variability observed in the H$\alpha$ P
Cygni line between 1998-2009 (Fig. 1) reflects a continuation of this
behaviour.

\subsection{The early-B HGs: HD~190603}

Unfortunately, photometric datasets for HD~190603 are somewhat sparse
in comparison to the previous two stars, with no photometric
observations reported over the last decade and a $>$16~yr gap in the
1980s-90s (Table A.1). Nevertheless, we find no evidence for secular
variability over the $\sim$46~yr period from 1952-1998. Conversely,
rapid photometric variability has been reported on several occassions
by both Percy \& Welch (\cite{percy}) and Koen \& Eyer (\cite{koen}).

Likewise comparatively few spectral observations have been reported, a
problem exacerbated by the lack of reduced spectra being
presented. Consequently we are forced to simply present the spectral
classifications reported for HD~190603 in Table A.6. These data
indicate a corresponding lack of spectroscopic evolution over a 50
year period. Indeed, the description of the spectrum by both Beals
(\cite{beals}) and Merrill \& Birwell (\cite{merrill33}) as,
respectively, that of a P Cygni supergiant and a Be star, suggest a
similar morphology prior to 1933. The latter authors further described
variable emission lines; similar findings being reported by Rosendhal
(\cite{rosendahl}) and Rivinius et al.  (\cite{rivinius}).

\subsection{The early-B HGs: HD~80077, HD~169454 and BP~Cru}

We next turn to the remaining early-B HGs. 
We refrain from tabulating their sparse photometric datasets but note
that there is no evidence for secular variability in any of the three
stars; for instance comparison of the optical (Hiltner \cite{hiltner},
Kilkenny et al. \cite{kilkenny}) and near-IR (Whittet et
al. \cite{whittet}, Skrutskie et al. \cite{skrutskie}) data for
HD~169454 reveal constancy over $\sim$35 and $\sim$26~yr intervals
respectively. However, following the previous discussions, rapid, low
amplitude photometric variability appears ubiquitous (van Leeuwen et
al. \cite{vanL98}, Sterken \cite{sterken77} and Hammerschlag-Hensberge
et al. \cite{HH76} respectively).

As with HD~190603, Table A.6 summarises their spectral types as
reported in the literature. No evidence for the long-term (secular)
evolution of spectral type over timescales in excess of 30~yrs was
found, with the description of HD~169454 as a B star with H\,{\sc i} P
Cygni emission lines (Merrill \& Burwell \cite{merrill33}) suggesting
spectral stability for nearly a century. In contrast line profile
variability (LPV) on a timescale of days-weeks is present in all 3
stars (Knoechel \& Moffat \cite{knoechel}, Rivinius et
al. \cite{rivinius} and Kaper et al. \cite{kaper06} respectively).

\subsection{The early-BHGs/WNVL stars Wd1-5 and 13}

The remaining early-B HGs are found within the massive young
cluster Wd1, 
and on the basis of a restricted wavelength range (5800-8900{\AA})
were classified as borderline BHG/very late WN stars, forming an
evolutionary sequence from Wd1-5 through -13 to the WN9h star Wd1-44
(Clark et al. \cite{clark08}). Unfortuntely, these stars have only
been observed over the past $\sim$decade, but no evidence for long
term variability has been found (Clark etal. \cite{clark10}). However,
Wd1-13 is a confirmed $\sim$9.27~day massive binary, while pronounced
LPV in Wd1-44 also argues for such an identification. Consequently, we
suspect that all three stars to have formed as the result of close
binary evolution.

\subsection{The late-B HGs}

Finally, we examine the eight BHGs with spectral types of B5 and later
that have been identified within the Galaxy. Three - W7, W33 (both B5
Ia$^+$) and W42a (B9 Ia$^+$) - are located within the massive young
cluster Wd~1 and are discussed in detail in Clark et al. (\cite{clark10}).
Unfortunately, the long-termspectroscopic and photometric datasets
for these stars are less complete than those of the early-B HGs
described above, although they are sufficient to confirm the presence
of rapid LPV. This behaviour, as well as short term low amplitude
photometric pulsations is also present in the other examples; 
HD~160529 (A9-B8 Ia$^+$; Stahl et al. \cite{160529}), HD~168607 (B9
Ia$^+$; Chentsov et al. \cite{chentsov}, Sterken et
al. \cite{sterken99}), HD~168625 (B8 Ia$^+$; Sterken et
al. \cite{sterken99}; Chentsov et al. \cite{chentsov}), HD~183143 (B7
Iae; Adelman et al. \cite{adelman00}, Chentsov et al. \cite{chentsov})
and HD~199478 (B8 Iae Percy et al. \cite{percy08}, Markova \& Valchev
\cite{mar00}), suggesting that the $\alpha$ Cygni instabilities are
ubiquitous across the complete temperature and luminosity range
spanned by BHGs.

This phenomenon is also found to extend to cooler temperatures, having
been identified in a number of early-A (A0-2.5 Ia$^+$/Iae) stars with
similar spectral morphologies to the BHGs - e.g.  HD~92207 (A0 Iae;
Sterken \cite{sterken77}, Kaufer et al. \cite{kaufer97}), HD~223385
(A2.5 Ia$^+$; Adelman \& Albayrak \cite{adelman97}, Chentsov
\cite{chentsov}) and AS~314 (A0 Ia$^+$; Miroshnichenko \cite{miro00}).

However, unlike the early-B HGs, HD~160529 and 168607 demonstrate
characteristic LBV photometric modulation, with the former also
exhibiting correlated spectral type variability, while both HD~168625
and HD92207 show evidence for a complex, dusty circumstellar
environment (Roberto \& Herbst \cite{rob}; Clarke et
al. \cite{clarke}) implying recent enhanced mass loss, possibly
associated with an LBV phase.

\begin{figure*}
\includegraphics[width=12cm,angle=270]{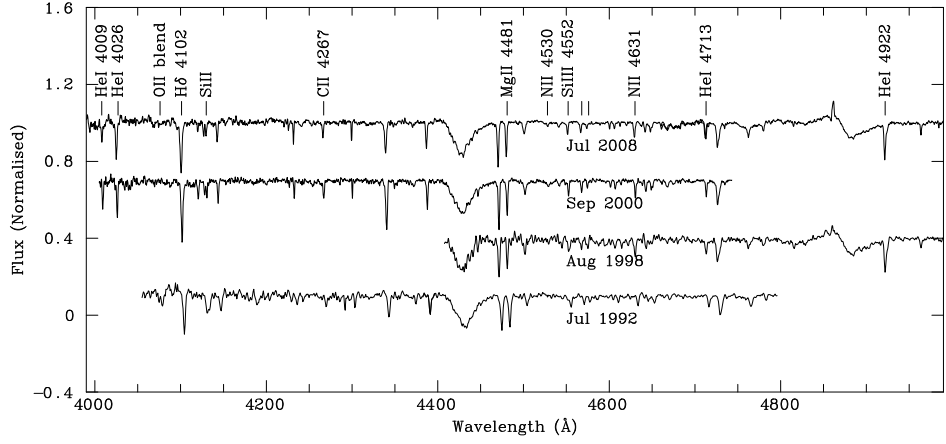}
\includegraphics[width=12cm,angle=270]{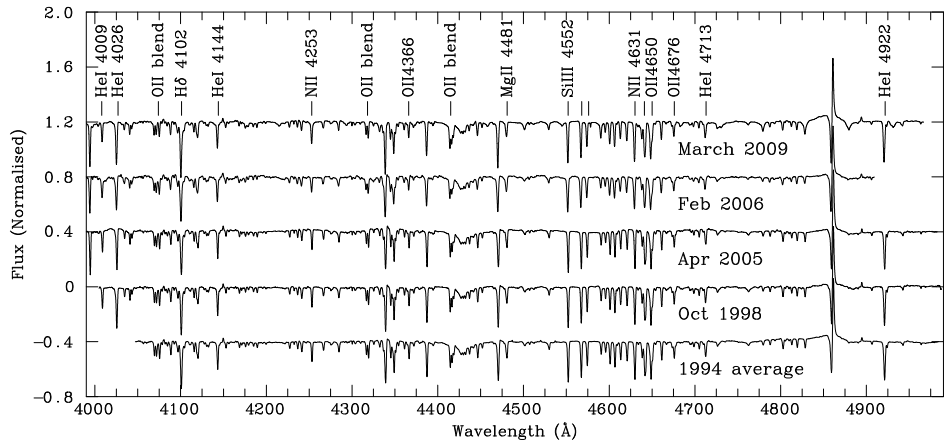}
\caption{Long-term evolution  of Cyg~OB2~\#12 (1992-2008; top panel) and $\zeta^1$~Sco (1998-2009; bottom panel) in the
$\sim$4000-4800{\AA} spectral region, encompassing the Si\,{\sc iii} 4552{\AA}/Si\,{\sc ii} 4128{\AA} temperature diagnostic.}
\end{figure*}

\begin{figure*}
\label{fig-ibandcomp}
\resizebox{\hsize}{!}{\includegraphics[angle=270]{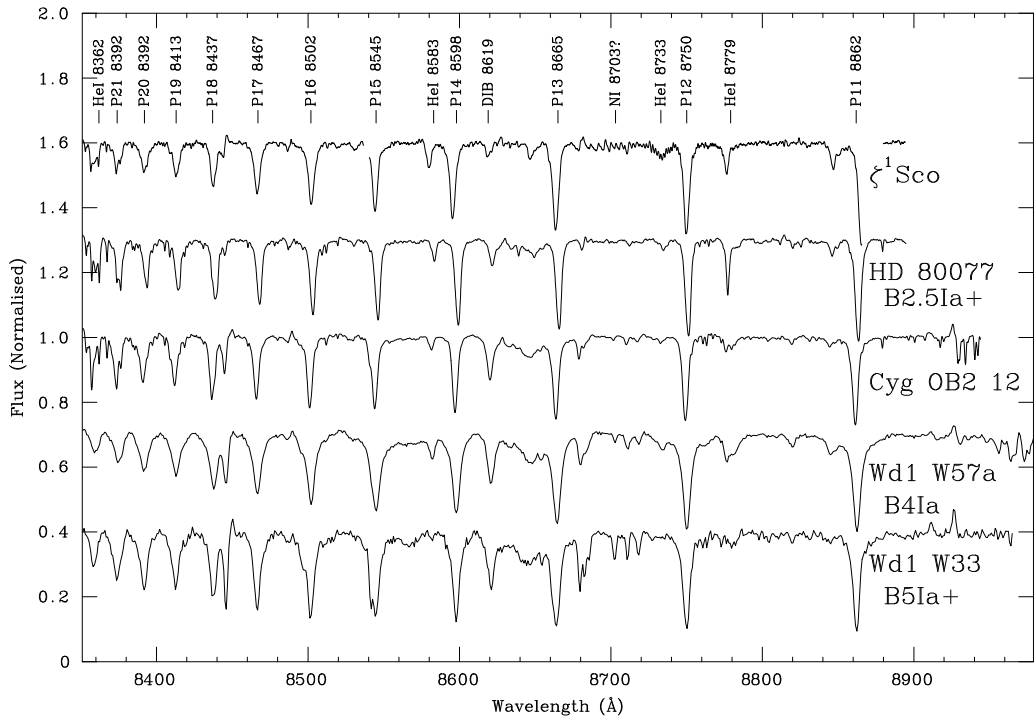}}
\caption{Comparison of the I band spectra  of Cyg~OB2~\#12 and $\zeta^1$~Sco 
with other galactic BHGs. Note the systematic weakening of He\,{\sc i} and strengthening
of N\,{\sc i} transitions with decreasing temperature, which permit spectral classification
in whis wavelength range.
}
\end{figure*}

\longtabL{1}{
\begin{landscape}
\begin{longtable}{lccccccccr}
\caption{Summary of historical photometric magnitudes used to construct the optical-near IR SEDs, ordered by time.}\\
\hline\hline
 Date & \multicolumn{8}{c}{Waveband} & Reference \\
  & $U$& $B$ & $V$ & $R$ & $I$& $J$& $H$ & $K$ &  \\
\hline
\endfirsthead
\caption{continued.}\\
\hline\hline
 Date & \multicolumn{8}{c}{Waveband} & Reference \\
  & $U$& $B$ & $V$ & $R$ & $I$& $J$& $H$ & $K$ &  \\
\hline
\endhead
\hline
\endfoot
 & & & & & & & & & \\
\multicolumn{10}{c}{\bf Cyg~OB2~\#12} \\
\smallskip
$<$1954 & 16.67$\pm$0.1 & 14.66$\pm$0.04 & 11.47$\pm$0.04 & - & - & - & - & - & Sharpless (\cite{sharpless}) \\
$<$1962 & 17.21 & 14.71 & 11.49 & 8.36 & 6.20 & 4.55 & - & 2.81 & Johnson \& Borgman (\cite{johnson63}) \\
 $<$1964&17.20$\pm$0.01 & 14.70$\pm$0.01 & 11.48$\pm$0.01 & 8.26$\pm$0.01 & 5.94$\pm$0.01  & 4.32$\pm$0.01 & -    & 2.66$\pm$0.1  & Johnson (\cite{johnson})\\
$<$1967 & 17.15 & 14.70 & 11.48 & 8.26 & 5.95 & 4.42 & - & 2.70 & Wisniewski et al. (\cite{wisniewski}) \\     
$<$1968 & - & - &- & 8.26 & 5.94 & 4.32 & - & 2.66 & Johnson (\cite{johnson68}) \\
$<$1971 & - & - & - & - & - & - & - & 2.60 & Stein \& Gillett (\cite{stein}) \\
$<$1973 & - & - & - & - & - & - & 3.34$\pm$0.06 & 2.77$\pm$0.05 & Chaldu et al. (\cite{chaldu}) \\
1972 Autumn &  -    &   -   & -     & -  & -   & 4.59$\pm$0.05 & 3.37$\pm$0.05    & 2.82$\pm$0.05 & Voelcker (\cite{voelcker}) \\
1975-6 &  -    &   -   & -     & -  & -   & 4.38$\pm$0.02 &3.28$\pm$0.02 & 2.72$\pm$0.02 & Harris et al. (\cite{harris}) \\ 
1980 Aug 26 &  -    &   -   &  & -  & -   & -    & -  & 2.80$\pm$0.01 & Abbott et al. (\cite{ab84}) \\
1981 July &  -    &   -   &  & -  & -   & -    & 3.33$\pm$0.01 & 2.72$\pm$0.01 & Leitherer et al. (\cite{leitherer}) \\
1989 May 26-30 & - & - & - & - & - & 4.39$\pm$0.04 & 3.34$\pm$0.04 & 2.75$\pm$0.04 & Torres-Dodgen et al. 
(\cite{torres}\\
1990 June 6 & 17.18$\pm$0.01 & 14.81$\pm$0.01 & 11.46$\pm$0.01 & -  & -   & -    &  -   & -    & Massey \& Thompson (\cite{massey}) \\
1998 June 21 & -     & -     & -     &  - & -   & 4.7$\pm$0.3 & 3.5$\pm$0.3 & 2.70$\pm$0.4 & Skrutskie et al.(\cite{skrutskie})  \\
& & & & & & & & & \\
\multicolumn{10}{c}{\bf $\zeta^1$~Sco} \\
$<$1924   & - & -               & 4.88 & - & - & - & - & -   & HD catalogue \\   
1949-1952$^1$ & - & -        & 4.88 & - & - & - & - & - & Schilt \& Jackson (\cite{schilt}) \\
$<$1954$^2$   & - & -         & 4.9  & - & - & - & - & - & Bidelman (\cite{bidelman}) \\
$<$1958   & - & -          & 4.86 & - & - & - & - & - & Code \& Houck (\cite{code}) \\
$<$1958 & - & 5.03  & 4.3 & - & - & - & - & -  & Evans et al. (\cite{evans}) \\
1958 March-April & 4.54 & 5.20 & 4.82 & - & - & - & - & - & Westerlund (\cite{westerlund}) \\
$<$1961   & - & -          & 4.80 & - & - & - & - & - & Buscombe (\cite{buscombe}) \\
1963 & - & - & 4.74 & - & - & - & - & - & Andrews et al. (\cite{andrews}) \\
1963 June & 4.57 & 5.19 & 4.72 & - & - & - & - & - & Feinstein (\cite{f68})\\
1964 March &4.64 & 5.20 & 4.72& - & - & - & - & - & Feinstein (\cite{f68}) \\
1967-68 & - & - & 4.71 & - & - & - & - & - &  Schild et al. (\cite{schild}) \\
1968 July & - & - & - & - & - & - & 3.30 & 3.11 & Schild et al. (\cite{schild71}) \\
$<$1973 & - & - & 4.73$\leq V \leq$4.83 & - & - & - & - & - & Jaschek \& Jaschek (\cite{jj}) \\
1974 April & - & - & - & - & - & 3.47$\pm$0.03 & 3.27$\pm$0.03 & 3.11$\pm$0.03 & Whittet et al. (\cite{whittet}) \\
1963 to 1976$^3$ & - & - & 4.73 & - & - & - & - & - & Feinstein \& Marraco (\cite{feinstein79}) \\
1979 Apr. to  1980 Sept$^4$& - & - & 4.66$\leq V \leq$4.78 &- & - & - & - & - &  Burki et al. (\cite{burki}) \\
1981 August 9 & - & - & - & - & - & - & - & 3.14$\pm$0.05 & Abbott et al. (\cite{ab84}) \\
1974-82 & 4.76 & 5.29 & 4.77 & 4.31 & 3.85 & - & - & - & Fernie (\cite{fernie}) \\
1982 August & - & - & - & - & - & 3.58$\pm$0.03 & 3.32$\pm$0.03 & 3.18$\pm$0.03 & Leitherer \& Wolf (\cite{lw}) \\
$<$1984 & - & - & - & - & - & 3.53$\pm$0.02 & 3.31$\pm$0.02 & 3.13$\pm$0.03 & Lopez \& Walsh (\cite{lopez}) \\ 
$<$2003 & 4.61 & 5.18 & 4.78 & - & - & - & - & - & Reed et al. (\cite{reed}) \\
1999 May 22 & -     & -     & -     &  - & -   & 3.6$\pm$0.3 & 3.3$\pm$0.2 & 3.3$\pm$0.2 & Skrutskie et al.(\cite{skrutskie})  \\
& & & & & & & & & \\

\multicolumn{10}{c}{\bf HD190603} \\

1952-4 & 5.72 & 6.18& 5.65 & -& -& -& -& -& Hiltner (\cite{hiltner})\\

$<$1958 &5.59 &  6.13 & 5.56 & - & - & - & - & - & Golay (\cite{golay}) \\

$<$1961 & - & 6.14 & 5.59 & - & - & - & - & - & Bouigue et al. (\cite{bouigue}) \\

1962-70 & 5.66 & 6.14 & 5.62  & - & - & - & - & - & Crawford et al. (\cite{crawford}) \\

 $<$1964&5.72$\pm$0.01 & 6.18$\pm$0.01 & 5.65$\pm$0.01 & 5.12$\pm$0.01 & 4.72$\pm$0.01  & 4.50$\pm$0.01 & -    & 4.11$\pm$0.1  & Johnson (\cite{johnson})\\

$<$1966 & 5.73 & 6.19& 5.65 & - & - & - & - & - & Johnson et al. (1996) \\

$<$1968 & 5.69& 6.15 & 5.60 & -& -& -& -& -& Blanco et al. (\cite{blanco})\\

$<$1968 & - & 6.14 & 5.62 & - & - & - & - & - & Lesh (\cite{lesh})\\

1968-9 & 5.81 & 6.25 & 5.70& - & - & - & - & - & Burnichon \& Garnier (\cite{burnichon}) \\

1972 Autumn &  -    &   -   & -     & -  & -   & 4.50$\pm$0.05 & 4.14$\pm$0.05    & 4.11$\pm$0.05 & Voelcker (\cite{voelcker}) \\

$<$1974 & 5.72 & 6.18& 5.64 &- & -& -& -& -& Nicolet (\cite{nicolet})\\

1974 Sept. & -& -& -& -& -& 4.43$\pm$0.03& 4.24$\pm$0.03& 4.10$\pm$0.03& van Breda \& Whittet (\cite{vb}) \\

1974-82 & 5.64 &  6.13 & 5.56 &  5.01 & 4.62 & - & - & - & Fernie (\cite{fernie}) \\

1975 Oct. & - & - & - & - & - & - & - & 4.1$\pm$0.2 & Sneden et al. (\cite{sneden}) \\

1976-78& - & 6.18$\pm$0.02& 5.63$\pm$0.02& 5.08$\pm$0.02& 4.71$\pm$0.02 & -& -& -& Moffett \& Barnes (\cite{moffett})\\

1979-82 & - & 6.18 & 5.64 & - & - & - & - & - & Percy \& Welch (\cite{percy}) \\

1998  May 11 & - & - & - & - & - & 4.42$\pm$0.26 & 4.26$\pm$0.20 & 4.04$\pm$0.03 & Skrutskie et al. (\cite{skrutskie}) \\
\hline
\end{longtable} {Note that where available errors and exact dates of
  observations are presented. For brevity we have not replicated the B
  band lightcurve of Gottlieb \& Liller (\cite{gottlieb}) for Cyg
  OB2\#12 here. Observations between 1945-1951 by Cousins
  (\cite{cousins}) suggest that $\zeta^1$~Sco was variable with low
  amplitude ($\Delta m=0.05$mag) between 1945-51, although details of
  the filter set used are not described. Sterken (\cite{sterken77})
  report 35 epochs of differential {\em ubvy} photometry between
  1973-1974 for $\zeta^1$~Sco, with a further 399 epochs between
  1982-95 presented by Sterken et al. (\cite{sterken}); throughout
  these periods minimal variability was observed ($\Delta$ {\em ubvy}
  $\leq$0.1mag.). Finally Percy \& Welch (\cite{percy}) report 52
  epochs of BV band photometry between 1979 May 30 and 1982 July 1
  indicating t to be a short period variable.  $^1$3 observations in
  this period; no details of band pass $^2$Phot. magnitude possibly
  sourced from HD catalogue?  $^3$3 observations over this period.
  $^4$ 43 observations over this period.}
\end{landscape}
}

\begin{table}
\caption{Summary of mid- to far-IR fluxes used to construct the SEDs}
\begin{center}
\begin{tabular}{lcr}
\hline
\hline
Wavelength & Magnitude & Reference \\
($\mu$m)   &           &           \\
\hline

\multicolumn{3}{c}{\bf Cyg~OB2~\#12} \\

2.30   &    2.72 $\pm$0.01    &           Leitherer et al. (\cite{leitherer}) \\
2.345  &     2.635$\pm$0.02   &            Harris  et al. (\cite{harris}) \\    
3.013  &    2.405$\pm$0.02    &           Harris  et al. (\cite{harris})  \\  
3.108  &     2.375$\pm$0.02   &           Harris  et al. (\cite{harris})  \\   
3.420  &     2.281$\pm$0.02   &           Harris  et al. (\cite{harris})  \\   
3.45  &      2.37$\pm$0.05   &            Abbott et al. (\cite{ab84}) \\  
3.50  &      2.28$\pm$0.05   &            Chaldu et al. (\cite{chaldu}) \\  
3.57  &      2.28$\pm$0.02   &            Leitherer et al. (\cite{leitherer}) \\  
3.57  &      2.170$\pm$0.05   &            Torres-Dodgen et al. (\cite{torres}) \\  
3.58  &     2.217$\pm$0.02    &           Harris  et al. (\cite{harris})  \\ 
3.820 &      2.178$\pm$0.02   &           Harris  et al. (\cite{harris})  \\  
4.29  &      2.057$\pm$0.118  &            MSX (Egan et al. \cite{egan})  \\  
4.80  &       2.37$\pm$0.10   &            Abbott et al. (\cite{ab84})\\   
4.97  &      2.06$\pm$0.02    &           Leitherer et al. (\cite{leitherer})  \\ 
8.28  &      1.946$\pm$0.044   &            MSX (Egan et al. \cite{egan})  \\ 
10.2  &       2.03$\pm$0.10    &           Abbott et al. (\cite{ab84}) \\  
10.9  &      1.95$\pm$0.07     &          Leitherer et al. (\cite{leitherer}) \\  
12.0  &       1.642$\pm$0.00    &           IRAS PSC (1985) \\   
12.13 &      1.900$\pm$0.056   &           MSX (Egan et al. \cite{egan})  \\  
14.65 &      1.713$\pm$0.067   &           MSX (Egan et al. \cite{egan}) \\   
20.0  &      1.54$\pm$0.10    &          Abbott et al. (\cite{ab84}) \\   
21.34 &     1.575$\pm$0.079    &           MSX (Egan et al. \cite{egan})\\   
25.0  &     1.204     &         IRAS PSC (1985) \\ 
 & & \\
\multicolumn{3}{c}{\bf $\zeta^1$~Sco} \\
3.45(L)& 2.98$\pm$0.03 & Leitherer \& Wolf (\cite{lw}) \\
4.7(M) & 2.90$\pm$0.03 & Leitherer \& Wolf (\cite{lw}) \\
8.28 & 2.624 & MSX (Egan et al. \cite{egan}) \\
12.13 &3.416 & MSX (Egan et al. \cite{egan}) \\
14.65 &3.560 & MSX (Egan et al. \cite{egan}) \\
 & & \\
\multicolumn{3}{c}{\bf HD190603} \\
3.6 & 3.96$\pm$0.18 & Sneden et al. (\cite{sneden}) \\
4.9 & 3.89$\pm$0.20 & Sneden et al. (\cite{sneden}) \\
8.28 & 3.68$\pm$0.04 & MSX (Egan et al. \cite{egan})\\
9.0 & 3.12$\pm$0.01 & AKARI \\
12.13 & 3.47$\pm$0.08 & MSX (Egan et al. \cite{egan})\\
14.65 & 3.46$\pm$0.09 & MSX (Egan et al. \cite{egan}) \\
\hline
\end{tabular}
\end{center}
\end{table}

\begin{landscape}
\begin{table}
\caption{Summary of mm \& radio fluxes in mJy  used to construct the SEDs, ordered by time.}
\begin{center}
\begin{tabular}{lcccccccccr}
\hline
\hline
 Date &  \multicolumn{9}{c}{Waveband} & Reference \\
         & 0.13cm     & 0.7cm        & 1.3cm    & 2cm          &  3.5cm        & 3.6cm          & 6.1cm         & 6.2cm      & 20.5cm       &  \\
\hline

\multicolumn{11}{c}{\bf Cyg~OB2~\#12} \\
1980 March 22 & -&     -        & 21$\pm$3 &     -         &    -           &     -           &         -      &          -  &          -    &  Wendker \& Altenhoff (\cite{wendker}) \\
1980 May 22+23& -&       -      &     -     &          -    &         -      &        -        &      -         &3.2$\pm$0.3 &   -           & Abbott et al. (\cite{abbott}) \\
1981 Oct. 16  & -&    -         &      -    &   6.0$\pm$2.0&       -        &       -         &       -        &3.4$\pm$0.2 &   -           & Bieging et al. (\cite{bieging}) \\
1982 March 27 & -&      -       &     -     &   4.5$\pm$0.3&       -        &       -         &          -     &      -      &           -   &  White \& Becker (\cite{white}) \\
1982 Aug. 26  & -&       -      &     -     &        -      &       -        &       -         &         -      &2.2$\pm$0.2 & 2.0$\pm$0.2  & Bieging et al. (\cite{bieging}) \\
1983 Aug./Oct.& -&     -        &    -      &        -      &        -       &      -          & 4.0$\pm$0.3   &      -      & 2.1$\pm$0.3  & Becker \& White (\cite{becker}) \\
1993 May 1    & -&       -      &      -    &         -     &        -       & 6.06$\pm$0.07  & 3.94$\pm$0.07 &            &    -          & Waldron et al. (\cite{waldron}) \\  
1994 Sept.    & -&    -         &      -    &   7.7$\pm$0.3& 4.74$\pm$0.14 &     -           &       -        &4.00$\pm$0.20&   -           & Scuderi et al. (\cite{scuderi}) \\     
1994 Oct.     & -&    -         &     -     &  12.0$\pm$0.2& 7.40$\pm$0.08 &      -          &       -        &4.00$\pm$0.20&   -           & Scuderi et al. (\cite{scuderi}) \\   
1995 April 27 & -&22.9$\pm$0.6 &   -       & 11.3$\pm$0.1 & 7.18$\pm$0.04 &     -           & 3.64$\pm$0.12 &   -         &      -        & Contreras et al. (\cite{contreras}) \\    
1999 June 28  & -& 22.9$\pm$0.6&      -    & 9.0$\pm$1.5  & 5.9$\pm$0.1   &        -        & 4.2$\pm$0.1   &     -       &   -           & Contreras et al. (\cite{contreras04}) \\   
& & & & & & & & & & \\
\multicolumn{11}{c}{\bf $\zeta^1$~Sco} \\
1981 May 7  & -  & - & - & -           & - & - & 1.2$\pm$0.3 & - & - & Bieging et al. (\cite{bieging}) \\
1984 March 9 & - & - & - & -           & - & - & 2.0$\pm$0.2 & - & - & Bieging et al. (\cite{bieging}) \\
1984 April 3 & - & - & - & 4.3$\pm$0.1 & - & - & 1.9$\pm$0.2 & - & - & Bieging et al. (\cite{bieging}) \\
1990 Sept. 12-14  & 23.0$\pm$2.4 & - & - & - & - & - & - & - & - & Leitherer \& Robert (\cite{leitherer91}) \\
$<$2006 & - & - & - & - & - & 2.4 & - & - & - & Benaglia et al. (\cite{benaglia}) \\
& & & & & & & & & & \\
\multicolumn{11}{c}{\bf HD~190603}\\
1979 July 13 & - & - & - & - & - & - & $<$0.5 & - & - & Abbott et al. (\cite{abbott}) \\
1994 Oct. & - & - & - & 0.7$\pm$0.2 & 0.7$\pm$0.1 & - & - & 0.58$\pm$0.06 & - & Scuderi et al. (\cite{scuderi}) \\
\hline
\end{tabular}
\end{center}
{Note that for brevity the wavelengths are presented to the nearest 0.1cm, with exact values given 
in the papers in question.}
\end{table}
\end{landscape}

\longtab{4}{
\begin{longtable}{lcccr}
\caption{Summmary of historical spectral observations of Cyg~OB2~\#12} \\
\hline
\hline
Date & Wavelength & Spectral Type & Comments & Reference \\
\hline
\endfirsthead
\caption{continued.}\\
\hline
\hline
Date & Wavelength & Spectral Type & Comments & Reference \\
\hline
\endhead
\hline
\endfoot

{\em $\leq$1954} & {\em 4100-6600{\AA}} & {\em $<$A0 I} & {\em H$\alpha$ in emission} & Morgan et al. (\cite{morgan}) \\[1mm]

{\em $\leq$1957} & 8-9000{\AA}    & {\em B5 Ia}&   -                      & Sharpless (\cite{sharpless}) \\[1mm]

{\em $\leq$1970} & {\em near IR}   & {\em B8 Ia$^+$} &    -               & Zappala  (\cite{zappala}) \\[1mm]   

{\em $\leq$1971} &      ?          &       -         & {\em H$\alpha$ in emission} & Bromage (\cite{bromage}) \\[1mm]

{\em $\leq$1973} & 5-8700{\AA}+    &       -         & H$\alpha$ in emission & Chaldu et al. (\cite{chaldu}) \\
                 & 8-11000{\AA}    &                 &                        &                             \\[1mm]

 1977 July       & 3840-4480{\AA}  &    B8 Ia$^+$    &  Variable H$\alpha$  &  Souza \& Lutz (\cite{souza}) \\
                 &    {\em +H$\alpha$}   &                 & in emission          &  \\[1mm]  

1980 Autumn      & {\em H$\alpha$}  &       -         & {\em H$\alpha$ in emission} & Leitherer et al. (\cite{leitherer}) \\[1mm]

{\em $\leq$1981} & {\em 4000-6600{\AA}} & -          & {\em H$\alpha$+H$\beta$ in emission} & Hutchings (\cite{hutchings}) \\[1mm]

1989-90 summer + & {\em encompasses H$\alpha$} & B5 Ia & H$\alpha$+H$\beta$ in emission, & Massey \& Thompson (\cite{massey}) \\
`few years earlier'    & {\em + $\sim$4089-4686{\AA}}        & B5 Ia & H$\gamma$+H$\delta$ infilled   &  \\[1mm]

1990 August  & 0.98-1.10$\mu$m & $>$B3 Ia,$<$B8 Ia & - & Conti \& Howarth (\cite{conti}) \\[1mm]

1992 July   & 0.98-1.10$\mu$m & $>$B3 Ia, $<$B8 Ia & - & Conti \& Howarth (\cite{conti}) \\[1mm]

{\bf 1992 July 17} & {\bf 4036-4836{\AA}} & {\bf B4-5 Ia} &{\bf H$\gamma$+H$\delta$ in absorption} & {\bf This work} \\
                 & {\bf 5765-9586{\AA}} & {\bf B3 Ia} & {\bf H$\alpha$ in emission}   & \\[1mm]

1994 Sept.    & 2-2.2$\mu$m     &  $>$B0.5 Ia, $<$B9 Ia & - & Hanson et al. (\cite{hansonK} ) \\[1mm]

{\bf 1995 July 17}       & {\bf 6366-6772{\AA}}  &   - & {\bf H$\alpha$ in emission} & {\bf This work} \\[1mm]

1995 Dec. 23       & 2.35--8$\mu$m &  $>$B3 Ia, $<$B9 Ia & Br$\alpha$, Pf$\alpha$+$\beta$ in emission & Whittet et al. (\cite{whittet}) \\[1mm]

{\bf 1996 April 4}       & {\bf 2.35--8$\mu$m} & {\bf $>$B3 Ia, $<$B9 Ia} & {\bf Br$\alpha$, Pf$\alpha$+$\beta$ in emission} & {\bf This work} \\[1mm]

1996 Oct. 17       & 2.35--8$\mu$m &  $>$B3 Ia, $<$B9 Ia & Br$\alpha$, Pf$\alpha$+$\beta$ in emission & Whittet et al. (\cite{whittet}) \\[1mm]

1997 June & 1.66-1.72$\mu$m &  $>$B1.5 Ia, $<$B8 Ia & - & Hanson et al. (\cite{hansonH}) \\[1mm]

1998 Apr-May       & 2.35--8$\mu$m &  $>$B3 Ia, $<$B9 Ia &  Br$\alpha$, Pf$\alpha$+$\beta$ in emission & Lenorzer et al. (\cite{lenorzer}) \\[1mm]

{\bf 1998 Aug. 09} & {\bf 3940-5737{\AA}}  & {\bf B4 Ia}& {\bf H$\beta$ in emission} & {\bf This work} \\
                 & {\bf 6366-6772{\AA}} & - & {\bf H$\alpha$ in emission}   & \\[2mm]

1999 July 10 & 5500-7700{\AA}  &  B Ia & H$\alpha$ in emission & Kiminki (2010, priv. comm.) \\[1mm]

2000 July 10 & 5500-7700{\AA}  &  B Ia & H$\alpha$ in emission & Kiminki (2010, priv. comm.) \\[1mm]

2000 Sept. 18       & 3600-5200{\AA}       & B3 Iae       & H$\beta$ infilled & Kiminki et al. (\cite{kiminki}) \\[1mm]

{\bf 2000 Sept. 23} & {\bf 4000-4750{\AA}} & {\bf B4 Ia} & {\bf H$\gamma$+H$\delta$ in absorption} & {\bf This work} \\
                 & {\bf 6340-6740{\AA}} & - & {\bf H$\alpha$ in emission}   & \\[1mm]

2001 June 21 & 4542-7939{\AA}  & B5 Ia & Balmer line emission  & Klochkova \& Chenstov (\cite{klochkova}) \\[1mm]

2001 August 24   & 3800-4500{\AA}  & B6 Iae &H$\gamma$+H$\delta$ in absorption & Kiminki et al. (\cite{kiminki}) \\[1mm]

2001 Sept. 01   & 3800-4500{\AA}  & B8 Iae & H$\gamma$+H$\delta$ in absorption &Kiminki et al. (\cite{kiminki}) \\[1mm]

2003 April 12 & 5273-6764{\AA}   & B5 Ia & H$\alpha$ emission & Klochkova \& Chenstov (\cite{klochkova}) \\[1mm]

{\bf 2007 Aug. 22}  & {\bf 7600-8900{\AA}} & {\bf B3 Ia}&-  & {\bf This work} \\[1mm]

2007 Aug. 30 & 5600-6800{\AA}  & B Ia & H$\alpha$ in emission & Kiminki (2010, priv. comm.) \\[1mm]

{\bf 2008 July 22}  & {\bf 4000-5270{\AA}} & {\bf B4 Ia}& {\bf H$\beta$ in emission} & {\bf This work} \\
                    &                      &              & {\bf H$\gamma$+H$\delta$ in absorption} &    \\
                    & {\bf 6450-7150{\AA}}      &  -   & {\bf H$\alpha$ in emission} & \\
                    & {\bf 8350-8900{\AA}}      & {\bf B3 Ia} &  & \\         
\hline
\end{longtable}
{Note that a parameter is listed in italics if not explicitly given (wavelength of observation)
or the  data from which it is derived was not presented. Parameters given in bold are from this work.  H$\alpha$ is
 variable between the two epochs of observations presented by Klochkova \& Chenstov (\cite{klochkova})
as H$\beta$ and $\gamma$ are between the 2000-8 spectra presented here, while Conti \& Howarth 
(\cite{conti}) report no variability between 1990-2 in the $\sim 1{\mu}$~m region. Selected spectra between 1992-2008 are plotted in Fig. 
A.1.}
}

\longtab{5}{
\begin{longtable}{lcccr}
\caption{Summary of historical spectral observations of $\zeta^1$~Sco}\\ 
\hline
\hline
Date & Wavelength & Spectral  & Comments & Reference \\
     &            & Type & & \\
\hline
\endfirsthead
\caption{continued.}\\
\hline
\hline
Date & Wavelength & Spectral  & Comments & Reference \\
& & Type & & \\
\hline
\endhead
\hline
\endfoot

$<$1896 & - & - & {\em H$\beta$ in emission} & Pickering (\cite{pickering}) \\[1.5mm]

1891-1899 & {\em 4000-4900{\AA}} & B1p & {\em P Cygni profiles in H$\beta$, $\gamma$}  & Cannon (\cite{cannon}) \\
                &                 &   & {\em all other lines in absorption} & \\
                &                 &   & {\em (He\,{\sc i} 4009, O\,{\sc ii} 4349, C\,{\sc ii} 4267)}       & \\[1.5mm]

1924            &  {\em 4100-4900{\AA}} & {\em B(2)eq} & {\em P Cygni profiles in H$\beta$, $\gamma$} & Merrill et 
 al. (\cite{merrill25}) \\
 & & & {\em H$\delta$ in absorption} \\[1.5mm]

1903-1929 & - & {\em B1e Ia} & - & Bidelman (\cite{bidelman}) \\[1.5mm]

1926-1930 & {\em 3950-4900{\AA}}   & {\em B1pe} & - &  Rimmer (\cite{rimmer}) \\[1.5mm]

30/07/34    & - & - & {\em P Cygni profiles in H$\alpha$ and D3 He\,{\sc i}} & Merril \& Burwell (\cite{merrill43})\\[1.5mm]

$<$1938      & -    & - & {\em Listed as variable} & Payne-Gaposchkin \& \\
                &      &   &                    & Gaposchkin (\cite{pg}) \\[1.5mm]

1954-58 &{\em 3800-6600{\AA}} & B1.5 Ia$^+$ & No appreciable difference  & Code \& Houck (\cite{code}) \\[1.5mm]
 & & & from  Cannon (\cite{cannon}) & \\[1.5mm]

26/04/58 to 03/09/60$^1$ & {\em 3900-4500{\AA}}     & {\em B1e Ia} && Buscombe et al. (\cite{buscombe}) \\[1.5mm]

22/08/61   & 6200-6600{\AA} & - & {\em H$\alpha$ in emission} & Jaschek et al. (\cite{jaschek}) \\[1.5mm]

7/08/66    & {\em 3800-4700{\AA}} & {\em B1 Iae} & {\em H$\beta$ in emission, H$\gamma$, $\delta$} & Buscombe \& Kennedy \\
          &     & & {\em \& He\,{\sc i} 4471 in absn.} & (\cite{buscombe68}) \\[1.5mm]

6/07/60 & 3100-6750{\AA} & B1 Ia & P Cygni profiles in  & Hutchings (\cite{hutchings68}) \\
9/08/60 &     &    & H$\alpha$, {\em $\beta$ \& He\,{\sc i}$\lambda\lambda$5876, 6678} \\ 
10/08/60      &  & & Remaining  lines in absorption & \\
28/06/66      & & & Variability in H$\alpha$ profile, but & \\
27/07/66      & & & no global spectral evolution & \\[1.5mm]

$<$1969 & {\em blue end} & B1.5Ia & - & Schild et al. (\cite{schild}) \\[1.5mm]

21/04/65, 20/05/67 & 3500-4900{\AA} & B1 Ia$^+$ & P Cygni profiles in H$\beta$, $\gamma$  & Jaschek \& Jaschek \\
18/03/68, 14/08/70 & & & Variable He\,{\sc i}, N\,{\sc ii} \& O\,{\sc ii} phot. lines & (\cite{jj}) \\[1.5mm]

6/08/68 & {\em 5800-6700{\AA}} & - &  P Cygni profiles in  &  Rosendhal (\cite{rosendahl}) \\ 
& & & H$\alpha$, He\,{\sc i}$\lambda\lambda$5876, 6678 & \\[1.5mm]

16/05/73 to 10/07/73 & {\em blue end} & - & {\em No apparent variability in spec. type} & Hutchings et al. (\cite{hutchings76}) \\[1.5mm]

28/04/74 to 1/05/74 & 3800-4400{\AA} & {\em B1.5 Ia$^+$} & {\em No variability above 20\% in} & Walborn (\cite{walborn}) \\ 
28/04/75      &                &             & {\em line strength reported} & \\[1.5mm]

1974 Oct. \& 1975 Mar. & $\sim$H$\alpha$ & - & Variation in H$\alpha$ P Cygni profile & Dachs et al. (\cite{dachs}) 
\\[1.5mm]

26/06/72 to 14/08/75$^2$ & 3400-6700{\AA} &{\em B1Ia} & Variable P Cygni profiles in  & Sterken \& Wolf  \\
& & & H$\alpha$, $\beta$ \& He\,{\sc i}$\lambda\lambda$5876, 6678 & (\cite{sterken78}) \\
                         &                &           & No secular evolution in spec. type & \\[1.5mm]

$<$1984  & 4300-4900{\AA}& B1.5 Ia$^+$ & Variable P Cygni profiles in  &  Lopez \& Walsh  \\
         & 5500-6800{\AA}&        & H$\alpha$, $\beta$ \& He\,{\sc i}$\lambda$6678  & (\cite{lopez})\\[1.5mm]

1995$^4$    & 3450-8630{\AA} & B1.5 Ia$^+$ & Variable P Cygni profiles in & Rivinius et al. 
(\cite{rivinius}) \\ 
1990 to 1994$^3$ & 4000-6740{\AA} & & H$\alpha$, $\beta$, He\,{\sc i}$\lambda$ 6678 \& Fe\,{\sc iii} & \\
                 &                &        &  H$\gamma$, $\delta$ in absorption & \\
                 &                &        & No changes in He\,{\sc i}:Si\,{\sc iii} ratio & \\                
& & & \& hence spec. type & \\[1.5mm]

{\bf 1998 Oct. 7} & {\bf 4000-8950{\AA}} & {\bf B1.5 Ia$^+$} & {\bf H$\alpha$, $\beta$, He\,{\sc i}$\lambda$ 6678 P Cygni, higher} & 
{\bf This work} \\
& & & {\bf Balmer series in absorption} & \\[1.5mm]

{\bf 1999 June-July}$^5$  & {\bf 4000-8950{\AA}} & {\bf B1.5 Ia$^+$} & {\bf H$\alpha$, $\beta$, He\,{\sc i}$\lambda$ 6678 P Cygni, higher} & {\bf This work} 
\\
& & & {\bf Balmer series in absorption} & \\[1.5mm]

2002 March-May$^6$ & 5810-7205{\AA} & - & Variable H$\alpha$ P Cygni & Morel et al. (\cite{morel}) \\[1.5mm]

2003 May-June & 4000-7000{\AA} &B1.5 Ia$^+$ & - & Crowther et al. (\cite{pacBSG}) \\[1.5mm]

{\bf 2005 April 24} & {\bf 3800-6800{\AA}} & {B1.5 Ia$^+$} & {\bf H$\alpha$, $\beta$ \& He\,{\sc i}$\lambda$6678 P Cygni, higher}  & {\bf This work} \\
& & & {\bf Balmer series in absorption} & \\[1.5mm]

{\bf 2006 Feb. 16} & {\bf 3933-7985{\AA}} & {\bf B1.5 Ia$^+$} & {\bf H$\alpha$,$\beta$ \& He\,{\sc i}$\lambda$6678 P Cygni, higher } & {\bf This work} \\ 
& & & {\bf Balmer series in absorption} & \\[1.5mm]

{\bf 2009 March 06} & {\bf 3060-5600{\AA}} & {\bf B1.5 Ia$^+$} &  {\bf H$\beta$ P Cygni, higher Balmer} & {\bf This work} \\
& & & {\bf series in absorption} & \\[1.5mm]

\end{longtable}
{Note that a parameter is listed in italics if not explicitly given (wavelength of observation)
or the  data from which it is derived was not presented. Selected spectra between 1992-2009 are  plotted 
in Fig. A.1.
$^1$6 observations in this period,
$^{2}$44 observations  in this period,
$^3$233 observations in this period,
$^4$57 observations in this period,
$^5$14 observations in this period.
$^6$ 6 observations in this period.}
} 

\begin{table}
\begin{center}
\caption[]{Summary of the dates of historic spectral classifications of the Galactic BHGs.}
\begin{tabular}{lc}
\hline
\hline
Publication & Observation \\
     &          Date         \\  
\hline

\multicolumn{2}{c}{\bf HD~80077 (B2.5 Ia$^+$)} \\

Morgan et al. (\cite{morgan55}) & 1942-55$^a$ \\
Buscombe \& Kennedy (\cite{buscombe69}) & 1963-67$^b$ \\
Nordstrom (\cite{nord}) & $<$1975$^a$ \\
Moffat \& Fitzgerald (\cite{moffat}) & $<$1977$^a$ \\
Knoechel \& Moffat (\cite{knoechel}) & 1977$^{a,b}$ \\
Negueruela et al (in prep.) & 2008 \\
& \\
\multicolumn{2}{c}{\bf HD~169454 (B1 Ia$^+$)} \\
Morgan et al. (\cite{morgan55}) & 1942-55$^a$ \\
Hiltner (\cite{hiltner}) & $<$1956 \\
Code \& Houck (\cite{code}) & $<$1958 \\
Botto \& Hack (\cite{botto}) & $<$1962\\
Hutchings (\cite{hutchings70}) & 1968 \\
Wolf \& Stahl (\cite{wolf85})& 1972-4\\
Walborn (\cite{walborn80}) & 1973-5 \\
Hutchings (\cite{hutchings76}) & 1973-5 \\
Rivinius et al. (\cite{rivinius}) & 1992-5$^b$ \\
Hanson et al. (\cite{hansonK}) & 1994$^c$ \\
Hanson et al. (\cite{hansonH}) & 1997$^c$ \\
Groh et al. (\cite{groh}) & 2001-4$^a$ \\
& \\
\multicolumn{2}{c}{\bf HD~190603 (B1.5 Ia$^+$)}\\
Morgan et al. (\cite{morgan55}) & 1942-55 \\
Ahmad (\cite{ahmad}) & $<$1952 \\
Hiltner (\cite{hiltner}) & $<$1956 \\
Slettebak (\cite{slettebak}) & $<$1956 \\
Lesh (\cite{lesh}) & $<$1968 \\
Hutchings (\cite{hutchings70}) & 1968 \\
Walborn (\cite{walborn71}) & 1969 \\
Hutchings (\cite{hutchings76}) & 1973-5 \\
Bisiacchi et al. (\cite{bisiacchi}) & 1975-6 \\
Andrillat et al. (\cite{andrillat}) & 1988-93 \\
Rivinius et al. (\cite{rivinius}) & 1990-1$^b$ \\
Lennon et al. (\cite{lennon}) & 1990 \\
Hanson et al. (\cite{hansonK}) & 1994$^{c,d}$ \\
Blum et al. (\cite{blum}) & 1996$^{c,d}$ \\
Hanson et al. (\cite{hansonH}) & 1998$^{c}$ \\
Crowther et al. (\cite{pacBSG}) & 1992-2003 \\
Markova \& Puls (\cite{markova}) & $<$2007 \\ 
& \\
\multicolumn{2}{c}{\bf BP~Cru (B1 Ia$^+$)} \\
Vidal (\cite{vidal}) & 1973$^d$ \\
Bord (\cite{bord}) & 1974 \\
Hammerschlag-Hensberge et al. (\cite{HH}) & 1975 \\
         & 1977 \\
         & 1978 \\
Parkes et al. (\cite{parkes})&  1977-78$^d$ \\ 
Kaper et al. (\cite{kaper95}) & 1984 \\
Kaper et al. (\cite{kaper06}) & 1996$^b$ \\
                              & 2002$^b$ \\
\hline
\end{tabular}
\end{center}
{All classifications made in the optical(4-6000{\AA}) band unless otherwise noted -
$^a$marginally earlier spectral type of B2 Iae given.
$^b$multiple spectra in this period.
$^c$observations within the near-IR ($\sim$1-2.2$\mu$m) window.
$^d$marginally later spectral type of B1.5 or B2 Iae given.
}
\end{table}

\section{Spectropolarimetry}

Considering a simple `core--halo' wind model for heuristic purposes,
electron scattering of photospheric radiation in an asymmetric outflow
will generate a grey intrinsic linear polarization, while emission lines
formed in the wind will see a smaller scattering optical depth, and so
are expected to be less polarized (e.g., McLean \cite{mclean79}).  This will
result in depolarization through the line, although the addition of an
interstellar-polarization vector means that this `line effect' often
manifests in other ways, including an observed increase in degree of
polarization, $P$.

Previously unpublished spectropolarimetric observations of
Cyg~OB2~\#12 and $\zeta^1$~Sco were obtained as part of the
investigations reported by Harries, Howarth \& Evans (\cite{harries02}).
Cyg~OB2~\#12 was observed on 1995 July 17 (using the WHT with ISIS
spectrograph; $R \simeq 3300$), and $\zeta^1$~Sco on 1997 Jun 5 (AAT
with RGO spectrograph; $R \simeq 5000$), with the data acquisition and
reduction essentially as in the manner described by Harries \& Howarth
(\cite{harries96}).  Results are displayed in Fig.~B.1.

For Cyg~OB2~\#12 there is no compelling evidence of a change in
polarization through the H$\alpha$ emission line (although there is a
hint of a possible increase in $P$), and hence no strong indication of
large-scale asymmetry in the electron-scattering envelope.  Similarly,
there is no strong evidence for temporal polarimetric variability;
although the spread in published $R$-band photopolarimetric
measurements is somewhat larger than their quoted \textit{formal}
errors, the results presented by Schmidt et al. (\cite{schmidt92};
 $P =
7.893{\pm}0.037$\%, $\theta = 116.23 \pm 0.14^\circ$) Whittet et al.\
(\cite{whittet92}; $7.97{\pm}0.05$\%, 117$^\circ$), and Kobulnicky, Molnar \&
Jones (\cite{kobulnicky94}; $8.35{\pm}0.21$\%, $117.8^\circ$) are all broadly
consistent with the present results, as are earlier O-band
measurements reported by Kruszewski (\cite{kruszewski71}) and by 
McMillan \& Tapia
(\cite{mcmillan}).  We note, however, an anomalous result reported by Schulz \&
Lenzen (\cite{schulz}; $P = 10.05{\pm}0.12$\%, $\theta = 125^\circ$); their
observations of two further stars agree well with measurements
presented by Schmidt et al.\ (\cite{schmidt92}).

In spite of its brightness, $\zeta^1$~Sco has been much less
extensively observed, with the only available point of direct
comparison being the narrow-band ($\Delta\lambda = 8.5$\AA) H$\alpha$
polarimetric measurement of $P = 2.39\pm0.03$\%, $\theta = 64.6^\circ$
reported by McLean \& Clarke (\cite{mcleanclarke}).\footnote{Matthewson \& Ford
  (\cite{mathewson}) report a blue-band measurement of $P = 2.29$\%, $\theta =
  67.3^\circ$.)}  This is in agreement with our results; however, our
spectropolarimetry clearly reveals a significant,
$\sim1.5^\circ$ position-angle rotation through the
emission line.

Physically, the line effect may be associated with large-scale,
axisymmetric structures (such as might result from rapid rotation), or
irregular `clumps'.  For a star with parameters similar to those of
$\zeta^1$~Sco, summarized in Table~3, the critical equatorial
rotation velocity\footnote{ The velocity at which the outward
  centrifugal force equals the inward gravitational force.} is $v_{\rm
  e}\simeq{210}$~\kms.  With a measured \vesini\ of $\sim$45~\kms\
(Table~3), the intrinsic rotation is probably substantially
subcritical ($v_{\rm e}/v_{\rm crit} \lesssim 0.69$ with 95\%\
confidence).  A rotationally-induced axisymmetric departure from
spherical symmetry (i.e. a `disc' or `polar' wind) 
therefore seems \textit{a priori} improbable as the
origin of the line effect in this star. A more likely cause of
the observed change in polarization is the presence of transient large-scale
inhomogenities, or `clumps', as has been proposed for P~Cygni and AG~Car on the basis of 
spectropolarimetric observations by
e.g. Nordsieck et al. (\cite{nordsieck}) and  Davies et al.\ (\cite{davies});
indeed  CMFGEN model atmosphere analysis  of both these stars confirms 
the presence of the significant  wind clumping (Najarro et al. \cite{paco01}, 
Groh et al. \cite{agcar}).
Therefore, by direct analogy, an additional  observational test of this hypothesis for $\zeta^1$ Sco would  be the detection  of time-dependent variations
 in polarization (under the assumption that  there is no
prefered geometry for the clumping), noting that spectroscopic observations are already strongly indicative
of the presence of transient wind structure (Appendix A).

\begin{figure}
\begin{center}
\resizebox{\hsize}{!}{\includegraphics[angle=0]{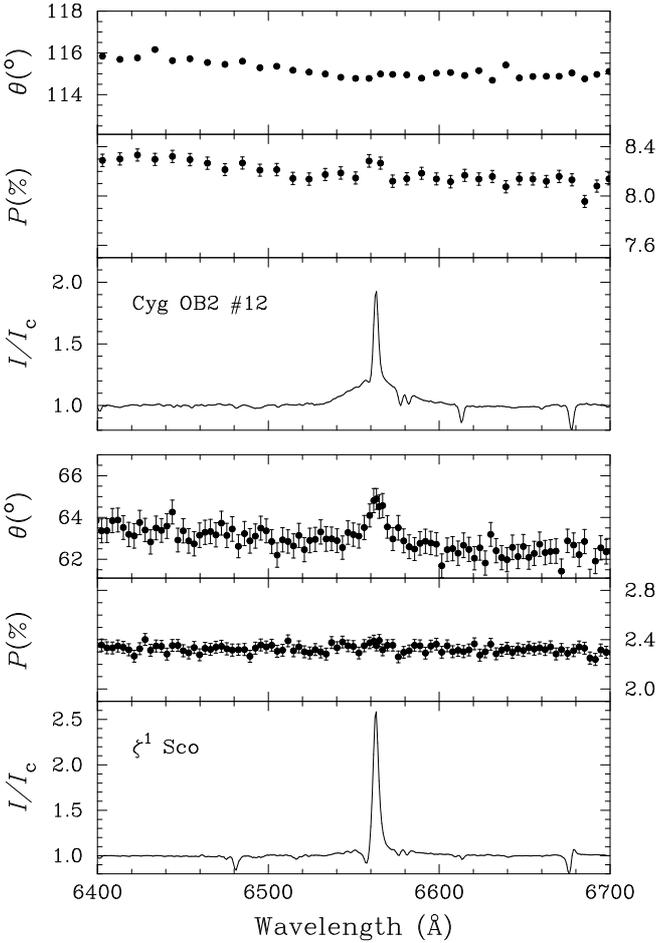}}
\caption{Linear spectropolarimetry of Cyg~OB2~\#12 and
$\zeta^1$~Sco.  Rectified spectra are shown at full resolution;
polarimetry is binned to an (almost) constant error of 0.05\%\ in
degree of polarization.}
\end{center}
\end{figure}

\section{The age of NGC~6231 and Sco~OB1}
Given its proximity ($\sim$1.64kpc; Sana et al. \cite{sana08}),
NGC~6231, its host association Sco~OB1, and the massive stars located
within both have been the subject of numerous multiwavelength studies.
Here we review these data in order to constrain the evolutionary
history of both cluster and association to better understand the
nature of $\zeta^1$~Sco.  Photometric studies by Baume et
al. (\cite{baume}) suggest a significant age spread within NGC6231,
with star formation apparently commencing $\sim 10$~Myr ago,
culminating in the formation of the massive stellar cohort
$\sim3-5$~Myr ago. The latter age is consistent with the findings of
Sana et al. (\cite{sana06}, \cite{sana07}), who report a similar value
of $2-4$~Myr from an analysis of the photometric data of (X-ray
selected) pre-MS stars. An analagous study of the wider Sco OB1
association by Perry et al. (\cite{per}) supports a non coeval star
formation history for this region, yielding an age of
$\sim 8^{+4.5}_{-3}$~Myr.

Catalogues of the OB stellar content of both NGC~6231 and Sco~OB1 are
provided by Sana et al. (\cite{sana06b}) and Ankay et
al. (\cite{ankay}) respectively, and allow for the individual
placement of stars on the HR diagram presented in Fig. 13. For those
stars in Sana et al. we adopted the luminosities given by these
authors and utilised the spectral type/temperature relation of Martins
et al. (\cite{martins05}). The temperatures and luminosities of the
stars in Ankay et al.  were again determined via the callibrations of
Martins et al., with reddening for individual stars calculated via the
intrinsic colours of Martins \& Plez (\cite{martins06b}).  Finally,
given their rather evolved nature, we adopted the results of the
tailored non-LTE analysis of Crowther \& Evans (\cite{pac6231}) for
HD~151804 and 152408.

Comparison of these data to non-rotating and rotating Geneva
isochrones clearly indicate that the region {\em as a whole} appears
non coeval, at first glance being consistent with a spread of ages of
between $2.5-5$~Myr.  Such a conclusion is supported by inspection of
the properties of individual stars in both association and cluster,
although interpretation is complicated by possible contamination of
the latter by the former if, as seems likely, Sco~OB1 hosts a younger
population than is found in NGC~6231.

A large number of Main Sequence (MS) stars are found within NGC~6231,
with the O8~V companion in the binary HD~152234 apparently being the
earliest and defining the MS turnoff; consistent with an age of
5~Myr. While an O6~V companion to the WC7 star WR79 has been reported
(Hill et al. \cite{hill}) the 126~day binary period of HD~152234 will
have ensured that neither component in this system will have
interacted, while the 8.89~day period of WR79 indicates that
significant mass transfer to the secondary may have occured
(e.g. Petrovic et al. \cite{pet}).  Encouragingly, the O9.7Ia primary
of HD152234 also lies on the 5~Myr isochrone for rapid initial
rotation, with the {\em current v}sin{\em i} also being consistent
with this placement (Fraser et al. \cite{fraser}).

Building on this approach, we find that HD~152219 (O9.5 III + B1-2
III-V) and HD~152134 (B0.5 Ia) lie upon the 5~Myr evolutionary track
for non-rotating stars, with - sequentially - HD~326331 (O8 III((f))),
HD~152247 (O9 III), HD~152249 (O9 Ib((f))) and HD~152234 (O9.7 Ia + O8
V) following the rotating track. In both cases the systematic
progression to later spectral types with increasing luminosity class
suggest these are real evolutionary sequences (cf. Cyg~OB2;
Negueruela et al.  \cite{iggy}).

However, there are a number of stars within NGC~6231 and Sco~OB1 that
appear incompatable with a $\sim 5$~Myr population: the O8 Iafpe/WN9ha
stars HD~151804 and 152048 and the O5.5-7.5 III stars HD~151515, 152723,
152233 and 152248. Regarding the former pair, Bohannan \& Crowther
(\cite{bohannan}) suggest a close physical kinship between these
objects, apparently precluding an evolution through the cooler
late-O/early-BSG phase present in NGC~6231.  Of the mid-O giants,
HD~151515 and 152723 appear somewhat subluminous for their spectral type
but the properties of the O5.5 III(f) + O7.5 and O7.5 III(f) + O7
III(f) binaries HD~152233 and 152248 (Sana et al. \cite{sana08},
\cite{sana01}) clearly support the assertion that a younger population
is present; although the dymanical masses for both components of the
latter system appear somewhat lower than expected ($29.6M_{\odot}$ and
$29.9M_{\odot}$ respectively).

We therefore conclude that comparison of NGC~6231 to Sco~OB1 {\em
  implies} that the respective stellar populations are non coeval.
However, with the exception of HD~152233 and HD~152248, the massive
stellar population of NGC6231 appears consistent with a single burst
of star formation $\sim$5~Myr ago. Indeed, no early-mid O MS stars
consistent with a younger population appears present within NGC~6231,
while the population of late-O/early-B MS stars that are present are
systematically displaced redwards from the $\leq$3~Myr
isochrones. Therefore, if they {\em are} members of NGC~6231, we cannot
exclude the possibility that, with spectral types $<$O8, HD~152233 and
HD~152248 are in fact {\em bona fide} blue stragglers; having evolved
via a different pathway from the majority of stars and hence that
NGC~6231 is truly coeval. In this respect it would therefore closely
resemble Cyg OB2, for which Negueruela et al. (\cite{iggy}) arrived at
a comparable conclusion.

\section{Spectral fits to $\zeta^1$~Sco and HD190603}
\label{app-spec-fits}
We present additional model fits to $\zeta^1$~Sco based on the FEROS
(3700-8850{\AA}) optical data (not
presented in the paper version due to reasons of space).

\begin{figure*}
\resizebox{\hsize}{!}{\includegraphics[angle=0]{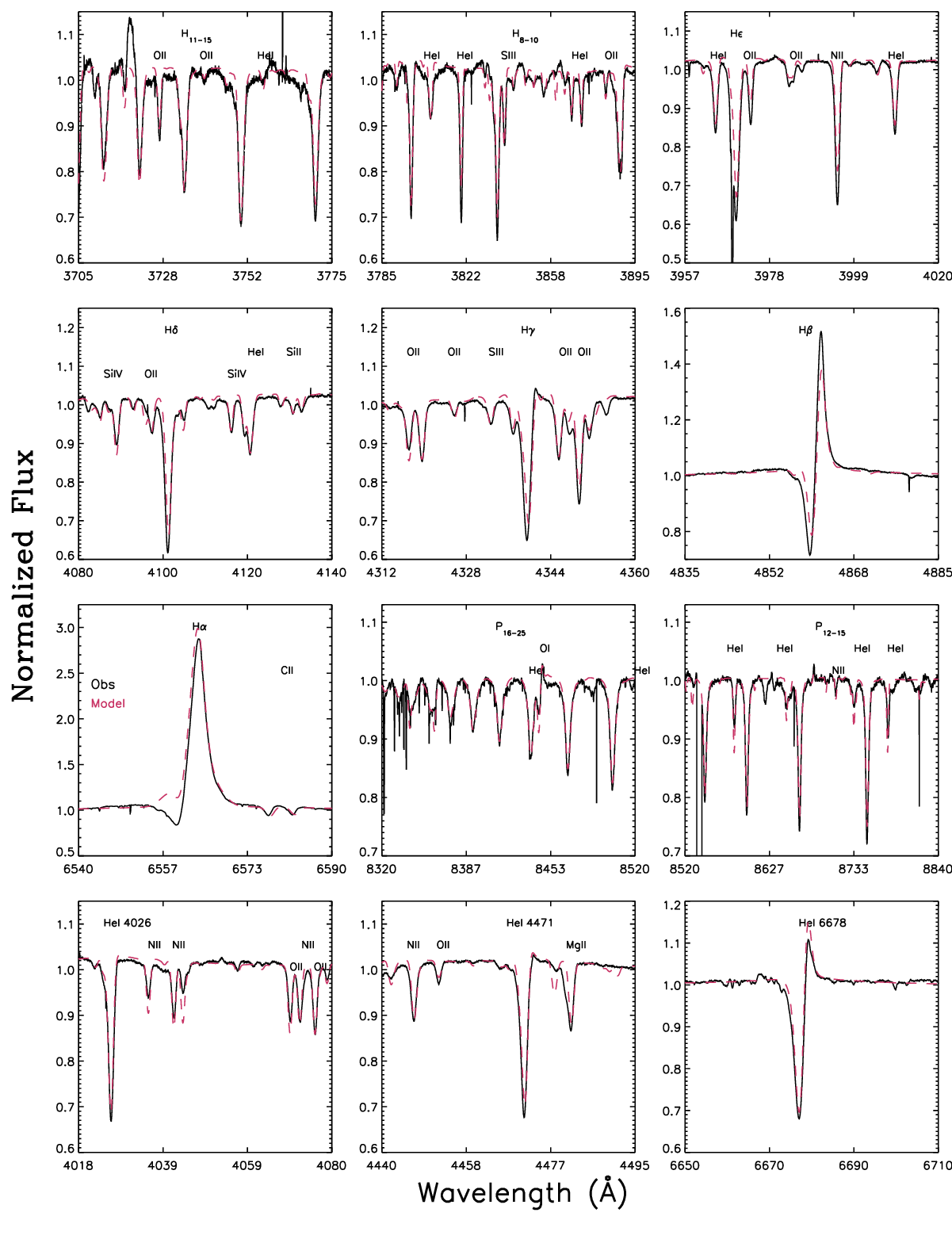}}
\caption{Comparison of the synthetic spectrum of $\zeta^1$~Sco (red
  dashed line) to observational data for various transitions between
  4090-22100{\AA}. Optical data correspond to the FEROS data kindly
  provided by Otmar Stahl. Note the lack of long-termvariability
  between these data and those presented in Fig. 7.}
\label{fit-zetopspec-stahl1}
\end{figure*}

\begin{figure*}
\resizebox{\hsize}{!}{\includegraphics[angle=0]{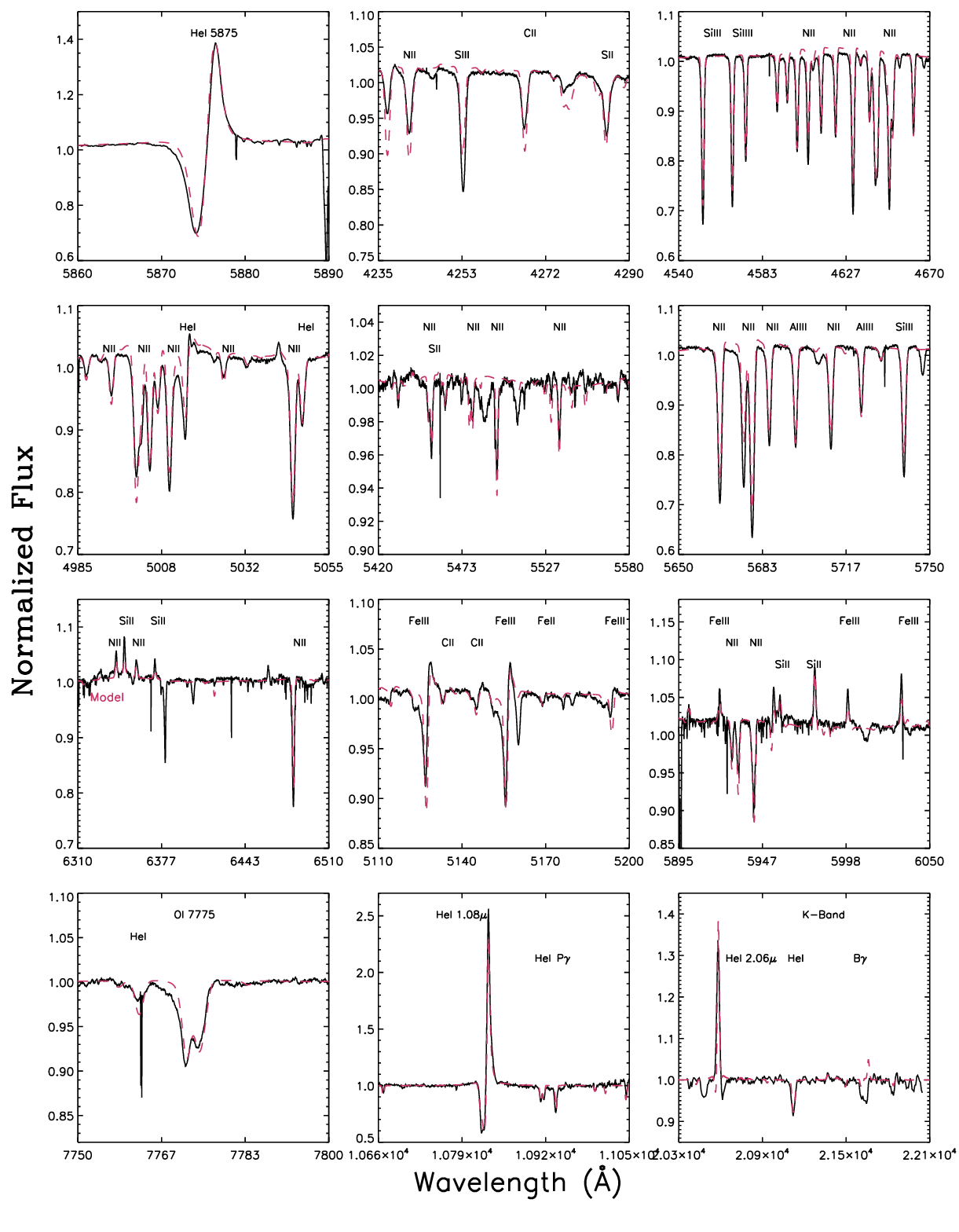}}
\caption{Same as Fig.\ref{fit-zetopspec-stahl1}, Cont.}
\label{fit-zetopspec-stahl2}
\end{figure*}

\end{document}